\documentclass{aastex}
\usepackage{emulateapj5}

\usepackage{epsfig}
\usepackage{amsmath}
\slugcomment{}
\shorttitle{}
\shortauthors{}

\begin{document}

\title{The Luminosity Function of Early-type Field Galaxies at $z\approx0.75$}

\author{
N.J.G. Cross\altaffilmark{1},
R.J. Bouwens\altaffilmark{2},
N. Ben\'{\i}tez\altaffilmark{1},
J.P. Blakeslee\altaffilmark{1},
F. Menanteau\altaffilmark{1},
H.C. Ford\altaffilmark{1},
T. Goto\altaffilmark{1},
B. Holden\altaffilmark{2},
A.R. Martel\altaffilmark{1},
A. Zirm\altaffilmark{3}, 
R. Overzier\altaffilmark{3},
C. Gronwall\altaffilmark{4},
N. Homeier\altaffilmark{1},
D.R. Ardila\altaffilmark{1},
F. Bartko\altaffilmark{5}, 
T.J. Broadhurst\altaffilmark{6},
R.A. Brown\altaffilmark{7},  
C.J. Burrows\altaffilmark{7},
E.S. Cheng\altaffilmark{8},
M. Clampin\altaffilmark{8},  
P.D. Feldman\altaffilmark{1},
M. Franx\altaffilmark{3},
D.A. Golimowski\altaffilmark{1},
G.F. Hartig\altaffilmark{7},  
G.D. Illingworth\altaffilmark{2},
L. Infante\altaffilmark{9}  
R.A. Kimble\altaffilmark{8},
J.E. Krist\altaffilmark{7},
M.P. Lesser\altaffilmark{10},
G.R. Meurer\altaffilmark{1},
G.K. Miley\altaffilmark{3},
M. Postman\altaffilmark{1,7},
P. Rosati\altaffilmark{11},  
M. Sirianni\altaffilmark{7}, 
W.B. Sparks\altaffilmark{7}, 
H.D. Tran\altaffilmark{12},
Z.I. Tsvetanov\altaffilmark{13},  
R.L. White\altaffilmark{1,7}
\& W. Zheng\altaffilmark{1}
}


\altaffiltext{1}{Department of Physics and Astronomy, Johns Hopkins
University, 3400 North Charles Street, Baltimore, MD 21218.}

\altaffiltext{2}{UCO/Lick Observatory, University of California, Santa
Cruz, CA 95064.}

\altaffiltext{3}{Leiden Observatory, Postbus 9513, 2300 RA Leiden,
Netherlands.}

\altaffiltext{4}{Department of Astronomy and Astrophysics, The
Pennsylvania State University, 525 Davey Lab, University Park, PA
16802.}

\altaffiltext{5}{Bartko Science \& Technology, P.O. Box 670, Mead, CO
80542-0670.}

\altaffiltext{6}{Racah Institute of Physics, The Hebrew University,
Jerusalem, Israel 91904.}

\altaffiltext{7}{STScI, 3700 San Martin Drive, Baltimore, MD 21218.}

\altaffiltext{8}{NASA Goddard Space Flight Center, Laboratory for
Astronomy and Solar Physics, Greenbelt, MD 20771.}

\altaffiltext{9}{Departmento de Astronom\'{\i}a y Astrof\'{\i}sica,
Pontificia Universidad Cat\'{\o}lica de Chile, Casilla 306, Santiago
22, Chile.}

\altaffiltext{10}{Steward Observatory, University of Arizona, Tucson,
AZ 85721.}

\altaffiltext{11}{European Southern Observatory,
Karl-Schwarzschild-Strasse 2, D-85748 Garching, Germany.}

\altaffiltext{12}{W. M. Keck Observatory, 65-1120 Mamalahoa Highway,
Kamuela, Hawaii 96743.}

\altaffiltext{13}{NASA Headquarters, Washington, DC 20546-0001.}

\begin{abstract}

We measure the luminosity function of morphologically selected E/S0
galaxies from $z=0.5$ to $z=1.0$ using deep high
resolution Advanced Camera for Surveys imaging data.  Our analysis
covers an area of $48\Box\arcmin$ (8$\times$ the area of the HDF-N)
and extends 2 magnitudes deeper ($I\sim24$ mag) than was possible in
the Deep Groth Strip Survey (DGSS). Our fields were observed as
part of the ACS Guaranteed Time Observations. At $0.5<z<0.75$, we find
$M_B^*-5\log\,h_{0.7}=-21.1\pm0.3$ and $\alpha=-0.53\pm0.2$, and at
$0.75<z<1.0$, we find $M_B^*-5\log\,h_{0.7}=-21.4\pm0.2$, consistent
with 0.3 magnitudes of luminosity evolution (from $0.5<z<0.75$). 
These luminosity functions are
similar in both shape and number density to the luminosity function
using morphological selection (e.g., DGSS), but are much steeper than
the luminosity functions of samples selected using morphological
proxies like the color or spectral energy distribution (e.g., CFRS,
CADIS, or COMBO-17).  The difference is due to the `blue',
$(U-V)_0<1.7$, E/S0 galaxies, which make up to $\sim30\%$ of
the sample at all magnitudes and an increasing proportion of faint
galaxies.  We thereby demonstrate the need for {\it both morphological
and structural information} to constrain the evolution of galaxies.

We find that the `blue' E/S0 galaxies have the same average sizes and
Sersic parameters as the `red', $(U-V)_0>1.7$, E/S0 galaxies at
brighter luminosities ($M_B<-20.1$), but are increasingly different at
fainter magnitudes where `blue' galaxies are both smaller and have
lower Sersic parameters. We find differences in both the
size-magnitude relation and the photometric plane offset for `red' and
`blue' E/S0s, although neither `red' nor `blue' galaxies give a good
fit to the size magnitude relation. Fits of the colors to stellar
population models suggest that most E/S0 galaxies have short
star-formation time scales ($\tau<1$ Gyr), and that galaxies have
formed at an increasing rate from $z\sim8$ until $z\sim2$ after which
there has been a gradual decline.

\end{abstract}


\section{Introduction}

The luminosity function of galaxies is the number density of galaxies as
a function of absolute magnitude. The shape of the luminosity function 
can be used to constrain galaxy formation models. The luminosity function
is often described by three numbers: $M^*$, the magnitude at which the 
number of bright galaxies rapidly decreases; $\phi^*$, the space density 
at $M^*$, and the faint end slope $\alpha$ which 
characterizes the ratio of dwarf galaxies to giant galaxies. Models of galaxy 
formation and evolution must be able to account for these parameters, which 
vary with galaxy type. Over the past few years, the luminosity function of 
high redshift ($z>0.5$) galaxies have been studied extensively through the 
use of deep, wide-area surveys.  Some of the more notable efforts include 
the Canada-France Redshift Survey (CFRS, Lilly et al. 1995), the Canadian 
Network for Observational Cosmology Field Galaxy Redshift Survey (CNOC2, Lin 
et al 1999), the Calar Alto Deep Imaging Survey (CADIS, Fried et al. 2001), 
the Deep Groth Strip Survey (DGSS, Im et al. 2002), the Subaru Deep Survey 
(Kashikawa et al. 2003), the 
Classifying Objects by Medium Band Observations (COMBO-17, Wolf et al. 
2003) and from a combination of Hubble Space Telescope (HST) and 
Very Large Telescope (VLT) images, Poli et al. (2003). Most of these 
use deep, ground-based images with spectroscopic or photometric redshifts to 
construct the luminosity function, but do not have the spatial resolution to 
measure the structural properties of galaxies at higher redshifts. 

Without information on the structural properties, ground-based surveys have 
resorted to using color information as a proxy for morphologies, whether this 
information comes in the form of a best-fit spectral energy distribution 
(e.g. Wolf et al. 2003), or a rest-frame color cut (e.g. Lilly et al. 1995). 
This 
can result in apparent discrepant results.  For example, Wolf et al. 
(2003) found that the elliptical/S0 (E/S0) galaxies that produce $\sim50\%$ 
cent of the current B-band luminosity density only contributed $\sim5\%$ 
at $z=1$. By contrast, using morphological classification, van den Bergh 
(2001) found that the fraction of elliptical galaxies has remained constant 
at $\sim17\%$, $0.25<z<1.2$, implying that either the luminosity of 
ellipticals has increased over time relative to other types of galaxies or 
that the differences in color-selection and morphological-selection have 
produced apparently inconsistent results between these surveys. 

Surveys using the Hubble Space Telescope (HST) such as the DGSS 
(Im et al. 2002; Simard et al. 2002) and the Medium Deep Survey 
(Griffiths et al. 1994) have been able to reliably morphologically classify 
and measure structural parameters for galaxies with $I_{AB}<22$ mag, but 
over much smaller areas of sky than the deep ground based surveys. 
These HST surveys have discovered a 
population of $0.3<z<1$ blue E/S0 galaxies (e.g. Menanteau et al. 1999, Im 
et al. 2001, Gebhardt et al. 2003) that have similar luminosities 
to standard red E/S0 galaxies. Im et al. (2001) find these make up $\sim15\%$
of the E/S0 sample whereas Menanteau et al. (1999) find a much 
higher fraction: $30-50\%$ of the sample. Objects such as these 
demonstrate the inherent weakness of using  color as a proxy for morphology. 
At low redshifts, almost all of the bright E/S0 
galaxies are red, with blue ellipticals (dwarf ellipticals) many magnitudes
fainter. From the work of Menanteau et al. (2004) and Im et al. 
(2001) it appears that most of these blue E/S0 galaxies have blue cores
and red exteriors, with the exteriors having the same colors as red E/S0 
galaxies, which have constant colors at all radii. Im et al. (2001) concluded that 
these blue E/S0 galaxies were less massive than the red E/S0 galaxies
based on the dynamical masses calculated from the velocity dispersions. 
However, because 
the velocity dispersions were measured much closer to the core of the galaxy for the 
low redshift red ellipticals, the high redshift blue ellipticals may be
more massive than the measurements suggest. Even if the measurements 
give accurate dynamical masses, the blue E/S0 galaxies have masses equivalent 
to the lower mass red E/S0s, so they may still yet evolve into high mass red E/S0s
through a combination of luminosity evolution that reddens the stellar population 
over time and mergers that increase the mass.

Luminosity evolution occurs when there is new star-formation, or when the 
stellar population ages, and does not necessarily imply any change in the
mass or number of stars in a galaxy. Structural parameters such as the size 
and shape are better indicators of the morphological evolution, since they 
are only weakly dependent on the age of the stellar population and are mainly
determined by dynamical characteristics such as total mass and angular 
momentum. Within the half-light radius of a giant elliptical galaxy the 
dynamical time-scale is very short, less than $10^8$ years, so dynamical 
equilibrium is reached very quickly. The size and shape of the galaxy will not 
change significantly unless mass is added via mergers or
accretion; a close encounter changes the angular momentum; tidal forces
disrupt the outer layers. Small changes in the apparent shape and size do occur when 
star-formation is localized in the center, in bars, rings or spiral arms, but these are 
much weaker changes than the variation in SED or color. Therefore morphology is a more
robust indicator of the nature of a galaxy, but it requires good resolution to
use.

Previous studies have differed in the way they have utilized size 
information to make inferences about evolution. Several surveys have assumed 
that galaxy size and shape are constant with redshift. 
Schade, Barrientos \& Lopez-Cruz (1997) showed that cluster ellipticals 
evolve as $\Delta\,M=-2.85\log_{10}(1+z)$, assuming they maintain a 
constant size, and Schade et al. (1999) demonstrated that field ellipticals show a 
similar evolution. Using a sample of 44 galaxies with $z<2$ Roche et al. 
(1998) discovered that ellipticals show significant luminosity evolution but 
little size evolution from $z=1.0$ to $z=0.2$. They found that most size 
evolution appears to happen at $z>1.5$. Graham (2002) compared the scatter 
in the `photometric plane' which only requires parameters measured from
galaxy images, to the scatter in the `fundamental plane' which requires
dispersion velocities measured from high resolution spectra. Graham showed 
that the photometric plane could be used to constrain distances to elliptical 
galaxies.

In Cross et al. (2001) and Cross \& Driver (2002), the
effects of surface brightness selection on the $z=0$ galaxy luminosity
function were discussed. In this paper we look at the LF of morphological early types at $0.5<z\le1$.
We then examine the effect that color selection has on this luminosity function. Finally, we use structural parameters to
test whether blue E/S0 galaxies are progenitors of red E/S0 galaxies and
what evolution has taken place from $z=1$ to $z=0.5$. 

The Advanced Camera for Surveys (ACS, Ford et al. 2002) significantly improves on WFPC2 in 
terms of sensitivity (a factor of 5), field of view (a factor of 2) and 
resolution (a factor of 2), giving well sampled PSFs in the i and z bands. 
This leads to significant improvements in both the accuracy of the size 
measurements and the overall sample size.

In this paper we use data from 5 fields observed as part of the ACS GTO program.
The total area is over 8 times the HDFN.
These fields were selected to observe very nearby ($z<0.03$) galaxies or
very distant ($z>4$) galaxies, so galaxies in the redshift range $0.5<z<1.0$ should 
be representative of the universe at that redshift. The fields are in 
various parts of the sky, sampling a large volume in each redshift range 
($\sim1.6\times10^4$ Mpc$^3$ $0.5<z<0.75$ and $\sim2.4\times10^4$ Mpc$^3$ 
$0.75<z<1.0$) so the effects of cosmic variance should be much 
smaller than in the Hubble Deep Fields. In fact, the relative independence 
of our fields makes this survey more competitive with larger surveys 
than one might think based upon the areal coverage alone. We express all 
magnitudes in the AB system and use a $\Omega_M=0.3$, $\Lambda=0.7$ cosmology 
with $H_0=70$ km s$^{-1}$ Mpc$^{-1}$. We define $h_{0.7}=H_0/70.$

\section{Data}

The data were extracted from 5 fields observed by the ACS Wide Field Camera 
(WFC) between April 2002 and June 2003. The fields were selected to give 
accurate photometric redshifts (3 or more filters), to not have any primary 
targets in the range $0.5<z<1.0$ and to not contain any clusters at lower 
redshifts. While the Hubble Deep Field North (HDFN) was only imaged in two 
ACS bands (F775W and F850LP), it has been imaged extensively in 7 optical and 
near infrared bands and has a large amount of spectroscopic follow-up. The 
combined area of these fields is $47.9\Box\arcmin$, over 8 times the area of the 
Hubble Deep Field North. The extinction values, E(B-V), are taken from the 
Schlegel, Finkbeiner \& Davis (1998) dust maps, and the total extinction in each 
filter, $A({\rm filter})$,  is calculated using the method described in 
Schlegel, Finkbeiner \& Davis (1998). A summary of the data properties in 
each field is given in Table~\ref{tab:fields} which lists the ACS filters, 
field-of-view, I-band exposure time, E(B-V), I-band extinction, I-band 
zeropoint and the number of E/S0 galaxies in our sample.  

\subsection{NGC 4676}

NGC 4676 is a low redshift pair of merging spiral galaxies and was observed 
as part of the ACS ``Early Release Observations'' (ERO) program (Ford et al. 
2002). We mask out NGC 4676 and use galaxies in the background field. It was 
observed for 6740s in the F475W (g) filter, 4000s in the F606W (V) filter and 
4070s in the F814W (I) filter. The area remaining after masking out the two 
prominent foreground galaxies is $7.8\Box\arcmin$.
 
\subsection{UGC 10214}

UGC 10214 is a low redshift spiral galaxy that is merging with a much smaller
dwarf galaxy and has an extended tidal tail as a result (Tran et al. 2003). 
As with NGC 4676 it was selected as part of the ERO program. We mask out UGC 
10214 and use galaxies in the background field (see Ben\'{\i}tez et al. 
2004). It was observed in 2 separate pointings giving a combined exposure of 
13600s in F475W (g), 8040s in F606W (V) and 8180s in F814W (I). The area 
remaining after masking out the prominent foreground galaxy is 
$10.7\Box\arcmin$.
 
\subsection{TN1338}

TN J1338 $-$1942 (TN1338) is a radio galaxy at $z=4.1$ that was observed as part
of our ACS/GTO program to study proto-clusters around high-redshift radio galaxies
(see Miley et al. 2004, Overzier et al., in prep). It was observed for 9400s in 
F475W (g), 9400s in F625W (r), 11700s in F775W (i) and 11800s in F850LP (z). The total 
observed area is $11.7\Box\arcmin$. 

\subsection{TN0924}

TN J0924 $-$2201 (TN0924),a radio galaxy at $z=5.2$, was also observed as part of the
high-redshift radio galaxy proto-cluster program (Overzier et al., in prep). It was 
observed for 9400s in F606W (V), 11800s in F775W (i) and 11800s in F850LP (z). The total 
observed area is $11.7\Box\arcmin$.

\subsection{HDFN}

The Hubble Deep Field North (HDFN) was observed with the ACS to find 
supernovae and test the ACS Grism (Blakeslee et al. 2003a).  It was observed 
for 5600s in the F775W (i) filter and 10300s in the F850LP (z) filter. 
We use the ACS i-band for measurements of the structural parameters, but we do 
not have enough ACS filters for accurate photometric redshifts. However there
is a deep 7-filter data available for the portion of the ACS image already
observed by WFPC2 (Williams et al. 1996). We use the photometric catalog from 
Fern\'andez-Soto, Lanzetta \& Yahil (1999, FLY99), which has very deep 
F300W (U), F450W (B), F606W (V), F814W (I) WFPC2 and Kitt Peak National 
Observatory (KPNO) J,H,K band
photometry. There are 146 spectroscopic redshifts from Cohen et al. 
(2000). We only use ACS data coincident with the deep WFPC2 image and take 
our photometric redshifts and colors from the FLY99 data. The observed area 
is $5.8\Box\arcmin$.

\subsection{Catalogs}
\label{sec:cat}

Each set of images was run through the ACS Science Data Analysis Pipeline 
(Blakeslee et al. 2003b). The data in each field were selected from 
the detection images produced from combining the filter images, weighted
by the inverse noise squared. This aids 
in the detection of extremely faint objects by combining the signal from the 
different filters to produce a more significant detection. Source Extractor 
(Bertin \& Arnouts 1996) was run first on the detection image and then in 
dual mode on the detection image and each filter image, to produce catalogs 
of the same objects, with photometry in matched apertures.
We use these source catalogs as the starting point for selecting our 
sample and measuring the photometric properties. 

\section{Measurements}

\subsection{Photometric Redshifts}
We use the Bayesian Photometric Redshift code (BPZ, Ben\'{\i}tez 2000) to 
calculate the photometric redshifts of galaxies in the fields of NGC 4676, 
UGC 10214, TN1338 and TN0924. This takes advantage of both the color 
information and a 
magnitude prior to constrain the redshift. The magnitude prior distinguishes
nearby red galaxies (e.g. giant ellipticals) from distant, redshifted 
blue galaxies, which while having similar colors when seen through
a small set of filters, will have very different magnitudes. We use the 
template spectra described in Ben\'{\i}tez et al. (2004), which 
are based upon a subset of the templates from Coleman, Wu \& Weedman (1980) 
and Kinney et al. (1996). The template set is: `El', `Sbc', `Scd', `Im',
`SB3' and `SB2'. These represent the typical
spectral energy distributions (SED) of elliptical, early/intermediate type spiral,
late type spiral, irregular and two types of starburst galaxies.
These templates have been modeled using Chebyshev 
polynomials to remove differences between the predicted colors and those of 
real galaxies. The final ``calibrated'' templates have been found to give better BPZ 
results on the HDFN (Ben\'{\i}tez et al. 2004). We use 
extinction-corrected isophotal magnitudes to maximize the signal-to-noise on the 
color input to BPZ. In each case, the aperture is the same for each filter. 
The magnitude prior is based on the Hubble Deep Field North 
database (Williams et al. 1996) which uses deep ($\sim 27$ mag arcsec$^{-2}$)
isophotal magnitudes. 

\subsection{Testing BPZ}

To test our photometric redshift catalogs for completeness,
contamination, and systematic and random errors we compare them to 
spectroscopic data in the HDFN and to simulations. Fig.~\ref{fig:filtspec} 
shows the spectral energy distribution of an elliptical galaxy against 
the throughput of the filters used. The lower panel shows the HDFN filter 
set, consisting of the UBVI WFPC2 filters and the JHK KPNO filters. The 
`El' SED is plotted 3 times, at $z=0.5$ (dotted line), at $z=0.75$ (short dashed line), 
and at $z=1.0$ (long dashed line). The main feature of this 
spectrum is the $4000{\rm \AA}$ break, which is indicated by the bold arrow at 
each of these redshifts. The $4000\,{\rm \AA}$ break is prominent in galaxies where 
there is very little ultraviolet radiation produced by hot, young stars,
compared to the optical flux produced by an older stellar population.
This break falls within the V or I filters at every redshift in the range 
that we use. The drop in flux per wavelength from one side of the break to 
the other side produces a significant change in magnitude from one filter to 
the next, leading to an accurate measurement of the photometric redshift.

The lower-middle panel shows the same plot for the ACS g, V and I filters used
in the UGC10214 and NGC4676 fields. The upper-middle panel shows the g,r,i 
and z filters used in the TN1338 field. The top panel shows the V,i and 
z filters used in the TN0924 field.

We use the HDFN photometric and spectroscopic redshifts to estimate 
the errors for 3-color BPZ measurements of real galaxies seen through the WFPC2 
filters and then use simulations to determine any biases in the BPZ 
measurements through ACS filters at the noise limits of our data. 
The g, V and I filters used in the UGC10214 and NGC4676 fields are similar 
in wavelength coverage to the B, V and I filters used in the HDFN dataset. 
Therefore we can test the accuracy of the photometric redshifts in these 
fields by calculating 3-color photometric redshifts for ellipticals
in the HDFN. In the upper panel of Fig.~\ref{fig:BPZ_hdfn}, we plot the 3-color
photometric redshifts calculated using the B, V and I filters against the 7 color 
photometric redshifts. The offset, $\frac{z_{3BPZ}-z_{7BPZ}}{1+z_{7BPZ}}=0.010\pm0.074$, 
is low and there are no outliers. We calibrate the 7-color photometric 
redshift to the spectroscopic sample and find a deviation 
$\frac{z_{7BPZ}-z_{spec}}{1+z_{spec}}=-0.045\pm0.026$, shown in the middle  
panel of Fig.~\ref{fig:BPZ_hdfn}. There is one outlier, a galaxy with
$z_{BPZ}=0.87$ and $z_{spec}=0.67$. As expected from the poor fit, this object has $(V-I)$ colors which are much redder and $(B-V)$ colors which are 
slightly bluer than one would expect for an elliptical galaxy at this 
redshift. The bottom panel shows the 3-color 
photometric redshifts corrected for this offset. The correction is described 
at the end of this section. The quoted error in the
above cases and for future BPZ measurements is for a single galaxy, so this offset is
significant.  Cohen et al. (2000) show 
that the errors in the spectroscopic data are $\Delta\,v=200$ km s$^{-1}$, implying $\Delta\,z=0.0007$. The final error is consistent with the typical 
scatter found in the overall analysis of all HDF redshifts
($\Delta\,z/(1+z)=0.06$). The offset between BPZ and spectroscopic
redshifts, implies some evolution in elliptical galaxies from $z=0.2$ (the redshift of 
the calibration cluster) and $z\sim0.75$. 

Given that all of the HDFN ellipticals have good 3-band photometric redshifts,
we expect that ellipticals in NGC4676 and UGC10214 should also have good
photometric redshifts. However, the noise in these fields are somewhat 
greater than the HDFN, so there may be some missing objects.

We test the reliability of BPZ in each of the fields using Bouwens' Universe 
Construction Set (BUCS, Bouwens, Magee \& Illingworth, in preparation; Bouwens, Broadhurst \& Illingworth 2003; Bouwens et al. 2004) simulations of 
$r^{1/4}$ elliptical galaxies with three different SEDs: `El', `Sbc' and `Scd' (Ben\'{\i}tez et al.\ 2004).
These simulations are designed to have the same noise characteristics as the observed ACS datasets and are processed in the same way as the data (\S~\ref{sec:cat}). Therefore, the UGC 10214 simulation, with double the exposure time, 
has $1.4\times$ the signal-to-noise of the NGC 4676 simulation.
We use the 3 SEDs to test the reliability of redshifts for early-type galaxies with a range of colors. All the simulations are made up of galaxies with 
elliptical morphologies ($\beta=4$) and a Schechter luminosity function with 
parameters $\phi^*=0.00475$, $M^*=-20.87$ and $\alpha=-0.48$. The density of 
galaxies was increased by a factor of 5 over the normal elliptical galaxy density to give a 
large sample of galaxies at each redshift. In these simulations elliptical galaxies are 
placed at random in 4 fields, each $2000\times2000$ pixels. Each of these fields is 
approximately the area of a single amplifier on the Wide Field Camera. 

Once the images had been processed 
we compared the simulation input catalog and the catalog of detected 
objects. The results are shown in Fig.~\ref{fig:sims1}. In each of the fields 
we find small differences between the measured redshift and the input 
redshift. The only major differences occur in the NGC4676 and UGC 10214 simulations,
in the $z_{\rm simulation}=0.95$ bin. In both cases $z_{\rm detection}$ is over 
estimated. Fig.~\ref{fig:filtspec} shows that at this redshift, the $4000{\rm \AA}$ break
is in the middle of the F814W filter with no redder filter to compare to. This is
also the redshift range at which there is increased scatter in 3-band photometric
redshifts in the HDFN, which had a similar combination of filters. The offsets are 
due to the increased scatter and are not a systematic effect. We find that the TN1338
simulation has a mean scatter $\sigma_z=0.023$, TN0924 has $\sigma_z=0.028$, NGC 4676 
has $\sigma_z=0.045$ and UGC 10214 has $\sigma_z=0.046$. Since the HDFN has similar filters
to NGC 4676 and UGC10214 and is deeper, we would expect $\sigma_z$ to be lower. The 
additional noise is due to the real galaxy spectral energy distributions varying from 
the ideal templates used in our simulations. There is a large increase in the scatter
for all galaxy types in the HDFN, UGC 10214 and NGC 4676 fields at $z>0.85$, with the 
rms in the HDFN increasing from $\sigma_z=0.029$ ($z<0.85$) to $\sigma_z=0.068$ ($z>0.85$)
and the rms in the UGC 10214 and NGC 4676 fields increasing from $\sigma_z=0.036$ ($z<0.85$)
to $\sigma_z=0.050$ ($z>0.85$). 

We can use the simulations to check for incompleteness. All of the 
galaxies with $B_{z=0}\le24.5$ mag ($B_{z=0}\le24.0$ mag at $z>0.75$) were 
detected apart from one or two galaxies close to the edge of each image, 
one or two with a nearby neighbor or a few galaxies at $z>1.2$ in TN0924.
At fainter magnitudes the errors become very large for galaxies in NGC4676 in
particular. Altogether $15\%$ of $0.5<z<1.0$ objects have
$-0.06<\Delta\,z/(1+z)>0.06$ and only $6\%$ have 
$-0.12<\Delta\,z/(1+z)>0.12$. There is also around $2\%$
contamination from lower or higher redshifts ($z<0.3$ and $z>1.2$).

We correct the BPZ redshift estimates to account for the difference between 
the spectroscopic and BPZ measurements for elliptical galaxies:

\begin{equation}
z_{\rm best}=\frac{z_{\rm BPZ}+0.045}{(1-0.045)}
\label{eq:zbest}
\end{equation}

$z_{\rm best}$ is plotted against $z_{\rm spec}$ in the lower panel of Fig. 
~\ref{fig:BPZ_hdfn}.

This changes the input BPZ redshift range to $0.43<z_{BPZ}<0.91$. It also
reduces the errors associated with $z_{BPZ}>0.85$ galaxies in UGC 10214 and 
NGC 4676 considerably. We use the Ben\'{\i}tez et al. (2004) 
errors ($\sigma_z=0.06$) for our BPZ measurements. We find that a few (7) of 
our objects have significantly broader probability density functions. 
The width of these PDFs are added in quadrature to the initial
$\sigma_z=0.06$. The objects in UGC 10214 and NGC 4676 with 
$z_{\rm BPZ}>0.85$ are given an uncertainty $\sigma_z=0.09$. This 
takes into account both template error (errors related to mismatches 
between the real and assumed templates) and random errors (due to the noise).

In summary, our final sample contains 72 galaxies, 10 of which have
spectroscopic redshifts.  The completeness is expected to be in excess
of 95\% ($\lesssim$3-4 missing galaxies), with a contamination of less than
2-3 galaxies (from redshift uncertainties).  We list the
properties of all our galaxies in Table~\ref{tab:gal_prop}, in two redshift
intervals ($0.5<z\le0.75$, $0.75<z\le1.0$). Within each interval they are 
listed in order of increasing restframe $(U-V)_0$ color (see Section 5.1).

\subsection{Measuring the Half-light Radius and Total Magnitude}

We calculate the half-light radius $r_e$ of each galaxy using GALFIT
(Peng et al. 2002). In each case we assume a single Sersic profile 
(see Eqn.~\ref{eq:ser}) and allow the Sersic parameter ($\beta$) to vary 
between 0 and 10. 

\begin{equation}
I(r)=I_{r_e}\exp\,\left\{-k\left[\left(\frac{r}{r_e}\right)^{\beta}-1\right]\right\}
\label{eq:ser}
\end{equation}

\noindent where $I_{r_e}$ is the surface brightness at the half-light radius,
$r_e$, and $k\sim1.9992\beta-0.3271$ (Capaccioli et al. 1989). The half-light 
radius is defined along semi-major axis. Since the 
shape and size of the galaxy can be strongly affected by the background,
we force the sky to the value calculated by Source Extractor.

An alternative way of measuring the half-light radius is through the growth curve.  The growth curve analysis uses a maximum likelihood fit to the 
measured flux in 14 circular apertures to estimate the Sersic parameter and 
half-light radius. We find that the correction from circular half-light 
radius to elliptical half-light radius is well fit by a Moffat profile, 
$r_e^{\rm ell}=r_e^{\rm cir}(1+(\frac{1}{a})^2)^b/(1+(\frac{\phi}{a})^2)^b$, 
where the $\phi$ is the ratio of semiminor axis to semimajor axis and the 
Moffat parameters $a$ and $b$ are only weakly dependent on the Sersic 
profile. The best fit parameters for an exponential profile ($\beta=1$) are 
$a=0.38$ and $b=0.28$, whereas a de Vaucouleur's profile ($\beta=4$) is well 
fit by $a=0.24$ and $b=0.21$. Therefore, if $\phi=0.8$, 
$\frac{r_e^{\rm ell}}{r_e^{\rm cir}}=1.11$ and 
$\frac{r_e^{\rm ell}}{r_e^{\rm cir}}=1.09$ for $\beta=1$ and $\beta=4$ 
respectively, and if $\phi=0.6$, $\frac{r_e^{\rm ell}}{r_e^{\rm cir}}=1.26$ 
and $\frac{r_e^{\rm ell}}{r_e^{\rm cir}}=1.22$ respectively. These two
examples demonstrate the weak dependency on $\beta$. Once the best fit 
parameters are found, a new total flux is calculated and the process is 
iterated until the new flux is no longer larger than the old flux.

We use the output from GALFIT for the rest of our analysis since it is corrected 
for the PSF, which is important for galaxies with $r_e<0.4\arcsec$, but 
use the growth curve to identify outliers. The scatter in the two measurements 
is linear with size:

\begin{equation}
\Delta\,r_e=0.25r_e-0.013
\end{equation}

Outliers are objects where the difference between the growth curve and GALFIT is greater 
than 2.0 times the standard error at that size. The few outliers found had nearby 
neighbors that affected the growth curve analysis or GALFIT. In each case the 
size was checked manually. In most cases GALFIT gave the best fit, but for the largest
object (number 30, in Table~\ref{tab:gal_prop}), we found that neither the 
growth curve or GALFIT yielded a good fit. We used 
the ELLIPROF task in Vista to get an ellipse fit model and IRAF PHOT procedure to continue 
the growth curve out to larger apertures. Both methods give an elliptical half light radius
$r_e=1.75\arcsec$ compared to $r_e=1.34\arcsec$ for the original growth curve method and
$r_e=2.05\arcsec$ for GALFIT. Once we got our best fit half-light radius, 
we ran GALFIT with this fixed half-light radius  to get the Sersic parameter and total 
magnitude.  

We convert the apparent half-light radius (in arcsec) to the intrinsic 
half-light radius (in kpc) using:

\begin{equation}
R_e=4.85\times10^{-3}\,r_e\,d_a(z,\Omega_m,\Lambda,H_0)
\label{eq:hlr}
\end{equation}

\noindent where $d_a$ is the angular-size distance (in Mpc) calculated from the redshift and cosmology.

\subsection{The Rest-frame B-band magnitude.}

The rest-frame Johnson B band has a mean wavelength $\sim4400{\rm \AA}$ which 
translates to $\sim7700{\rm \AA}$ at $z=0.75$. This puts it into either the 
F775W-band or F814W-band, available for our datasets. Most of the 
rest-frame B flux falls within these bands, so the k-corrections from these 
bands should be the smallest and most accurate, and similarly for the 
structural parameters. For convenience, we let the I-band refer to either 
F775W or F814W throughout this section. Converting from these filters to the 
$z=0$ Johnson B removes any differences particular to the passband. 
Correcting to total magnitudes removes any differences particular to the
depth. Once these corrections have been made the data from
all fields should be homogeneous and the only differences should come from
cosmic variance and the field-of-view.

The rest-frame B-band magnitude is calculated using the k-correction of the 
best fit BPZ SED from the I band to the z=0 Johnson B-band.

\begin{equation}
B_{\rm z=0,iso}=I_{\rm iso}+k({\rm SED},I,z_{\rm BPZ},B,0)
\end{equation}

\noindent where the k-correction $k({\rm SED},I,z_{\rm BPZ},B,0)$ is the difference
in magnitude between the integrated flux through an I-band filter at $z_{\rm BPZ}$
and the B-band filter at $z=0$. The SEDs fit our colors best at $z_{\rm BPZ}$ rather
than $z_{\rm best}$, so we must use the $z_{\rm BPZ}$ to calculate the k-correction.
$B_{z=0,{\rm iso}}$ and $I_{\rm iso}$ are the restframe
B band and measured I band isophotal magnitudes, which have a strong dependence on the
surface brightness limit and the redshift (Cross et al. 2001), so a 
correction must be made for the missing flux.  GALFIT calculates the total 
flux of each galaxy in the I-band, which we then trivially convert to a total
I-band magnitude $I_T$. We can transform this to the total rest-frame 
B magnitude, $B_{z=0,T}$.

\begin{equation}
B_{z=0,T}=B_{z=0,iso}+I_{T}-I_{iso}
\label{eq:tot}
\end{equation}

\noindent The total magnitude is between $0.1$ and $0.7$ mag brighter than the 
isophotal magnitude, with a mean aperture correction of $0.34$ mag.
Finally we convert to absolute magnitudes. Since we have already k-corrected
and extinction corrected the data, the equation is simply:

\begin{equation}
M_{B,T,z=0}=B_{z=0,T}-5\log(d_L)-25.
\label{eq:AbM}
\end{equation}

\noindent where the luminosity distance $d_L$ is in Mpc. The effective surface brightness
 of the galaxy is defined as the mean surface brightness within the half-light radius. 
The intrinsic effective surface brightness is calculated from the absolute magnitude and 
half-light radius to remove the $(1+z)^4$ redshift dependence:

\begin{equation}
\mu_e=M_B+5\log_{10}R_e+38.57
\label{eq:MRmu}
\end{equation}

\noindent where the constant converts from magnitudes per kpc to mag 
arcsec$^{-2}$.

\section{Sample Selection}

We select elliptical and S0 (E/S0) galaxies on the basis of morphology to 
a rest-frame B magnitude limit that gives us the largest sample with 
reliable redshifts and 
morphologies. We select over a redshift range $0.5<z_{\rm best}<1.0$ since the 
$4000\,{\rm \AA}$ break is outside our range of filters for $z<0.25$ and 
$z>1.25$. For redshifts close to these limits it will be increasingly 
difficult to estimate an accurate photometric redshift. The 
k-corrections between I-band and rest-frame B-band have a very weak dependence
on the SED across this range and have the weakest dependence at $z=0.75$. At
$z<0.5$, the errors in the k-correction increase (the standard deviation 
across the range of SEDs is 0.16 mag at $z=0.5$ and 0.28 mag at $z=0.3$), 
and the additional volume over which galaxies can be seen is relatively 
small. For $z>1.0$, the errors in the k-corrections increase, and the range 
of absolute magnitudes that can be sampled decreases. At $z=1$, it is 
possible to see $M_B<-20.1$ galaxies, by $z=1.2$, the combination of distance 
modulus and k-correction reduces the range to $M_B<-21.6$, so only the 
very brightest galaxies $\sim M_B^*$ will be sampled.

Initially, galaxies are selected with $0.5\le z<1.0$ and $B_{z=0,iso}\le25.5$
mag. Stars are removed by selecting and removing objects with the SExtractor 
stellaricity flag $>0.8$. This sample was 
morphologically classified using a semi-automated method. The first part of 
the classification was by eye. For an object  to be selected as an early-type 
galaxy, it had to be axi-symmetrical, centrally concentrated and must not 
have any spiral features. This removes spiral galaxies, chain galaxies, 
mergers, irregulars and most starbursts.  The galaxies that were 
selected as early types were then run through GALFIT as described above to 
determine the half-light radius, Sersic parameter and total magnitude. 
Objects with $\beta<2$ or $r_e<0.1$ were removed from the sample. The 
$\beta\gtrsim2$ criterion is effective at removing any residual irregular 
or starburst galaxies which were not caught by the first test.  Removing 
$r_e<0.1$ galaxies, eliminates those objects where the
errors on $r_e$ and $\beta$ will be large, dominated by the seeing and pixel
scale. These objects may not really be E/S0 galaxies, even if we measure
$\beta>2$. We find that morphological classification is easy for $I<24$ mag, 
but becomes progressively more difficult at fainter magnitudes until it 
becomes almost impossible at $I>25$ mag. 

Since we are interested in the rest-frame B-band properties of our galaxies,
our magnitude limit should be the total rest-frame B magnitude.
The main criterion for sample selection is the magnitude at which photometric 
redshifts and morphological classification become unreliable.

Fig. ~\ref{fig:BI_z} shows the difference between the total rest-frame 
$B_{z=0,T}$ magnitude and the $I_{iso}$ magnitude. For $z<0.75$, there is a 
fairly constant offset $B_{z=0,T}-I_{iso}=0.61$ mag with a scatter of $0.2$ 
mag, and the offset for $z\ge0.75$ is $B_{z=0,T}-I_{iso}=0.17$ mag with
a scatter of $0.3$ mag. Using a limit $B_{z=0,T}=24.5$ mag at $z<0.75$ is equivalent
to a limit of $I_{iso}=23.89$ and $B_{z=0,T}=24.0$ mag at $z>0.75$ is equivalent
to a limit of $I_{iso}=23.83$.

The data can be used to test these limits, using the odds value that is calculated in BPZ. 
The odds value is the integration of the probability density function (PDF) between 2 
standard deviations of the Bayesian redshift.

\begin{equation}
{\rm Odds}=\int_{z_{\rm BPZ}-2\sigma}^{z_{\rm BPZ}+2\sigma}{\rm pdf}(z)dz
\label{eq:rfodds}
\end{equation}

Ben\'{\i}tez et al. (2004, in preparation) determined the standard deviation 
to be $\sigma=(1+z)\sigma_z$ where $\sigma_z=0.06$. Thus a galaxy with a 
well-defined PDF, a single peak with a small standard deviation should have an 
odds value $\ge0.95$. $75\%$ of galaxies of $z_{\rm best}\le0.75$ and $85\%$ 
of $z_{\rm best}>0.75$ have odds $\ge0.95$. If the magnitude limits are increased by
0.5 mag, only $67\%$ of the new $0.5\le z<0.75$ galaxies have 
odds $\ge0.95$ and only $33\%$ of the new $0.75\le z<1.0$ have odds 
$\ge0.95$. Both the BPZ results from the simulations and the data suggest the 
best limits are $B_{z=0,T}<24.5$ for $z<0.75$ and $B_{z=0,T}<24.0$ for 
$z\ge0.75$, both roughly equivalent to $I_{iso}<24.0$.

In summary, the final selection criteria are: Morphologically
elliptical galaxies, defined by a centrally-concentrated, axisymmetric
profile with $r_e\ge0.1\arcsec$ and $\beta\ge2$. Objects in our lower
redshift sample $0.5<z<0.75$ have a rest-frame $B$-band magnitude
limit of 24.5 mag (observed $I \lesssim23.5$) while objects in our
higher redshift sample $0.75<z_{\rm best}\le1.0$ have a $B$-band
magnitude limit of 24.0 mag (observed $I \lesssim23.5$)

The $0.5\le z<0.75$ sample contains 32 galaxies and the $0.75\le z<1.0$ 
sample contains 40 galaxies. Since our samples are morphologically selected 
rather than color or SED selected we will be able to study the color 
evolution of the galaxies. In Fig.~\ref{fig:cutouts} we show all of the 
galaxies in our data set. These are ordered in the same way as Table~\ref{tab:gal_prop}.

\subsection{Errors}

The final redshift errors are calculated from the BPZ of Ben\'{\i}tez et al. (2004). This gives 
$\Delta\,z=0.06(1+z)$ as the final photometric redshift errors that we use. The
errors in the spectroscopic redshifts are $\Delta\,z=0.0007$ from Cohen et al.
(2000). The error bars in absolute magnitude, half-light radius and 
surface brightness are calculated from the errors in magnitude, half-light 
radius and redshift:

\begin{equation}
\Delta\,M_B=\sqrt{(\Delta\,B)^2+(\frac{\partial\,M}{\partial\,z}\Delta\,z)^2}
\end{equation}

\begin{equation}
\Delta\,R_e=R_e\,\sqrt{(\frac{\Delta\,r_e}{r_e})^2+(\frac{\partial\,d_a}{\partial\,z}\Delta\,z)^2}
\end{equation}

\begin{equation}
\Delta\,\mu_e=\sqrt{(\Delta\,B)^2+(\frac{4.3\Delta\,z}{(1+z)})^2+(\frac{2.2\Delta\,r_e}{r_e})^2}
\end{equation}

\noindent where $\Delta\,B$ is the final error in the restframe B-band magnitude. This 
includes the measured error in the F775W or F814W magnitude calculated in the ACS pipeline,
which ranges from $\Delta\,m=0.002$ mag to $\Delta\,m=0.08$ mag, the error in the 
k-correction from the F775W/F814W to rest-frame B ($\Delta\,k\sim0.05$ mag), the 
uncertainty in the zeropoint ($\Delta\,zp\sim0.02$ mag) and the uncertainty in the 
isophotal to total magnitude correction ($\Delta\,m_{tot}\sim0.05$ mag).
For objects with photometric redshifts, the errors are dominated by the redshift error.

\section{Properties of Early Type Galaxies}

\subsection{Colors of Early Type Galaxies}

An unbiased look at the colors of E/S0 galaxies is important, not only
for understanding their star formation history, but also for
understanding the role that color selection has in isolating large
samples of these objects at high redshift.  Such color (or SED)
selections have already been employed in the CFRS, CADIS, and COMBO-17
surveys and are relatively cheap to perform, requiring only
ground-based imaging over large areas of the sky.  Morphologies and
structural properties are, by contrast, much more expensive to
acquire, requiring the unique high resolution capabilities of HST.
But, the cheaper route may not be the best route as selecting by color
can result in contamination from particularly red later types
(non-E/S0s) or incompleteness to blue E/S0s.

Fig.~\ref{fig:Mcol} plots the absolute B-band magnitude against the 
rest-frame $(U-V)_{z=0}$ (AB) color. This is calculated in the same way as 
the rest-frame B magnitude:

\begin{equation}
\begin{split}
(U-V)_{z=0}&=\left[m_{\rm F1}+k({\rm SED},{\rm F1},z_{\rm BPZ},U,0)\right] \\
&-\left[m_{\rm F2}+k({\rm SED},{\rm F2},z_{\rm BPZ},V,0)\right]
\label{eq:rfcol}
\end{split}
\end{equation}

\noindent where $m_{\rm F1}$ and $m_{\rm F2}$ are the magnitudes in the 
closest filters (Filter1 and Filter2) to the redshifted rest-frame 
$U$ and $V$ filters. Filter1 and Filter2 are defined in Table~\ref{tab:filts}
for each field. The SED is the best-fit SED (or linear combination of SEDs) 
from BPZ and $k({\rm SED},m_{\rm F1},z_{\rm BPZ},U,0)$ is the k-correction 
between the observed filter at $z_{\rm BPZ}$ and the Johnson U band
filter at $z=0$.

We find a significant range in colors of early-type galaxies, with the 
majority having $(U-V)_0>1.7$. Those with $(U-V)_0>1.9$ have colors similar 
to the classic red ellipticals 
that form the red sequence seen in both clusters (Blakeslee et al. 2003c)
and the field (Bell et al. 2004). The red colors are consistent with an old 
coeval population of stars. While there is a slight color magnitude 
relationship for $(U-V)_0>1.9$ galaxies, the red sequence is blurred by a 
combination of the wide redshift range and errors in the photometric 
redshift. For the remainder of the paper we define galaxies with 
$(U-V)_0>1.7$ as `red' and galaxies with $(U-V)_0<1.7$ as `blue'.

There is a set of early-type galaxies with  $(U-V)_0<1.7$. 
These have a broad color distribution, implying a wide range in age or 
metallicity, with some ongoing star-formation. There is also a wide range
in absolute magnitude for $(U-V)_0>1.1$, $-22.5<M_B<-18$. The lower panel of 
Fig.~\ref{fig:UVodds} shows the reliability of the redshift with color. 

The redshift odds are good for more than $80\%$ of the galaxies 
with a small dependence on color (see \S4, Eqn~\ref{eq:rfodds}). The 
fraction of galaxies with just a 
single peak or a narrow dominant peak (total probability greater 
than $90\%$) in the probability density function is shown by 
the solid histogram in the middle panel of Fig.~\ref{fig:UVodds}. The 
fraction is $>85\%$ for all but the bluest, $(U-V)_0<1.12$, galaxies
where it is reduced to $\sim60\%$. Some of these galaxies have a
slightly wider probability density function with a few nearby peaks that
overlap. These can be used, but have a large uncertainty in their 
photometric redshifts. We also have two galaxies with a secondary peak
at $z\sim4$. In the top panel of Fig.~\ref{fig:UVodds} we show the uncertainty
as determined from the PDF. The open squares denote the galaxies with a 
dominant narrow peak and the filled squares represent those with overlapping
peaks. The filled circles are the mean uncertainty calculated in the 
simulations for the 3 different SEDs. Galaxies with $1.4<(U-V)_0<1.9$ have
the lowest uncertainties and the greatest chance of having a single peaked
probability distribution function. Bluer galaxies have larger 
uncertainties and greater chance of a multipeaked distribution.  We take 
into account the increased uncertainty resulting from the broader probability
distributions.

We will use the $(U-V)_0$ colors calculated here
to test the effect of color selection on the luminosity 
function and structural properties, but note that one color by itself is not 
enough to put constraints on both the age and star-formation timescale of 
these galaxies.

To estimate the ages of galaxies, we use the methodology of Menanteau, 
Abraham \& Ellis (2001) to fit exponentially decaying starburst models to
the colors of early type galaxies. But instead of fitting to the radial variations
in color, 
we use a simpler method, fitting to the overall colors. We use the Bruzual \& 
Charlot (2003) models to calculate the expected colors for a set of 
exponentially decaying continuous 
starburst models. These models assume a Salpeter Initial Mass Function and
solar metallicity. The time scales for the exponential decay ($\tau$) were 
allowed to have the
following values: $0.1, 0.2, 0.4, 1.0, 2.0, 5.0, 9.0$ Gyr. Elliptical 
galaxies are generally expected to have very short timescales ($\tau\le1$ 
Gyr), while late-type galaxies are expected to have much longer timescales 
($\tau>2$ Gyr). The models calculate the expected magnitudes in each filter 
at a given redshift for a variety of galaxy ages ($T$). For each field we 
produced the models for redshifts from $0.4$ to $1.1$ at intervals of $0.05$. 

Using the set of models where the redshift most closely matches our redshift 
estimate $z_{\rm best}$ for each object,
the model colors were compared to the measured colors, after converting from 
Vega to AB magnitudes. We calculated the maximum likelihood for the 
different values of $\tau$ and ($T$) using the following equation:

\begin{equation} 
\ln{\cal L}=\sum_{i}\left[\ln (2\pi\sigma_{C,i}^2)+\left(\frac{C_{mod,i}(\tau,T)-C_{i}}{\sigma_{C,i}}\right)^2\right]+\ln(p)
\end{equation}

\noindent where $C_i$ are the measured colors (e.g. [g-V] and 
[V-I] in the case of NGC4676), $\sigma_{C,i}$ are the errors in the 
measured colors and  $C_{mod,i}(\tau,T)$ are the model colors, a function of 
the timescale $\tau$ and the age $T$. A prior $p$ is used such that the 
combination of age and lookback-time does not exceed the age of the universe
in the adopted cosmology.

\begin{equation} 
\begin{split}
p&=1; t_{\rm frm} < T_{\rm uni}-1.0 \,{\rm Gyr}  \\
&=\frac{(T_{\rm uni}-t_{\rm frm}-0.5\,{\rm Gyr})}{(0.5\,{\rm Gyr})}; 0.5\,{\rm Gyr}\,<T_{\rm uni}-t_{\rm frm}<1.0\,{\rm Gyr} \\
&=0; t_{\rm frm} > T_{\rm uni}-0.5\,{\rm Gyr}
\end{split} 
\end{equation}

\noindent where $t_{\rm frm}=T+t[z,(\Omega_M,\Lambda,H_0)]$ is the formation time
of the galaxy and $T_{\rm uni}=13.5$ Gyr is 
the age of the universe in the adopted cosmology. 

We used the following combinations of adjacent 
filters for each field. UGC10214 / NGC4676: $(g-V)$, $(V-I)$; TN0924:
$(V-i)$, $(i-z)$; TN1338: $(g-r)$, $(r-i)$, $(i-z)$; HDFN: $(U-B)$, $(B-V)$,
$(V-I)$.
In Fig.~\ref{fig:Ages} we plot the star formation timescale ($\tau$) against
the galaxy age (bottom left hand panel), against the lookback time (bottom
middle hand panel) and formation time (bottom right hand panel) as square
points (both open and filled). The top
panels show the histogram of galaxy age, lookback time and formation time.
The error bars are calculated using a Monte Carlo simulation assuming 
Gaussian errors in the redshift and colors. 

Most galaxies have $\tau\le1$ Gyr suggesting an intense period of star 
formation that then rapidly decreased. We see a strong peak in galaxy ages of 
2 Gyrs, but a wide spread with $T<1$ Gyr and
$T>7$ Gyr in some cases. The formation times show a peak at $1.5<z<2$ 
($8.5<t_{\rm frm}<10$ Gyr), consistent with the star formation history seen in 
Heavens et al. (2004), with a rapid falloff at high redshift and a lower limit
at the lookback time of our sample. 

Fig.~\ref{fig:Mcol} also shows the expected evolutionary tracks of galaxies 
with different masses and decay timescales. Galaxies undergoing pure 
luminosity evolution with an exponentially decaying star-formation rate 
as described above will move along these tracks from blue to red. The tracks 
show that these galaxies, regardless of the decay timescale reach a maximum 
B-band luminosity at $(U-V)_0<0.7$ and then they gradually fade as they 
redden. The Bruzual-Charlot models are calculated for a 1 stellar mass 
object, so the evolutionary tracks are calculated by scaling 
$M_B^{z=0}$ by $10^{11}$ and $10^{12}$. While most of the $(U-V)_0>1.7$ 
E/S0s have $M>10^{11}{\cal M_{\odot}}$ and some have $M>10^{12}{\cal 
M_{\odot}}$, the bluer E/S0s, $1.2<(U-V)_0<1.7$, have $10^{10}<M<10^{11}{\cal 
M_{\odot}}$ and those with $(U-V)_0<1.2$ have only $M<10^{10}{\cal 
M_{\odot}}$.  Note that these results should be treated with caution given their obvious dependence on our simple exponentially decaying model.  The very brightest of the blue E/S0s will end up amongst the red sequence that has 
already formed, but most will end up extending
the sequence to fainter absolute magnitudes, given pure luminosity evolution.
For reference, the expected color of a `red' elliptical at $z=0$ is $(U-V)_0=2.18$ assuming the Coleman, Wu \& Weedman (1980) SED, so even the reddest 
galaxies in the sample will undergo an additional $0.1-0.2$ magnitudes of 
reddening to $z=0$.  

Galaxies at lower redshifts can be seen to fainter luminosities, so it is best to compare
objects over a volume limited sample (i.e., where all objects are seen over 
the same absolute magnitude and intrinsic size ranges). The dashed histograms
and filled points in Fig.~\ref{fig:Ages} show galaxies with $M_B<-20.1$ mag 
and $R_e>0.8$ kpc. The peak 
formation age is slightly higher at $2<z<2.5$ ($10<t_{\rm frm}<11$ Gyr). This
is consistent with the Heavens et al. (2004) results that show that more massive
galaxies form earlier.

It is useful to compare the rest-frame $(U-V)_0$ colors to the 
Bruzual \& Charlot (2003) model results since these were calculated using 
very different methods. In Fig.~\ref{fig:BCUV} we plot $(U-V)_{0}$ against 
the number of decay timescales ($N_{\tau}=T/\tau$). As expected, there is 
a strong correlation between these numbers over the range $1<N_{\tau}<10$.

The strong correlation between the $(U-V)_{0}$ color
and the more complicated modeling that leads to age and decay timescale
indicates that many past surveys for early-type galaxies (e.g., CFRS, COMBO-17 and CADIS) will preferentially miss the younger versions of these galaxies.  For COMBO-17 and CADIS, the $(U-V)_{0}>1.7$ selection eliminated objects that have undergone star-formation for less than 7 decay timescales, and for the CFRS, the $(U-V)_{0}>1.38$ selection eliminated objects at less than 4 decay timescales. 

Early-type galaxies must have gone
through a period of high star-formation, and the youngest of these galaxies 
are being systematically missed by ground-based surveys that select by
color or SED, rather than morphology. 

Menanteau et al. (2004) looked at the color gradients of galaxies in the 
UGC10214 field and showed that $30-40\%$ of field ellipticals are 
from $0.3<z<1.2$ are blue and that the blue colors occur preferentially in the
cores. We find that $39\%$ of all our galaxies have $(U-V)_{0}<1.7$ and $33\%$
of our volume limited
sample have $(U-V)_{0}<1.7$, in complete agreement. The Menanteau et al. 
(2004) results show that the ongoing star-formation is localized to the 
core. 

\subsection{Structural Properties}

When we look at the structural properties of galaxies it is important to 
understand the selection effects. Fig.~\ref{fig:BBD} shows the 
distribution of galaxies in absolute magnitude and surface brightness, the
Bivariate Brightness Distribution (BBD, Cross et al. 2001).
Galaxies that are in the unshaded region meet the selection criteria (in
magnitude, half-light radius and surface brightness) at all redshifts in
the ranges prescribed. In this plot and most of the following plots, we
use shading to highlight the visibility of galaxies. No shading is used when 
objects have the maximum visibility, i.e. they can be seen right across the
redshift range. Cross-hatching is used when, given our selection criteria, no 
galaxies can be seen (visibility is zero). Light shading denotes
parts of parameter space where a galaxy can be seen at the minimum redshift 
but not all the way to the maximum redshift. The visibility function shows
us when a correlation is real or is due to a selection effect. It also helps
us to properly weight our data.

The absolute selection limits are calculated from the apparent $z$, $r_e$,
$B_{z=0}$ and $\mu_{lim}^{app}$ selection limits, using Eqns~\ref{eq:hlr},\ref{eq:AbM} 
 and $\mu_{lim}=\mu_{lim}^{app}-10\log_{10}(1+z)$. The low 
surface brightness boundary is the limit at which the mean surface
brightness of a galaxy within the half-light radius is lower than the 
threshold of the shallowest survey (i.e. NGC4676, $\mu_{lim,I}^{app}=25.2$ 
mag arcsec$^{-2}$) and so it becomes difficult to accurately measure the 
half-light radius. The following absolute selection limits are used:
$0.5<z<0.75$ range ($z=0.5$), $M_B^{z=0}=-17.8$ mag, $\mu_{lim,B}=23.9$
mag arcsec$^{-2}$, $R_e=0.61$ kpc; $0.5<z<0.75$ range ($z=0.75$), 
$M_B^{z=0}=-18.8$ mag, $\mu_{lim,B}=23.3$ mag arcsec$^{-2}$, $R_e=0.73$ kpc;  
$0.75<z<1.0$ range ($z=0.75$), $M_B^{z=0}=-19.3$ mag, $\mu_{lim,B}=22.8$ mag 
arcsec$^{-2}$, $R_e=0.73$ kpc; $0.75<z<1.0$ range ($z=1.0$), 
$M_B^{z=0}=-20.1$ mag, $\mu_{lim,B}=22.2$ mag arcsec$^{-2}$, $R_e=0.80$ kpc.

The $0.5<z<0.75$ sample has a narrow range in surface brightness ($18.5<\mu_e<21$ mag 
arcsec$^{-2}$), with one outlier. This is the galaxy found earlier to have 
$r_e=1.75\arcsec$, corresponding to $R_e=11.0$ kpc. The $0.75<z<1.0$ sample has a 
wider range in surface brightness ($17.5<\mu_e<21.5$ mag arcsec$^{-2}$). 

In the unshaded region where galaxies can be seen over the whole range
of redshifts, the volume over which a galaxy can be seen is constant, and
so the space density is proportional to the number of galaxies. This 
'volume-limited' sample is useful for comparing galaxies over a range of
magnitudes. To compare galaxies within each redshift range, we use 
a sample that is volume-limited from $0.5<z\le1.0$, with $M_B\le-20.1$ mag 
and $R_e>0.8$ kpc.  Only 20 of the 32 galaxies from the $0.5<z<0.75$ subsample makes it into this volume-limited sample, and 34 of the 40 galaxies in the $0.75<z<1.0$ subsample (Table~\ref{tab:gal_prop}).  The ratio of galaxies in these two samples is $1:1.7$ (20:34), which is very similar to the ratio of comoving volume: $1:1.5$.

In Fig.~\ref{fig:histbeta} we compare the histogram of the Sersic parameters
in each redshift range. The lower panel shows the $0.5<z\le0.75$ sample and 
the upper panel shows the $0.75<z\le1.0$ sample. The long-dashed line 
represents the full distribution at each redshift and the solid line 
represents the equivalent volume limited samples, selected at 
$M_B\le-20.1$ mag and $R_e\ge0.8$ kpc. To compare each distribution
we calculate the biweight and biweight-scale (Beers, Flynn \& Gebhardt 1990).
These are equivalent to the mean and standard deviation in the case of
a Gaussian distribution and a large number of data points. The biweight is more
robust in the case of a non-Gaussian distribution with small-number 
statistics. The biweight and biweight-scale values of $\beta$ in the 
volume limited sample are tabulated in Table~\ref{tab:stprop}, for both
redshift ranges and for `red' galaxies, $(U-V)>1.7$, and `blue' galaxies
$(U-V)<1.7$. The biweight, $<\beta>\sim4.4$ for the present sample and does
not vary significantly with redshift or color. The $(U-V)>1.7$ have slightly 
lower values,
closer to the de Vaucouleur's value ($\beta=4.0$), and the bluer galaxies
have larger values on average, although with a larger biweight-scale, 
indicating a wider range of values. The larger values of $\beta$ are 
consistent with the bluer galaxies having a starburst in the cores: the 
central regions will be slightly brighter, making the
galaxies appear more concentrated.  However, at fainter luminosities $-20.1<M_B<-18.8$, the distributions become significantly different.  There is a 
significant increase in the number of low $\beta$, `blue' galaxies, leading
to a biweight of $<\beta>=2.7$ compared to $<\beta>=4.1$ for the `red' 
galaxies. Kolmogorov-Smirnov tests demonstrate that the `blue' and `red'
galaxies with $M_B\le-20.1$ mag and $R_e\ge0.8$ kpc are equivalent to
each other at 85\% and 55\% confidence in the redshift ranges $0.5<z<0.75$ and $0.75<z<1.0$, respectively.  At fainter magnitudes ($M_B\ge-18.8$ mag), this probability decreases to 1\%, thus implying a split in the properties between `red' and `blue' early-type galaxies.  This latter comparison is only possible in our lower redshift slice $0.5<z<0.75$, for objects with $R_e\ge0.73$ kpc.

Fig.~\ref{fig:histRe} shows the histogram of half-light radii. The 
results for the volume limited sample are summarised in 
Table~\ref{tab:stprop}. The distribution of $R_e$ appears uneven, considering 
the smooth distribution of $\beta$. Again there is no change between the two 
redshift ranges with the biweight size $\sim2.6$ kpc and there is no 
significant difference in the biweight sizes of red or blue early-types in 
either redshift range. The Kolmogorov-Smirnov test gives high probabilities
$73\%$ and $79\%$ that the `red' and `blue' galaxies have equivalent size
distributions with $M_B\le-20.1$ mag and $R_e\ge0.8$ kpc at $z<0.75$ and
$z\ge0.75$, respectively. At lower luminosities, the biweight size is lower,
and the difference between the `red' and `blue' distributions is greater
with a K-S probability of $52\%$. However the discrepancy is much lower than
with the Sersic parameters.

Comparing samples in the same luminosity range can be 
misleading, since ellipticals are expected to show luminosity evolution simply as a result of passive evolution in the stars.  Ideally one would like 
to compare objects of the same mass, but without dynamical information we 
compare the luminosity of objects of a similar size, since size is expected 
to change more slowly. Schade et al. (1999) looked at the relationship 
between half-light radius and B-band absolute magnitude of field ellipticals. 
They 
used 17 ellipticals in the range $0.5<z\le0.75$ (15 had spectroscopic redshifts) and 20 in the range $0.75<z\le1.0$ (11 had spectroscopic redshifts). 
We have a larger sample, which extends to fainter absolute 
magnitudes, but fewer spectroscopic redshifts (4 out of 32 and 6 out of 
40, respectively, for our two subsamples). 
Schade et al. (1997) find that cluster ellipticals are fit 
well by $M_B=-3.33\log(R_e)-18.65+\Delta\,M_B$ where 
$\Delta\,M_B=s\,\log(1+z)$ and in Schade et al. (1999) they show that 
$\Delta\,M_B=-0.56\pm0.3$ for ellipticals in the range $0.5<z\le0.75$ and 
$\Delta\,M_B=-0.97\pm0.14$ 
for ellipticals in the range $0.75<z\le1.0$. In Fig.~\ref{fig:Re_M} we 
measure the relationship between $M$ and $R_e$ for our galaxies. 
Schade et al. (1999) used a cosmology with $\Omega_M=1.0,\Lambda=0.$ and 
$H_0=50$ km s$^{-1}$ Mpc$^{-1}$. Converting to the cosmology used in this 
paper, we now have the relationship $M_B=-3.33\log(R_e)-18.56+\Delta\,M_B$ at
$0.5<z\le0.75$ and $M_B=-3.33\log(R_e)-18.60+\Delta\,M_B$ at $0.75<z\le1.0$. 
The results are tabulated in Table~\ref{tab:stprop}.

The solid lines in Fig.~\ref{fig:Re_M} show our best fit results for 
$\Delta\,M_B$ in the volume limited sample (i.e the parameter space not 
shaded). We find $\Delta\,M_B=-0.78\pm0.07$ ($\chi^2_{\nu}=3.0$) for 
galaxies in the range $0.5<z\le0.75$ and $\Delta\,M_B=-1.32\pm0.06$ 
($\chi^2_{\nu}=2.4$) for E/S0s in the range $0.75<z\le1.0$. Since the 
$\chi^2_{\nu}$ values are so high, the fits are poor. We find poor fits for 
both red and blue galaxies and both redshift ranges. Furthermore, there is no obvious correlation between the half-light radius and absolute magnitude.  We 
find many more compact luminous E/S0 galaxies (both red and blue) than 
Schade found. As with the Sersic parameter, we find a significant change
in $\Delta\,M_B$ for fainter `blue' galaxies and a much greater variance in
$\Delta\,M_B$ for `blue' galaxies. One effect that is difficult to 
take into account is the `color-selection effect'. The most rapidly
evolving `blue' galaxies at $0.75<z<1.0$ will become `red' at $0.5<z<0.75$. 
This could increase the evolution seen amongst red galaxies and decrease the 
evolution seen amongst `blue' galaxies. 

While we find a poor fit to the magnitude size relationship, we find a much 
better fit to the photometric plane. Graham (2002) demonstrated that the 
`photometric plane' variables $R_e, \mu_e, \beta$ are correlated with an 
rms scatter of 0.170, compared to the `fundamental plane' variables 
$R_e, \mu_e, \sigma_0$ which are 
correlated with an rms scatter of 0.137 for a selection of elliptical and 
S0 galaxies in the Fornax and Virgo clusters. M\'arquez et al. (2001) 
demonstrate that the photometric plane naturally emerges for relaxed Sersic
profile systems as a scaling relation between potential energy and mass.
The photometric plane:

\begin{equation} 
\log(R_e)=a(\log(\beta)+0.26\mu_e)+b 
\end{equation}

\noindent is plotted in Fig.~\ref{fig:photplan} and compared to the Graham 
(2002) result. The two redshift samples are fit by constraining the slope 
such that $a=0.86$, the same as in Graham (2002)
and then the offset $b$ is found. We use $\Delta\beta=0.5$ in our error 
assessment. This is the scatter found when comparing the $\beta$ from GALFIT
to the $\beta$ from the growth curve analysis. The values of $b$ for each of 
the samples 
are tabulated in Table~\ref{tab:stprop}. There is a significant shift in the 
offset 
from the Graham (2002) result ($b=-4.85$) to our results suggesting 
evolution in the photometric plane. There is also a small but insignificant 
change in offset between the different colored galaxies, and the fits for the 
$(U-V)_0>1.7$ galaxies are better for brighter galaxies. The increased 
variance in the `blue' galaxies at $M_B<-20.1$ is consistent with earlier
results. Since the earlier results showed that there is no significant 
variation in $<R_e>$ or $<\beta>$ at $M_B<-20.1$ with redshift, the shift is 
caused by a change in surface brightness. 

At fainter absolute magnitudes, the offset changes slightly. 
The offset for `blue' galaxies with $M_B<-20.1$ at $z<0.75$ is similar
to that of `red' galaxies with $M_B<-18.8$ at $z<0.75$. Unfortunately 
the differences are not significant so this does not demonstrate 
evolution.

Solving the photometric plane equation for $\mu_e$, we find the evolution
in surface brightness (for $M_B<-20.1$) and compare our results to the 
change in surface 
brightness found in ellipticals in the Sloan Digital Sky Survey 
(Bernardi et al. 2003) from $z=0.06$ to $z=0.2$ and in Schade et al. (1999), 
at $z=0.35$ and $z=0.78$. These are shown in the top panel of 
Fig.~\ref{fig:photplan}. The variation is linear with redshift, 
$\Delta\mu=-1.74z$. Using the fundamental plane
results for E/S0 galaxies in the DGSS, Gebhardt et al. (2003)
find an evolution in surface brightness $\Delta\mu_e=-3.38z+4.97z^2-4.011z^3$.
Our results are consistent with both the DGSS and the SDSS results.

No evolution is apparent in the structural properties of early-type 
galaxies over the redshift interval $0.5<z\le1.0$, as shown by the lack of 
variation in the biweight and biweight-scale of
$\beta$ and $R_e$ at $M_B<-20.1$ with redshift (Table~\ref{tab:stprop}). 
There are also no significant differences between the sizes of `red' and 
`blue' galaxies.  There is however some indication that
the Sersic parameter is slightly larger for bright ($M_B<-20.1$) `blue' 
galaxies with a wider dispersion (as demonstrated by the larger 
biweight-scale), as well as increased variance found in the size-magnitude 
and photometric plane measurements for `blue' galaxies. 
When fainter galaxies ($-20.1<M_B<-18.8$) are added into the sample at
$z<0.75$, a significant decrease is found in the $\beta$ values of `blue' 
galaxies and $R_e$ values for both samples, as well as a small decrease
in $\beta$ for the `red' sample. 

The fainter `blue' galaxies have a significant effect on the offset measured 
in the size-magnitude relation relative to `red' E/S0 galaxies, but they do 
not significantly effect the offset in the photometric plane. 

The weaker correlations amongst bluer galaxies could be due to 
three different effects: the bluer galaxies have larger photometric 
variations; the bluer galaxies haven't reached dynamical equilibrium;
the errors in photometric redshifts are greater for the bluer galaxies.
While there is a small color dependency on redshift, it only affects the
very bluest ($(U-V)_0<1.2$) galaxies and these are not particularly abundant 
at $M_B<-20.1$.
The dynamical time for elliptical galaxies is very low: even if `blue' 
E/S0 galaxies are 10 times less massive than `red' E/S0 galaxies
as suggested by Im et al. (2001) the dynamical time would only be a few times
$10^8$ yrs, much lower than the formation timescales and estimated ages. 
The surface brightness in these galaxies is not as tied to the Sersic 
parameter and half-light radii as it is for the `red' E/S0s.

We recalculated the above results using the geometric mean half-light radius 
instead of the semi-major axis, and find that this makes no significant 
difference to our conclusions.

\section{The Space Density of E/S0 Galaxies}

We calculate the space density over the BBD (see Fig.~\ref{fig:BBD}) using the 
bivariate (in $M_B$ and
$\mu_e$) Stepwise Maximum Likelihood (SWML) method of Sodr\'e \& Lahav 
(1993), modified to incorporate photometric redshift errors using the method 
of Chen et al. (2003).
The bivariate SWML takes into account limits in both magnitude and size, and 
outputs the correctly weighted space density of galaxies as a function of 
both absolute magnitude and effective surface brightness (Cross et al. 2001). 
The luminosity function can be calculated by summing this distribution in
the surface brightness direction. Our limits are $B<24.5$ mag, 
$r_e\ge0.1\arcsec$, $\mu_e^{app}=25.7$ mag arcsec$^{-2}$ for $0.5<z<0.75$ 
and $B<24.0$ mag, $r_e\ge0.1\arcsec$, $\mu_e^{app}=25.2$ mag arcsec$^{-2}$ for 
$0.75<z<1.0$. SWML is found to be give unbiased results even in very 
inhomogeneous samples
(Willmer 1997; Takeuchi, Yoshikawa \& Ishii 2000). SWML gives the shape
of the luminosity function, but needs to be normalized independently. We 
normalize by calculating the number of galaxies in each magnitude bin (after 
distributing each galaxy by the probability density function in redshift) in 
the volume limited region of the luminosity function and dividing by the
known volume. This is divided by the luminosity function calculated by the
SWML method to give a normalization factor in each bin and the mean is found.

\subsection{The Luminosity Function of E/S0 Galaxies}
\label{sec:lf}

Fig.~\ref{fig:LF} shows the luminosity functions of both samples, 
calculated by summing the two-dimensional space density produced above along 
the surface brightness direction. The bottom panel shows the full 
morphologically selected luminosity functions for both redshift ranges and 
the middle and top panels show the $(U-V)>1.38$ and the $(U-V)>1.7$ 
luminosity functions respectively. The square points and solid error bars 
show the luminosity function for the $0.5<z<0.75$ sample and the triangular 
points and dashed error bars show the $0.75<z<1.0$ sample. The solid and 
dashed lines show the best fit Schechter function to each redshift range
respectively. Since we cannot determine the luminosity function of the 
$0.75<z<1.0$ fainter than $M_B=-19.3$ mag, the faint end slope cannot be 
properly constrained. Therefore we fit the Schechter function using the 
faint end slope calculated from the $0.5<z<0.75$ sample in each case. The best fit 
parameters for all the Schechter functions are tabulated in 
Table~\ref{tab:lf}.

For the $0.5<z<0.75$ sample, we find $\phi^*=(1.61\pm0.18)\times10^{-3}h_{0.7}^3$
Mpc$^{-3}$mag$^{-1}$, $M^*-5\log\,h_{0.7}=(-21.1\pm0.3)$ mag and 
$\alpha=-0.53\pm0.17$ and we find that the $0.75<z<1.0$ sample has 
$\phi^*=(1.90\pm0.16)\times10^{-3}h_{0.7}^3$ Mpc$^{-3}$mag$^{-1}$, 
$M^*-5\log\,h_{0.7}=(-21.4\pm0.2)$mag and $\alpha=-0.53$ (fixed). The 
evolution in
the luminosity function can be accounted for by a decrease in 
luminosity of $0.36\pm0.36$ mag from $0.75<z<1.0$ to $0.5<z<0.75$ and a decrease
of $(15\pm12)\%$ in the number density. 

In the $0.5<z<0.75$ range, removing the $(U-V)_0<1.38$ galaxies significantly
reduces the number of low luminosity galaxies, changing the faint end 
slope from $\alpha=-0.53$ to $\alpha=0.24$. $M^*$ for the $0.5<z<0.75$ sample is 
$0.26\pm0.5$ mag fainter and has decreased in space density by $(39\pm11)\%$ 
compared
to the $0.75<z<1.0$ sample. The $M_B-5\log\,h_{0.7}=-20$ point in the $0.75<z<1.0$ sample
suggests a steeper faint end slope for this population.  However, the present data are not deep enough to properly constrain it.  The 
$(U-V)_0>1.7$ population is similar in character to the $(U-V)_0>1.38$ population, with an evolution
of $0.22\pm0.5$ mag in luminosity and a  $(34\pm12)\%$ decrease in number density 
 
In Fig.~\ref{fig:LF_0.5z0.75} we compare the $0.5<z<0.75$ LFs with each other and 
with other rest-frame B luminosity functions for early-type galaxies in this redshift
range. In each case, we have converted from the given cosmology to 
$\Omega_M=0.3$, $\Lambda=0.7$, $H_0=70\,$km s$^{-1}$Mpc$^{-1}$. In the
case of the COMBO-17 data (Wolf et al. 2003) we have converted from Vega to
AB magnitudes by subtracting 0.13 mag (Johnson-B band). All the results are
summarised in Table~\ref{tab:lf}. The points with errorbars are those for the 
morphologically selected luminosity function. The solid lines show our ACS LF, with the 
thick line showing the full color range, the medium line showing the 
$(U-V)_0>1.38$ sample and the thin line showing the $(U-V)_0>1.7$ sample.
The main difference is between the morphologically-selected sample and the 
color-selected sample, with little difference between the $(U-V)_0>1.38$ sample and
the $(U-V)_0>1.7$ sample. This difference occurs at the faint end, where most of the
very blue galaxies are. The thin lines 
with long dashes and short dashes show the COMBO-17 LFs (Wolf et al. 2003) at 
$z=0.5$ and $z=0.7$, respectively, while the thin dotted line shows the CADIS 
LF (Fried et al. 2001). The medium thick long-dashed line represents the 
CFRS (Lilly et al. 1995) LF, and the thick long-dashed line shows the DGSS
(Im et al. 2002) luminosity function. The COMBO-17 and CADIS LFs 
are for objects classified as E-Sa from SED templates and should be best 
matched to the $(U-V)_0>1.7$ sample. The CFRS should be best matched to the
$(U-V)_0>1.38$ sample and the DGSS is morphologically selected and so can be 
compared to the full sample. Fig.~\ref{fig:LF_0.75z1.0} shows the equivalent 
plot for $0.75<z\le1.0$.
 
The present study has somewhat different parameters from the DGSS, with 
a fainter $M^*_B$ and higher $\phi^*$, but the luminosity functions 
are similar, see Fig.~\ref{fig:LF_0.5z0.75}, with the main differences due 
to the correlations between $M^*_B$ and $\alpha$. At $M_B=-21.1$ ($M^*_B$), 
our $0.5<z\le0.75$ LF has a space density that is $33\%$ larger. There is
closer agreement at other magnitudes. Our LF at $0.75<z\le1.0$ does not
have such close agreement. At $M_B=-21.4$ ($M^*_B$), our LF has a 
space density that is $89\%$ larger. The DGSS was not able to constrain the 
faint end slope, so
they used a value $\alpha=-1.0$, based on the morphologically selected
low redshift luminosity functions calculated from the Second Southern
Sky Redshift Survey (SSRS2, Marzke et al. 1998) and the Nearby Optical
Galaxy Sample (NOG, Marinoni et al. 1999). Our sample goes almost 2
magnitudes deeper than the DGSS and hence we are able to constrain the
faint end slope: $\alpha=-0.53\pm0.17$ is shallower than the DGSS LF,
but is much steeper than the color-selected luminosity functions.  We
can achieve this greater depth ($I\sim24$ vs $I\sim22$) due to
a combination of improved pixel scale / point spread function and
images that are $1-1.5$ mag deeper, giving higher resolution images
with better signal-to-noise than the DGSS.

When we compare the samples selected with $(U-V)_0>1.38$, we find that the
CFRS is not a good match to the ACS LF. While both have similar values of $M^*$,
and the values of $\alpha$ are consistent given the shallow depth of the CFRS, the
space density is about twice as high in the CFRS as our measurement.
 In the higher redshift range, the CFRS luminosity
function is a closer match for $M_B<-21$ mag, but again overestimates the
number of galaxies for lower luminosities. This suggests some contamination
by late-type galaxies such as Sa/Sbc spirals which the morphological 
selection removes, as well missing the bluer early-type galaxies.

The $0.5<z\le0.75$ CADIS LF and $0.4<z\le0.6$ COMBO-17 LF both closely resemble the ACS 
$0.5<z\le0.75$, $(U-V)_0>1.7$ LF, with offsets of $\sim0.25$ magnitudes either way, which 
is well within the errors. However the $0.6<z\le0.8$ COMBO-17 LF has a much lower space 
density, which is also much lower than the ACS $0.75<z\le1.0$, $(U-V)_0>1.7$ LF.
In fact all of the SED selected luminosity functions in the $0.75<z\le1.0$ 
range find
a much lower space density than the ACS $0.75<z\le1.0$, $(U-V)_0>1.7$ LF. The 
ACS $0.75<z\le1.0$ LFs have much better agreement with the DGSS and the CFRS surveys.

The $0.75<z\le1.04$ CADIS LF, the $0.8<z\le1.0$ COMBO-17 LF and the 
ACS $0.75<z\le1.0$, $(U-V)_0>1.7$ LF all have a mean redshift $<z>$ of $\sim0.9$, 
so it is expected that the luminosity functions should be the same. 
At $M_B=-21.1$ ($M^*$ for the ACS LF), the space density measured in 
the CADIS LF is 0.43 that of the ACS LF and the space density measured in the 
COMBO-17 is only 0.13 that of the ACS LF. It appears that many more E/S0 
galaxies are missing from the high redshift COMBO-17 and CADIS luminosity functions than 
expected even considering the color selection in $(U-V)_0$. However,
there is an additional color selection for COMBO-17, which is the $R$-band selection.
Galaxies were initially selected to have $R<24$, so the selection is in the rest-frame
UV at $z>0.75$. Finally we calculate the luminosity function for our blue E/S0 galaxies. 
Since we have fewer blue, $(U-V)_0<1.7$, E/S0 galaxies, we combine all the 
$B_{z=0}\le24$ mag galaxies together to calculate a LF from $0.5<z\le1.0$, 
see Fig.~\ref{fig:LF_blue}. 
It has a much steeper faint end slope than the red E/S0 with 
$M^*-5\log\,h_{0.7}=(-22.1\pm0.4)$ and $\alpha=-1.19\pm0.15$. 

These results show that there is a wide variation in the luminosity
functions reported and that selection effects have a systematic effect
on the results.  In particular, for all color-selected samples, we
noted a significant underestimate of the faint end slope compared with
morphologically selected samples. The space density of $M^*$ galaxies
also varied greatly from survey to survey.

\subsection{The Surface Brightness Distribution}

The luminosity function of galaxies can be calculated as above by summing the 
space density in the surface brightness direction, as long as there are not 
any galaxies missing from the sample due to surface brightness dependent 
selection criteria (see Cross et al. 2001, Cross \& Driver 2002). 
Fig.~\ref{fig:BBD} demonstrates that we are not missing a significant 
population of low surface brightness ellipticals. However, the compact
($r_e<0.1\arcsec$) E/S0 galaxies that we removed from the sample do affect
the faint end of the luminosity function.

The surface brightness distribution for galaxies with $0.5<z\le0.75$ is 
shown in Fig.~\ref{fig:Sbd1}, for all the galaxies and galaxies in
different luminosity ranges. It is apparent that the surface brightness
distribution peaks at $\mu_e=20$ mag arcsec$^{-2}$ for bright galaxies and 
that any effects of missing galaxies are negligible for $M_B<-20$ mag and
small for $-20<M_B<-19$ mag. However they are important for $M_B>-19$. 
We estimate the effects by adding in all galaxies with $r_e<0.1$, regardless
of $\beta$, since $\beta$ will be difficult to accurately measure for such 
compact objects. At $z<0.75$ there are three additional objects, with
$M_B=-19.8,-19.3,-18.6$. The new luminosity function parameters are shown in 
Table~\ref{tab:lf}. The faint end slope is steeper, with $\alpha=-0.75$. 
$M_B^*$ is slightly brighter and $\phi^*$ is slightly reduced, but these
effects are due to the dependency of $M_B^*$ on $\alpha$. It must be 
emphasized that the additional compact objects may not meet the selection
criterion $\beta>2$ if observed by a telescope with better resolution, 
so this new luminosity function is an upper limit.

Fig.~\ref{fig:Sbd2} shows the surface brightness distributions of the 
$0.75<z\le1.0$ sample. At bright $M_B$ the surface brightness distribution
peaks at $\mu_e\sim19.5$ mag arcsec$^{-2}$, brighter than the $0.5<z\le0.75$ 
sample by $0.5$ mag arcsec$^{-2}$. This is consistent with the change in 
surface brightness measured in Fig.~\ref{fig:photplan}. The surface
brightness distributions suggest that there are no missing compact galaxies 
with $M_B<-21$. There are likely to be some missing galaxies in the
range $-21<M_B<-20$, but these only make up a small fraction of the total. 
For $M_B>-20$, the missing fraction is likely to be significant. Amongst 
galaxies that were discarded, there are two additional galaxies 
($M_B=-20.8$, $-20.2$) that have $r_e<0.1$ arcsec. This could increase the 
space density at $M_B=-20$ by
up to $10\%$. The low surface brightness limit also becomes important at $M_B>-20$. However since this is the limit where our measurements of the half-light radius begins to break down, rather than a detection limit (E/S0 have very high central surface brightnesses), it is unlikely that we are missing any
LSBGs.
 
\section{Discussion}

We find that `blue' E/S0 galaxies make up $30-50\%$ of $M_B<-20.1$
E/S0 galaxies at $0.5<z<0.75$ and $20-40\%$ of $M_B<-20.1$
E/S0 galaxies at $0.75<z<1.0$. Our results are consistent with both the 
Menanteau
et al. (1999) sample which found similar numbers to these depths and the 
Im et al. (2001) sample which found only $\sim15\%$.  Illustrating this agreement with the latter sample requires that we select galaxies to $I<22$ using the Im et al.\ (2001) $(V-I)$ color criteria (Figure 1 from that paper).  For the present sample this works out to a blue fraction of $23\pm11$ per cent, consistent with the above numbers.

From the analysis of the colors and structural properties of E/S0 
galaxies at $0.5<z<1.0$, it is apparent that bright ($M_B<-20.1$), 
`blue' $(U-V)_0<1.7$ E/S0 galaxies are not significantly different from
bright `red' $(U-V)_0>1.7$ E/S0 galaxies in terms of their structural 
parameters. When the stellar population has aged, these galaxies will be 
only slightly less luminous than the current `red' galaxies, and there
will be significant overlap, see Fig.~\ref{fig:Mcol}. They just have higher 
current star-formation rates, as measured by the $N_{\tau}$ from the
Bruzual \& Charlot (2003) models. However, these same models indicate that 
$(U-V)_0<1.7$ E/S0 galaxies are less massive than $(U-V)_0>1.7$ galaxies at 
the same luminosity. Fainter ($M_B>-20.1$) `blue' E/S0 galaxies are smaller 
with lower Sersic parameters than their `red' counterparts. These galaxies 
often have extremely blue colors $(U-V)_0<1.2$ and are likely to be less 
massive. The evolution tracks Fig.~\ref{fig:Mcol} suggests that these will 
fade by $\sim3$ mag as their stellar populations age. This is consistent 
with these galaxies becoming present day dwarf ellipticals. 

The best fits to these models give an increasing rate of formation from 
$z\sim8$ all the way down to $z\sim2$ with a short star-formation timescale 
$\tau\le1$ Gyr. There are only a few objects with longer timescales. The 
caveat in this modeling is that we have used simple, exponentially decaying 
star-formation models at a fixed metallicity, with no 
internal dust corrections since we are comparing observations in only 3 or 4 
filters. With 3 broadband filters, there are only 2 color constraints that 
one can apply to the models, so it is impossible to test for anything 
beyond a simple variation in age and timescale.
Observations of massive ellipticals at low redshifts and modeling support a 
single main burst, with a short timescale $0.1<\tau<0.3$ and a Salpeter
IMF (Pipino \& Matteucci 2004). 

While there is not enough data to 
find the best fit solution for a range of metallicities and dust models, it
is instructive to estimate the effect that different metallicities or dust 
will have on the result. As a test for these effects, we recalculated the 
ages and timescales using metallicities $Z=0.008$ and $Z=0.05$ to contrast 
with 
the results obtained with  the solar metallicities ($Z=0.02$). For each of 
these metallicities we calculated the ages and
star-formation timescales with no dust and using the dust model of
Charlot \& Fall (2000) assuming a V-band optical depth, $\tau_V=1$ and the 
fraction of light contributed by the `ambient' interstellar medium $\mu=0.3$. 
These values are the default values used in the Bruzual \& Charlot (2003) 
code, and are close to the ``standard'' values discussed in Charlot \& Fall 
(2000) and Bruzual \& Charlot (2003) for objects with $T>10^7$ yr. 
We recalculated models for 
star-formation timescales of 
$\tau=0,1,0.4,2.0$ only. We found that the addition of this dust 
model reduces the average age of each galaxy by $\sim10\%$, depending on
the metallicity. Since both age and dust tend to redden a galaxy, the 
same colors can reflect an old, dust free galaxy or a young dusty galaxy.
Age increases with decreasing metallicity and vice-versa due to the absorption
of blue light by metals in the atmospheres of stars. We find that 
$(\frac{\partial\ln t}{\partial\ln Z})\sim -0.4$, lower than the 
Worthey (1994) value $(\frac{\partial\ln t}{\partial\ln Z})\sim -1.5$. 
The lower value may be due to the different assumptions. Worthey assumed
a single-burst model, whereas we have a continuous exponentially decaying
star-formation rate. If these galaxies do have
a lower metallicity (as one might expect for intrinsically 
fainter, likely less massive, bluer galaxies: Tremonti 
et al. 2003), then they may be older than we estimate.

While the structural properties ($R_e$ and $\beta$) of bright `red' or `blue' 
E/S0s do not change significantly with redshift, there is a change in the
photometric plane offset, the size-magnitude relation and the luminosity
functions demonstrating significant luminosity evolution ($\sim0.4$ mag) from 
$0.75<z<1.0$ to $0.5<z<0.75$. There are some variations in structural 
properties between `red' and `blue' galaxies with the `red' galaxies 
having a smaller variance in the Sersic parameter than `blue' galaxies.

The luminosity evolution measured from the size-magnitude relation is
$0.6$ mag for `red' galaxies and $0.3$ mag for `blue' galaxies. However,
there is only a $0.1$ mag change in the $M^*$ point for the luminosity
function of `red' galaxies and $0.3$ mag overall. While these values are 
quite different, the size-magnitude relation does not give a good fit, 
so one should be careful with the interpretation. $(U-V)_0=2.0$ galaxies 
are expected to fade by $\sim0.25$ mag, regardless of the star-formation
timescale, 
over the $\sim1.5$ Gyr time span that separates the median redshift in
each bin. Over this same time, $(U-V)_0=1.8$ galaxies are expected to fade 
by $\sim0.55$ mag, so $\sim0.4$ mag of evolution is expected. To complicate
matters, some objects that were previously considered to be `blue' will
have aged sufficiently to be classified as `red'. The lower variation in
the $M^*$ point may relate to the smaller variation amongst the
very reddest galaxies ($(U-V)_0=2.0$), which are also generally the 
brightest. The expected variation
of `blue' galaxies is much wider ranging $0.2<\Delta\,M<1.5$, and 
shows a much greater dependence on the star-formation timescale, so any
offset is difficult to
predict, especially given that `blue' galaxies will eventually evolve to 
become `red' galaxies and other new `blue' galaxies may form. Since the
evolutionary tracks on Fig~\ref{fig:Mcol} suggest that the redder galaxies 
are more massive than the bluer galaxies, the luminosity evolution is 
particularly difficult to predict, as more massive `blue' galaxies will
become `red'. Indeed the galaxies with the most rapid evolution in
the restframe $M_B$ magnitude will also redden the most rapidly, so there is
a selection effect operating that will reduce the apparent luminosity 
evolution observed. 

There is a decrease in number density for $(U-V)_0>1.38$ E/S0 galaxies of 
($40\pm10\%$) from $z=0.89$ to $z=0.64$, which is very small 
compared to the factor 3 increase seen in the COMBO-17 and CADIS LFs over the 
same redshift range. This argues against hierarchical merging being an 
important evolutionary driver between $z=1$ and $z=0.5$, although it could be 
an important feature at higher redshifts or in lower luminosity objects. 

Using deep high resolution optical data we are able to measure the 
morphological E/S0 luminosity function almost 2 magnitudes deeper than the 
DGSS and constrain the faint end slope of the $0.5<z\le0.75$  LF. We find a 
fairly flat faint end slope $\alpha=-0.75\pm0.13$, slightly shallower than 
low redshift luminosity functions for morphologically selected E/S0s but
much steeper than color-selected samples. Our values for $M_B^*$ are 
consistent with the DGSS but our $\phi^*$ is larger by $\sim40\%$. 
This could be due to cosmic variance since both samples are small or due 
to the
differences in morphological selection. To address the latter point, we note 
that the DGSS sample is selected using
the the bulge-to-total ratio ($B/T>0.4$) and the residual parameter 
($R<0.06$). Fig. 9 of Im et al. (2002) demonstrates that varying the selection
criteria a little ($B/T>0.3$, $R<0.08$) can increase the sample size by 
$50\%$. Changing our selection criteria to $\beta<2.5$ reduces our 
sample size by $15\%$. These changes are expected to have more of an effect
at faint absolute magnitudes, where galaxies have flatter (i.e. lower
Sersic number) profiles (Graham \& Guzm\`an 2003). Thus different 
morphological selection criteria could explain the variation seen. 

Using purely photometric information (color, SED) to select the galaxy 
sample misses the bluer early types, and may lead to 
contamination from Sa/Sbc spiral galaxies or other red galaxies. As shown 
in Figs~\ref{fig:LF_0.5z0.75} \& ~\ref{fig:LF_0.75z1.0} there is a large 
variation in the measurement of the luminosity function. Indeed using the 
COMBO-17 results, one would be drawn to the conclusion that there were very 
few faint early-types ($\alpha=0.52$) and that there is strong number 
evolution in the luminosity function, suggesting that many spiral or other 
galaxies must have become ellipticals over time, e.g. via a high merger rate.
The COMBO-17 luminosity functions only sample the brightest luminosities 
at $z\sim1$, while the ACS and CADIS LFs reach 1.5 mag deeper. 

The luminosity function of blue E/S0s is steeper ($\alpha=-1.19\pm0.15$) with
a bright $M_B^*-5\log\,h_{0.7}=-22.1\pm0.4$, but a much lower space density 
$\phi^*=2.5\pm0.5\times10^{-4}\,h_{0.7}^3$ Mpc$^{-3}$mag$^{-1}$. 
Low luminosity systems have a greater proportion of
young star-forming systems, suggesting that the more massive galaxies
formed earlier or underwent more rapid star-formation, so they now contain
only older stars. These results provide a good fit to the models of Pipino 
\& Matteucci (2004).

\section{Acknowledgements}

ACS was developed under NASA contract NAS5-32865 and this research has 
been supported by NASA grant NAG5-7697. The STScI is operated by AURA Inc., 
under NASA contract NAS5-26555. We are grateful to Ken Anderson, Jon 
McCann, Sharon Busching, Alex Framarini, Sharon Barkhouser, and Terry Allen 
for their invaluable contributions to the ACS project at JHU.
We thank Jon McCann for his general computing support, including the 
development of FITSCUT, that we used to produce the color images.
We would like to thank the anonymous referee for his/her useful comments.

\clearpage

\begin{deluxetable}{lccccccc}
\tablecolumns{8}
\tablewidth{0pt}
\tablecaption{Summary of data from different fields.\label{tab:fields}}
\tablehead{
  \colhead{Field} &
  \colhead{Filters} &
  \colhead{Area ($\Box^{\arcmin}$)} &
  \colhead{$T_{\rm exp,I}$(s)\tablenotemark{a}} &
  \colhead{E(B-V)} &
  \colhead{A(I)\tablenotemark{a}} &
  \colhead{ZP(I)\tablenotemark{a}\,\,\,\tablenotemark{b}} &
  \colhead{N(E/S0)} 
}
\startdata
\tableline
NGC4676  & g,V,I & 7.8 & 4070 & 0.017 & 0.030 & 25.947 & 12 \\
UGC10214 & g,V,I & 10.7 & 8180 & 0.009 & 0.017 & 25.947 & 17 \\
TN1338   & g,r,i,z & 11.7 & 11700 & 0.096 & 0.193 & 25.655 & 14 \\
TN0924   & V,i,z & 11.7 & 11800 & 0.057 & 0.115 & 25.655 & 19 \\
HDFN     & i+FLY99 & 5.8 & 5600 & 0.012 & 0.024 & 25.655 & 10 \\ 
\tableline
\enddata
\tablenotetext{a}{The exposure time, extinction and zeropoint are given for the 
F775W or F814W filter, since this was used for measurements of the structural 
parameters.}
\tablenotetext{b}{This is the zeropoint for a 1s exposure.}
\end{deluxetable}

\begin{deluxetable}{lrrcccrc}
\tablecolumns{8}
\tablewidth{0pt}
\tablecaption{Summary of galaxy properties for all 72 E/S0s in the
sample. The galaxies in bold are those in the volume limited sample 
$M_B<-20.1$, $R_e>0.8$ kpc. The horizontal line separates those with
$0.5<z\le0.75$ from those with $0.75<z\le1.0$. Within each redshift range 
they are in order of their $(U-V)_0$ color.\label{tab:gal_prop}}
\tablehead{
  \colhead{No} &
  \colhead{RA} &
  \colhead{Dec} &
  \colhead{z\tablenotemark{a}} &
  \colhead{$M_B$} &
  \colhead{$(U-V)_0$} &
  \colhead{$\beta$} &
  \colhead{$R_e$ / kpc} 
}
\startdata
\tableline
01 & 16 06 16.9 &  55 26 53.4 &  0.65 & -19.14 &  0.61 &  2.56 &  0.96 \\ 
02 & 16 06 06.5 &  55 26 50.9 &  0.66 & -18.88 &  0.75 &  2.57 &  1.46 \\ 
{\bf 03} & {\bf 09 24 19.1} & {\bf -22 01 28.1} & {\bf 0.73} & {\bf -20.44} & {\bf 0.83} & {\bf 2.09} & {\bf 2.88} \\ 
04 & 16 06 16.5 &  55 23 44.2 &  0.52 & -19.26 &  0.84 &  2.61 &  1.63 \\ 
{\bf 05} & {\bf 12 46 18.1} &  {\bf 30 42 53.5} &  {\bf 0.66} & {\bf -20.75} &  {\bf 0.84} &  {\bf 7.18} &  {\bf 2.09} \\ 
06 & 13 38 30.0 & -19 44 20.0 &  0.61 & -19.36 &  0.96 &  3.50 &  0.74 \\ 
07 & 09 24 27.4 & -22 01 37.9 &  0.72 & -18.81 &  0.97 &  2.38 &  1.15 \\ 
{\bf 08} & {\bf 09 24 21.4} & {\bf -22 01 15.9} &  {\bf 0.73} & {\bf -20.30} &  {\bf 1.16} &  {\bf 5.18} &  {\bf 1.92} \\ 
09 & 09 24 25.4 & -22 02 46.9 &  0.73 & -19.33 &  1.17 &  2.32 &  1.45 \\ 
10 & 12 46 14.7 &  30 45 32.7 &  0.72 & -18.83 &  1.17 &  2.39 &  1.63 \\ 
{\bf 11} & {\bf 16 06 14.9} &  {\bf 55 26 52.8} &  {\bf 0.53} & {\bf -20.51} &  {\bf 1.22} &  {\bf 2.19} &  {\bf 1.36} \\ 
{\bf 12} & {\bf 09 24 23.2} & {\bf -22 03 04.3} &  {\bf 0.71} & {\bf -20.32} &  {\bf 1.27} &  {\bf 2.25} &  {\bf 2.47} \\ 
{\bf 13} & {\bf 12 46 18.3} &  {\bf 30 43 25.5} &  {\bf 0.63} & {\bf -21.26} &  {\bf 1.29} &  {\bf 5.26} &  {\bf 4.04} \\ 
14 & 09 24 17.4 & -22 01 38.5 &  0.73 & -19.55 &  1.40 &  4.86 &  1.61 \\ 
{\bf 15} & {\bf 09 24 20.0} & {\bf -22 03 15.2} &  {\bf 0.69} & {\bf -20.23} &  {\bf 1.40} &  {\bf 9.01} &  {\bf 2.08} \\ 
{\bf 16} & {\bf 12 46 12.3} &  {\bf 30 45 21.2} &  {\bf 0.70} & {\bf -22.83} &  {\bf 1.41} &  {\bf 5.03} &  {\bf 3.30} \\ 
{\bf 17} & {\bf 12 46 08.6} &  {\bf 30 41 52.7} &  {\bf 0.66} & {\bf -20.75} &  {\bf 1.42} &  {\bf 4.29} &  {\bf 2.89} \\ 
{\bf 18} & {\bf 13 38 31.5} & {\bf -19 45 05.7} &  {\bf 0.66} & {\bf -20.97} &  {\bf 1.71} &  {\bf 5.25} &  {\bf 1.44} \\ 
{\bf 19} & {\bf 09 24 24.5} & {\bf -22 00 42.3} &  {\bf 0.70} & {\bf -21.37} &  {\bf 1.74} &  {\bf 4.01} &  {\bf 2.49} \\ 
{\bf 20} & {\bf 09 24 15.7} & {\bf -22 01 33.4} &  {\bf 0.71} & {\bf -21.73} &  {\bf 1.74} &  {\bf 6.47} &  {\bf 2.07} \\ 
{\bf 21} & {\bf 16 06 02.0} &  {\bf 55 24 51.3} &  {\bf 0.59} & {\bf -20.97} &  {\bf 1.76} &  {\bf 3.22} &  {\bf 2.91} \\ 
{\bf 22} & {\bf 12 36 57.1} &  {\bf 62 12 10.8} &  {\bf 0.67\tablenotemark{a}} & {\bf -21.18} &  {\bf 1.93} &  {\bf 4.34} &  {\bf 2.72} \\ 
23 & 13 38 21.8 & -19 44 31.7 &  0.66 & -19.62 &  1.94 & 10.00 &  1.43 \\ 
24 & 12 46 17.3 &  30 42 28.2 &  0.55 & -19.40 &  1.96 &  3.40 &  0.88 \\ 
{\bf 25} & {\bf 12 46 15.2} &  {\bf 30 43 57.4} &  {\bf 0.53} & {\bf -20.45} &  {\bf 1.97} &  {\bf 3.00} &  {\bf 2.96} \\ 
{\bf 26} & {\bf 09 24 19.2} & {\bf -22 03 00.2} &  {\bf 0.64} & {\bf -21.77} &  {\bf 1.98} &  {\bf 2.73} &  {\bf 6.36} \\ 
{\bf 27} & {\bf 12 37 00.2} &  {\bf 62 12 35.0} &  {\bf 0.56\tablenotemark{a}} & {\bf -20.32} &  {\bf 1.99} &  {\bf 3.16} &  {\bf 2.07} \\ 
28 & 13 38 22.0 & -19 44 38.7 &  0.68 & -20.00 &  1.99 &  5.13 &  1.29 \\ 
29 & 12 36 46.2 &  62 11 51.4 &  0.50\tablenotemark{a} & -19.43 &  2.00 &  3.24 &  0.88 \\ 
{\bf 30} & {\bf 09 24 20.1} & {\bf -22 03 11.4} &  {\bf 0.53} & {\bf -21.30} &  {\bf 2.00} &  {\bf 6.59} & {\bf 11.03} \\ 
{\bf 31} & {\bf 12 46 17.8} &  {\bf 30 43 38.7} &  {\bf 0.55} & {\bf -20.87} &  {\bf 2.01} &  {\bf 3.14} &  {\bf 1.48} \\ 
{\bf 32} & {\bf 12 36 49.9} &  {\bf 62 12 46.0} &  {\bf 0.68\tablenotemark{a}} & {\bf -21.08} &  {\bf 2.02} &  {\bf 5.86} &  {\bf 4.40} \\
\tableline 
33 & 16 06 18.0 &  55 25 03.2 &  0.89 & -19.80 &  1.23 &  2.33 &  2.06 \\ 
{\bf 34} & {\bf 16 06 20.3} &  {\bf 55 24 44.2} &  {\bf 0.99} & {\bf -21.68} &  {\bf 1.24} &  {\bf 3.66} &  {\bf 1.56} \\ 
{\bf 35} & {\bf 13 38 20.7} & {\bf -19 43 30.9} &  {\bf 0.97} & {\bf -21.23} &  {\bf 1.30} &  {\bf 5.35} &  {\bf 1.74} \\ 
{\bf 36} & {\bf 09 24 21.9} & {\bf -22 02 54.8} &  {\bf 0.76} & {\bf -21.21} &  {\bf 1.39} &  {\bf 2.02} &  {\bf 1.32} \\ 
{\bf 37} & {\bf 09 24 16.3} & {\bf -22 01 17.1} &  {\bf 0.77} & {\bf -20.52} &  {\bf 1.39} &  {\bf 9.42} &  {\bf 2.42} \\ 
{\bf 38} & {\bf 09 24 26.1} & {\bf -22 03 31.0} &  {\bf 0.88} & {\bf -20.47} &  {\bf 1.40} &  {\bf 2.16} &  {\bf 2.77} \\ 
{\bf 39} & {\bf 12 36 45.8} &  {\bf 62 12 46.7} &  {\bf 0.90\tablenotemark{a}} & {\bf -21.52} &  {\bf 1.46} &  {\bf 6.17} &  {\bf 2.79} \\
{\bf 40} & {\bf 09 24 20.9} & {\bf -22 03 08.2} &  {\bf 0.75} & {\bf -20.99} &  {\bf 1.55} &  {\bf 4.49} &  {\bf 3.45} \\
41 & 16 06 03.6 &  55 24 54.0 &  0.84 & -19.78 &  1.56 &  4.79 &  3.42 \\ 
{\bf 42} & {\bf 13 38 22.4} & {\bf -19 42 52.8} &  {\bf 0.90} & {\bf -20.94} &  {\bf 1.58} &  {\bf 2.73} &  {\bf 4.68} \\ 
{\bf 43} & {\bf 13 38 24.4} & {\bf -19 41 34.0} &  {\bf 1.00} & {\bf -21.17} &  {\bf 1.66} &  {\bf 8.75} &  {\bf 2.55} \\ 
{\bf 44} & {\bf 13 38 30.5} & {\bf -19 42 44.1} &  {\bf 0.97} & {\bf -22.12} &  {\bf 1.70} &  {\bf 7.13} &  {\bf 7.46} \\ 
{\bf 45} & {\bf 13 38 31.4} & {\bf -19 44 36.8} &  {\bf 0.94} & {\bf -21.21} &  {\bf 1.72} & {\bf  4.29} &  {\bf 1.66} \\ 
{\bf 46} & {\bf 16 06 04.6} &  {\bf 55 24 44.5} &  {\bf 0.84} & {\bf -21.52} &  {\bf 1.74} &  {\bf 2.64} &  {\bf 2.92} \\ 
47 & 13 38 21.5 & -19 44 39.8 &  0.85 & -20.07 &  1.74 &  9.86 &  0.92 \\ 
{\bf 48} & {\bf 13 38 27.6} & {\bf -19 42 12.6} &  {\bf 0.98} & {\bf -20.35} &  {\bf 1.81} & {\bf  2.04} &  {\bf 3.79} \\ 
{\bf 49} & {\bf 13 38 30.1} & {\bf -19 44 14.7} &  {\bf 0.93} & {\bf -20.68} &  {\bf 1.86} &  {\bf 3.21} &  {\bf 1.53} \\ 
{\bf 50} & {\bf 09 24 27.3} & {\bf -22 02 54.3} &  {\bf 0.90} & {\bf -21.39} &  {\bf 1.91} & {\bf  6.33} &  {\bf 1.81} \\ 
{\bf 51} & {\bf 16 06 01.0} &  {\bf 55 24 57.6} &  {\bf 0.82} & {\bf -21.52} &  {\bf 1.93} &  {\bf 5.00} &  {\bf 3.33} \\ 
{\bf 52} & {\bf 13 38 25.4} & {\bf -19 43 33.3} &  {\bf 0.96} & {\bf -21.00} &  {\bf 1.96} &  {\bf 3.05} &  {\bf 1.42} \\ 
{\bf 53} & {\bf 09 24 16.4} & {\bf -22 01 11.6} &  {\bf 0.85} & {\bf -21.52} &  {\bf 1.97} &  {\bf 5.21} &  {\bf 3.96} \\ 
{\bf 54} & {\bf 16 05 59.4} &  {\bf 55 26 16.0} &  {\bf 0.81} & {\bf -22.47} &  {\bf 1.97} &  {\bf 8.38} &  {\bf 5.18} \\ 
55 & 16 06 13.5 &  55 24 45.8 &  0.84 & -19.68 &  1.99 &  2.43 &  1.50 \\ 
{\bf 56} & {\bf 16 06 08.6} &  {\bf 55 26 04.0} &  {\bf 0.79} & {\bf -20.52} &  {\bf 1.99} &  {\bf 2.52} &  {\bf 1.52} \\ 
{\bf 57} & {\bf 16 06 14.9} &  {\bf 55 27 12.7} &  {\bf 0.84} & {\bf -20.78} &  {\bf 1.99} &  {\bf 4.29} &  {\bf 1.06} \\ 
{\bf 58} & {\bf 09 24 28.0} & {\bf -22 01 38.2} &  {\bf 0.98} & {\bf -20.11} &  {\bf 1.99} &  {\bf 4.82} &  {\bf 1.43} \\ 
{\bf 59} & {\bf 12 46 17.1} &  {\bf 30 44 09.0} &  {\bf 0.86} & {\bf -20.25} &  {\bf 2.01} &  {\bf 2.73} &  {\bf 3.03} \\ 
{\bf 60} & {\bf 12 46 20.5} &  {\bf 30 42 54.8} &  {\bf 0.88} & {\bf -21.00} &  {\bf 2.01} &  {\bf 2.81} &  {\bf 4.15} \\ 
{\bf 61} & {\bf 12 36 54.8} &  {\bf 62 13 03.9} &  {\bf 0.95\tablenotemark{a}} & {\bf -20.26} &  {\bf 2.01} &  {\bf 2.98} &  {\bf 0.89} \\ 
62 & 16 06 00.7 &  55 24 59.5 &  0.76 & -19.93 &  2.02 &  2.90 &  2.42 \\ 
63 & 12 46 07.7 &  30 44 41.7 &  0.86 & -20.03 &  2.02 &  3.28 &  1.21 \\ 
{\bf 64} & {\bf 16 06 11.3} &  {\bf 55 27 45.3} &  {\bf 0.86} & {\bf -22.38} &  {\bf 2.02} &  {\bf 5.56} &  {\bf 3.72} \\ 
{\bf 65} & {\bf 12 36 42.8} &  {\bf 62 12 42.3} &  {\bf 0.85\tablenotemark{a}} & {\bf -21.64} &  {\bf 2.03} &  {\bf 4.49} &  {\bf 2.35} \\ 
{\bf 66} & {\bf 09 24 26.7} & {\bf -22 01 01.7} &  {\bf 0.97} & {\bf -20.40} &  {\bf 2.04} &  {\bf 4.19} &  {\bf 2.03} \\ 
{\bf 67} & {\bf 12 46 16.6} &  {\bf 30 44 48.2} &  {\bf 0.79} & {\bf -20.13} &  {\bf 2.04} &  {\bf 4.93} &  {\bf 1.64} \\ 
{\bf 68} & {\bf 12 36 55.1} &  {\bf 62 13 11.5} &  {\bf 0.97\tablenotemark{a}} & {\bf -21.90} &  {\bf 2.05} &  {\bf 3.90} &  {\bf 3.82} \\ 
{\bf 69} & {\bf 12 36 56.3} &  {\bf 62 12 20.4} &  {\bf 0.93\tablenotemark{a}} & {\bf -21.39} &  {\bf 2.06} &  {\bf 5.38} &  {\bf 3.48} \\ 
{\bf 70} & {\bf 13 38 31.0} & {\bf -19 44 41.9} &  {\bf 0.79} & {\bf -21.47} &  {\bf 2.06} &  {\bf 6.36} &  {\bf 3.67} \\ 
{\bf 71} & {\bf 12 36 43.5} &  {\bf 62 11 43.0} &  {\bf 0.77\tablenotemark{a}} & {\bf -22.09} &  {\bf 2.07} &  {\bf 5.15} &  {\bf 3.55} \\ 
{\bf 72} & {\bf 16 05 59.5} &  {\bf 55 26 13.3} &  {\bf 0.88} & {\bf -20.43} &  {\bf 2.08} &  {\bf 3.28} &  {\bf 1.27} \\ 
\tableline
\enddata
\tablenotetext{a}{The redshifts are $z{\rm best}$ in all cases apart from 
those marked which are spectroscopic redshifts in the HDFN ($RA\sim189.2$, 
$DEC\sim62.2$). }
\end{deluxetable}

\begin{deluxetable}{lcccc}
\tablecolumns{5}
\tablewidth{0pt}
\tablecaption{Summary of closest band to rest-frame U and rest-frame V. These bands are used
to give the best estimate of the restframe $(U-V)_0$, which is used to compare to the 
color or SED selected samples of CFRS, CADIS and COMBO-17.\label{tab:filts}}
\tablehead{
  \colhead{Field} &
  \colhead{Band1 ($z<0.75$)} &
  \colhead{Band2 ($z<0.75$)} &
  \colhead{Band1 ($z\ge0.75$)} &
  \colhead{Band2 ($z\ge0.75$)}
}
\startdata
\tableline
UGC10214 & F475W & F814W & F606W & F814W \\
NGC4676  & F475W & F814W & F606W & F814W \\
TN1338   & F475W & F850LP & F625W & F850LP \\
TN0924   & F606W & F850LP & F606W & F850LP \\
HDFN     & F450W & F814W & F606W & J$_{\rm KPNO}$ \\ 
\tableline
\enddata
\end{deluxetable}

\begin{deluxetable}{lcccccc} 
\tablecolumns{7}
\tablewidth{0pt}
\tablecaption{Comparison of structural properties of early-type galaxies.\label{tab:stprop}}
\tablehead{
  \colhead{Sample} &
  \colhead{$<\beta>$\tablenotemark{a}} &
  \colhead{$<R_e>$\tablenotemark{b}} &
  \colhead{$R_e$ vs $M_B$\tablenotemark{c}} &
  \colhead{$\chi^2_{\nu}$} &
  \colhead{$R_e$ vs $(\log(\beta)+0.26\mu_e)$\tablenotemark{d}} &
  \colhead{$\chi^2_{\nu}$}
}
\startdata
\tableline
$M<-20.1$, $R_e>0.8$ & & & & & &  \\
$z<0.75$ all & $4.4\pm0.4$ & $2.5\pm0.2$ & $-0.78\pm0.07$ & 3.0 & $-4.56\pm0.03$ & 1.2 \\
$z<0.75$ red & $4.2\pm0.5$ & $2.5\pm0.4$ & $-0.74\pm0.08$ & 4.1 & $-4.54\pm0.04$ & 0.6 \\
$z<0.75$ blue & $4.7\pm0.8$ & $2.5\pm0.3$ & $-0.96\pm0.17$ & 1.8 & $-4.60\pm0.06$ & 1.9  \\
$z>0.75$ all & $4.4\pm0.3$ & $2.6\pm0.2$ & $-1.32\pm0.06$ & 2.4 & $-4.49\pm0.03$ & 0.9 \\
$z>0.75$ red & $4.3\pm0.3$ & $2.7\pm0.3$ & $-1.34\pm0.06$ & 2.3 & $-4.47\pm0.03$ & 0.6 \\
$z>0.75$ blue & $4.7\pm1.0$ & $2.5\pm0.3$ & $-1.29\pm0.12$ & 3.0 & $-4.55\pm0.05$ & 1.7 \\ 
$M<-18.8$, $R_e>0.73$ & & & & & & \\
$z<0.75$ all & $3.7\pm0.3$ & $2.0\pm0.2$ & $-0.72\pm0.06$ & 2.3 & $-4.63\pm0.03$ & 1.4 \\
$z<0.75$ red & $4.1\pm0.4$ & $2.0\pm0.3$ & $-0.76\pm0.08$ & 2.8 & $-4.61\pm0.04$ & 1.6 \\
$z<0.75$ blue & $2.7\pm0.4$ & $1.8\pm0.2$ & $-0.60\pm0.12$ & 1.8 & $-4.67\pm0.04$ & 1.1  \\
\tableline
\enddata
\tablenotetext{a}{$\beta$ is the Sersic parameter, see Eqn~\ref{eq:ser}. The average in this case is the biweight and the error is the biweight-scale.}
\tablenotetext{b}{$R_e$ is the intrinsic half-light radius in kpc. The average in this case is the biweight and the error is the biweight-scale.}
\tablenotetext{c}{The offset in the size-magnitude relation compared to $z=0$.}
\tablenotetext{d}{The offset in the photometric plane at $z=0$ is $-4.85$.}
\end{deluxetable}

\begin{deluxetable}{lccccc}
\tablecolumns{6}
\tablewidth{0pt}
\tablecaption{Summary of $B$-band luminosity function Schechter parameters
for early type galaxies. All values have been corrected to $\Omega_m=0.3$,
$\Lambda=0.7$ and $H_0=70$ km s$^{-1}$Mpc$^{-1}$.\label{tab:lf}}
\tablehead{
  \colhead{Sample} &
  \colhead{Selection} &
  \colhead{Redshift} &
  \colhead{$M_B^*-5\log\,h_{0.7}$} &
  \colhead{$\phi^*$ / $10^{-4}h_{0.7}^3$} &
  \colhead{$\alpha$} 
}
\startdata
\tableline
ACS & Morph & $0.5<z<0.75$ & $-21.1\pm0.3$ & $16.1\pm1.8$ & $-0.53\pm0.17$ \\
ACS & Morph & $0.75<z<1.0$ & $-21.4\pm0.2$ & $18.9\pm1.6$ & $-0.53$ \\
DGSS & Morph & $0.6<z<1.2$ & $-21.75\pm0.15$ & $7.7\pm2.2$ & $-1.0$ \\
\tableline
ACS & $(U-V)>1.38$ & $0.5<z<0.75$ & $-20.6\pm0.5$ & $12.9\pm1.5$ & $+0.24\pm0.49$ \\
ACS & $(U-V)>1.38$ & $0.75<z<1.0$ & $-20.8\pm0.2$ & $21.2\pm1.8$ & $+0.24$ \\
CFRS & $(U-V)>1.38$ & $0.5<z<0.75$ & $-20.74$ & $31.5$ & $-0.37$  \\
CFRS & $(U-V)>1.38$ & $0.75<z<1.0$ & $-22.84$ & $1.9$ & $-2.01$ \\
\tableline
ACS & $(U-V)>1.7$ & $0.5<z<0.75$ & $-20.6\pm0.5$ & $10.3\pm1.4$ & $+0.35\pm0.59$ \\
ACS & $(U-V)>1.7$ & $0.75<z<1.0$ & $-20.7\pm0.2$ & $15.5\pm1.5$ & $+0.35$ \\
COMBO-17 & SED (E+Sa) & $0.4<z<0.6$ & $-20.69\pm0.16$\tablenotemark{a} & $9.8\pm4.1$ & $+0.52$ \\
COMBO-17 & SED (E+Sa) & $0.6<z<0.8$ & $-21.10\pm0.16$\tablenotemark{a} & $4.6\pm1.2$ & $+0.52$ \\
COMBO-17 & SED (E+Sa) & $0.8<z<1.0$ & $-20.96\pm0.21$\tablenotemark{a} & $1.6\pm1.4$ & $+0.52$ \\
CADIS & SED (E+Sa) & $0.5<z<0.75$ & $-20.65\pm0.27$ & $10.8\pm1.1$ & $-0.05\pm0.22$ \\
CADIS & SED (E+Sa) & $0.75<z<1.04$ & $-20.48\pm0.32$ & $4.9\pm1.0$ & $+0.63\pm0.58$ \\
\tableline
ACS & $(U-V)<1.7$ & $0.5<z<1.0$ & $-22.1\pm0.4$ & $2.5\pm0.5$ & $-1.19\pm0.15$ \\ 
\tableline
ACS & SB Corr & $0.5<z<0.75$ & $-21.3\pm0.3$ & $13.9\pm1.8$ & $-0.75\pm0.13$ \\
ACS & SB Corr & $0.75<z<1.0$ & $-21.6\pm0.2$ & $16.7\pm0.5$ & $-0.75$ \\
\tableline
\enddata
\tablenotetext{a}{The COMBO-17 magnitudes have been corrected by -0.13 to convert 
$M_B$(Vega) to $M_B$(AB).}
\end{deluxetable}

\clearpage
\includegraphics[width=150mm,height=150mm]{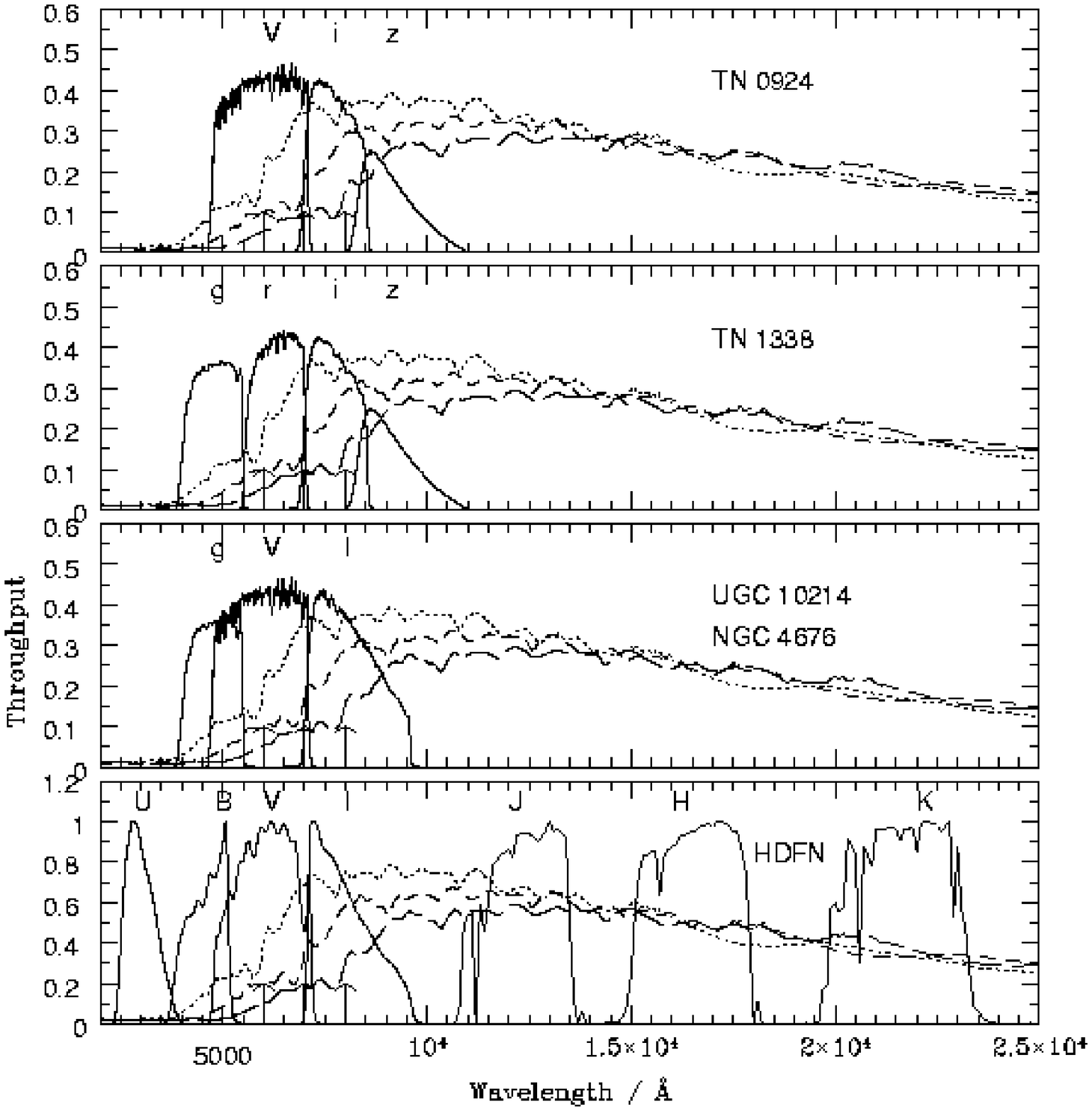}
\figcaption[flt_spec.ps]{The filter sets used in these observations. The top panel shows 
the three filters used in TN0924, $Viz$, the upper-middle panel shows the $griz$ 
filters used in TN1338, the lower-middle panel shows the $gVI$ filters
used in UGC10214 and NGC4676 and the bottom panel shows the HDFN, with 7-bands from 
$U$ to $K$. 
The dotted, short-dashed and long-dashed curves show the `El' SED (Ben\'{\i}tez et al.
2004, in preparation) at $z=0.5,0.75,1.0$ respectively. The arrows mark the position of the 
$4000\,{\rm \AA}$ break at these three redshifts. The $4000\,{\rm \AA}$ break is well 
within our filter coverage at all redshifts. \label{fig:filtspec}}

\includegraphics[width=150mm,height=150mm]{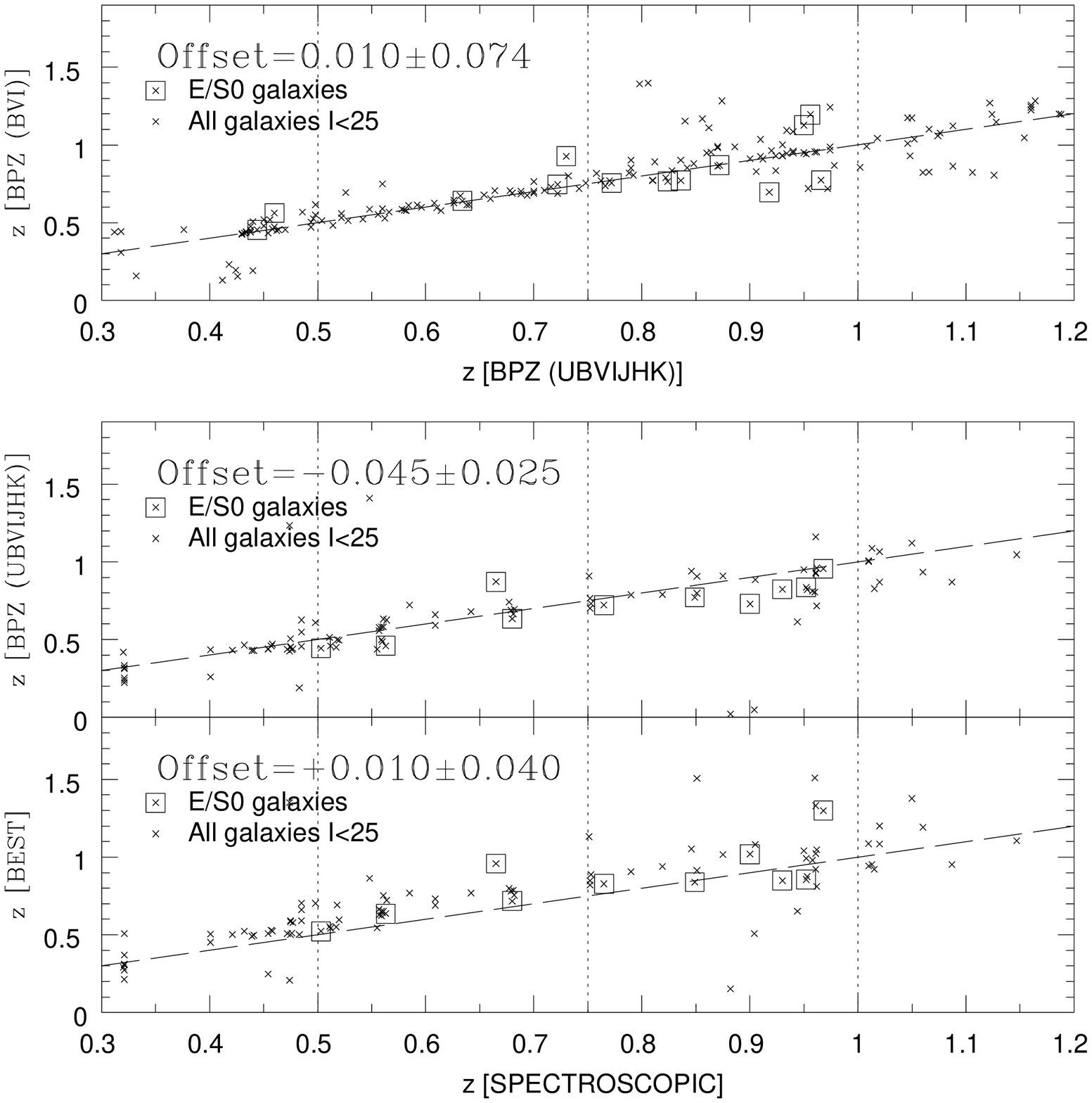}
\figcaption[test_hdfn_BPZ.ps]{The errors in BPZ derived from the HDFN. 
In the top panel we show the 3-color BPZ redshifts plotted against the 
7-color BPZ redshifts for all $I_{AB}<25$ galaxies with $0.3<z<1.2$. The 
squares surround the galaxies which are morphological elliptical galaxies and 
have $0.5<z_{\rm spec}<1.0$ or $0.5<z_{\rm BPZ}<1.0$ in the 7-color BPZ 
catalog. There are no outliers in our 
sample and the systematic offset and error in the redshift each galaxy 
are small: $\Delta\,z/(1+z)=0.010$ and $\sigma(\Delta\,z/(1+z))=0.074$ 
respectively. We then compare 7-color BPZ photometric redshifts to the 
smaller sample of objects with spectroscopic redshifts in the middle panel. 
We find that there is a significant offset between the 7-color BPZ and the 
spectroscopic redshifts. We correct for this offset, see Eqn~\ref{eq:zbest} 
and calculate $z_{\rm best}$, which is plotted in the lower panel.
\label{fig:BPZ_hdfn}}

\includegraphics[width=150mm,height=150mm]{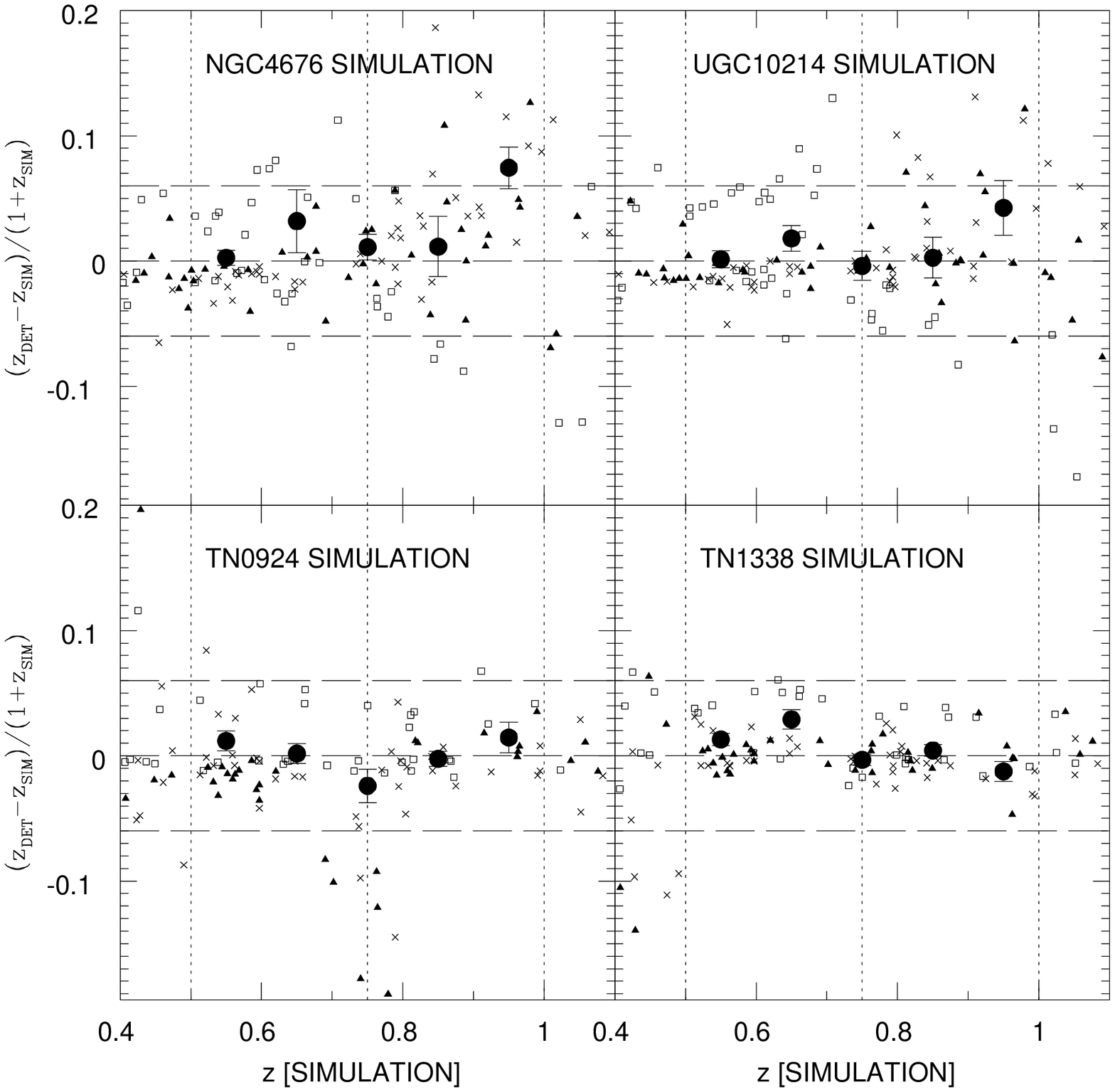}
\figcaption[test_BPZ1.ps]{The 4 panels show the results of the 4 simulations, TN0924, 
TN1338, NGC4676 and UGC10214 as labeled. The y axis is the difference in the 
redshift, $(z_{\rm detection}-z_{\rm simulation})/(1+z_{\rm simulation})$. The 
open squares
represent objects with an `El' SED, the crosses represent objects with a `Sbc' SED and 
the filled triangles represent objects with a `Scd' SED. The circles with 
errorbars represent the $3-\sigma$ clipped mean for the `El' and `Sbc' SEDs. 
The dotted lines mark out the samples at $z=0.5, 0.75, 1.0$. The dashed lines show the 
expected mean and standard deviations based on the measurements against 
spectroscopic data (Ben\'{\i}tez et al. 2004, in preparation). There are no significant 
systematic errors, but galaxies in UGC 10214 and NGC 4676 have large random errors 
for $z>0.85$. \label{fig:sims1}}

\includegraphics[width=150mm,height=150mm]{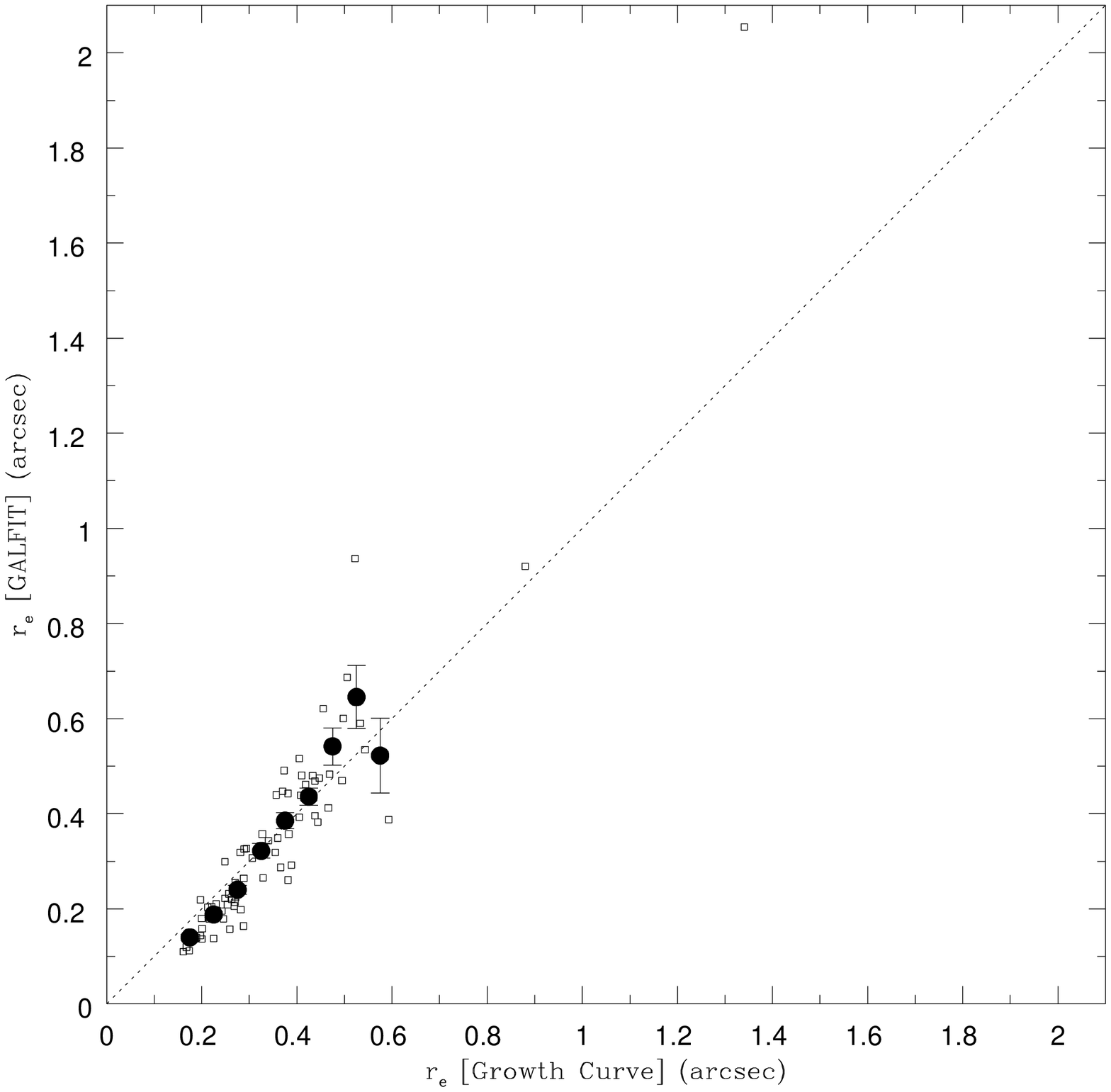}
\figcaption[comp_Re_ell.ps]{This shows a comparison of the half-light radius 
obtained via a growth curve analysis and the 2-D fitting program GALFIT 
(Peng et al. 2002) 
for the ellipticals that we have selected. The PSF correction becomes
important for $r_e<0.4\arcsec$. The typical error is $\Delta\,r_e
\sim0.05\arcsec$ for each galaxy, comparable to the pixel size.
The much larger errors at $r_e=0.55\arcsec$ are largely the result of nearby neighbors that
affect the photometry. All large outliers were checked manually to see if GALFIT or the 
growth curve was the source of error. All necessary changes were made.
\label{fig:recmp}}

\includegraphics[width=150mm,height=150mm]{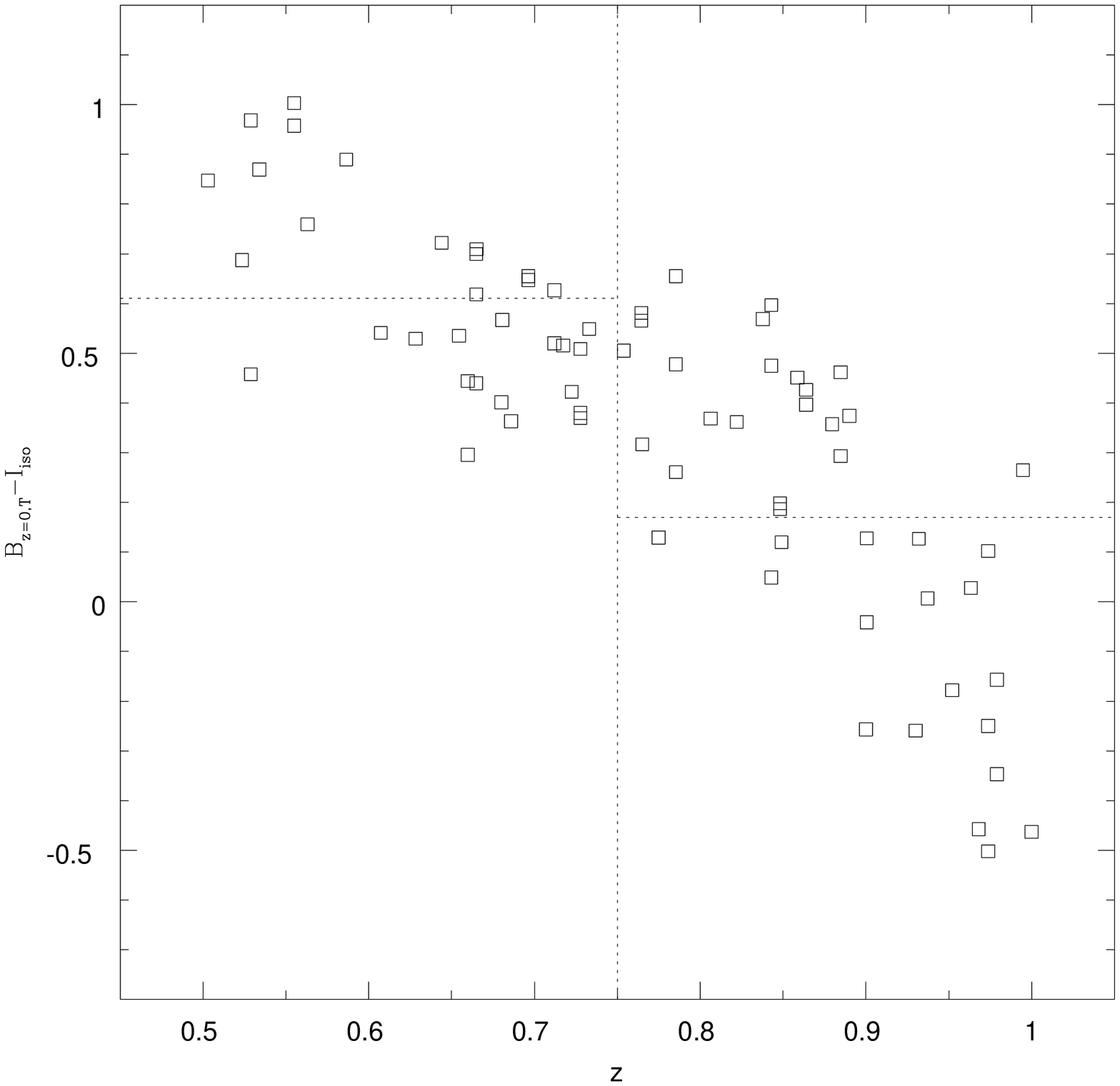}
\figcaption[corr_B_i.ps]{This plot shows the generalized k-correction from the measured 
$I_{\rm F814W}$ or $i_{\rm F775W}$ band magnitude to the total rest-frame $B$ magnitude 
for early type galaxies in our sample. 
Objects in the interval $0.5\le z<0.75$ have an approximate k-correction ($B_{z=0,T}-I_{iso}$) of 0.6 while for objects in the range
$0.75\le z<1.0$, the k-correction is closer to 0.2. \label{fig:BI_z}}

\includegraphics[width=176mm,height=198mm]{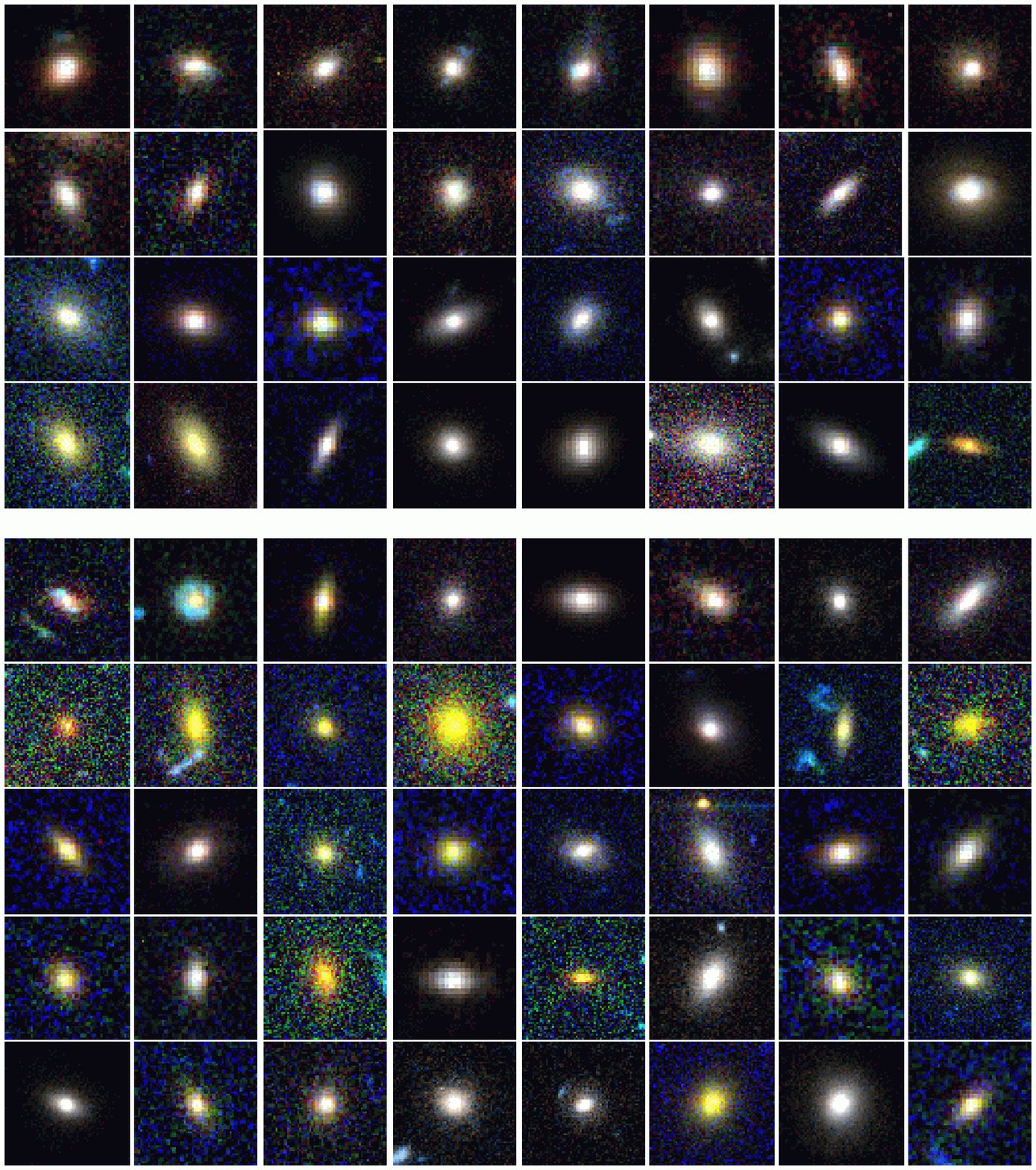}
\figcaption[mkfig3.ps]{This image displays 3-color postage stamps for all
of our galaxies. These use an asinh stretch (Lupton et al. 2004) that 
preserves the colors of bright regions of the galaxy while also showing
the fainter regions of these same objects. They are divided into the two 
redshift samples that we
use thoughout the analysis and then ordered by $(U-V)_0$ color, going from 
bluest (top left) to reddest (bottom right). This is the same order as Table
~\ref{tab:gal_prop}. In the case of the HDFN, we display a combination of the 
ACS i and z only. 
\label{fig:cutouts}}

\includegraphics[width=150mm,height=150mm]{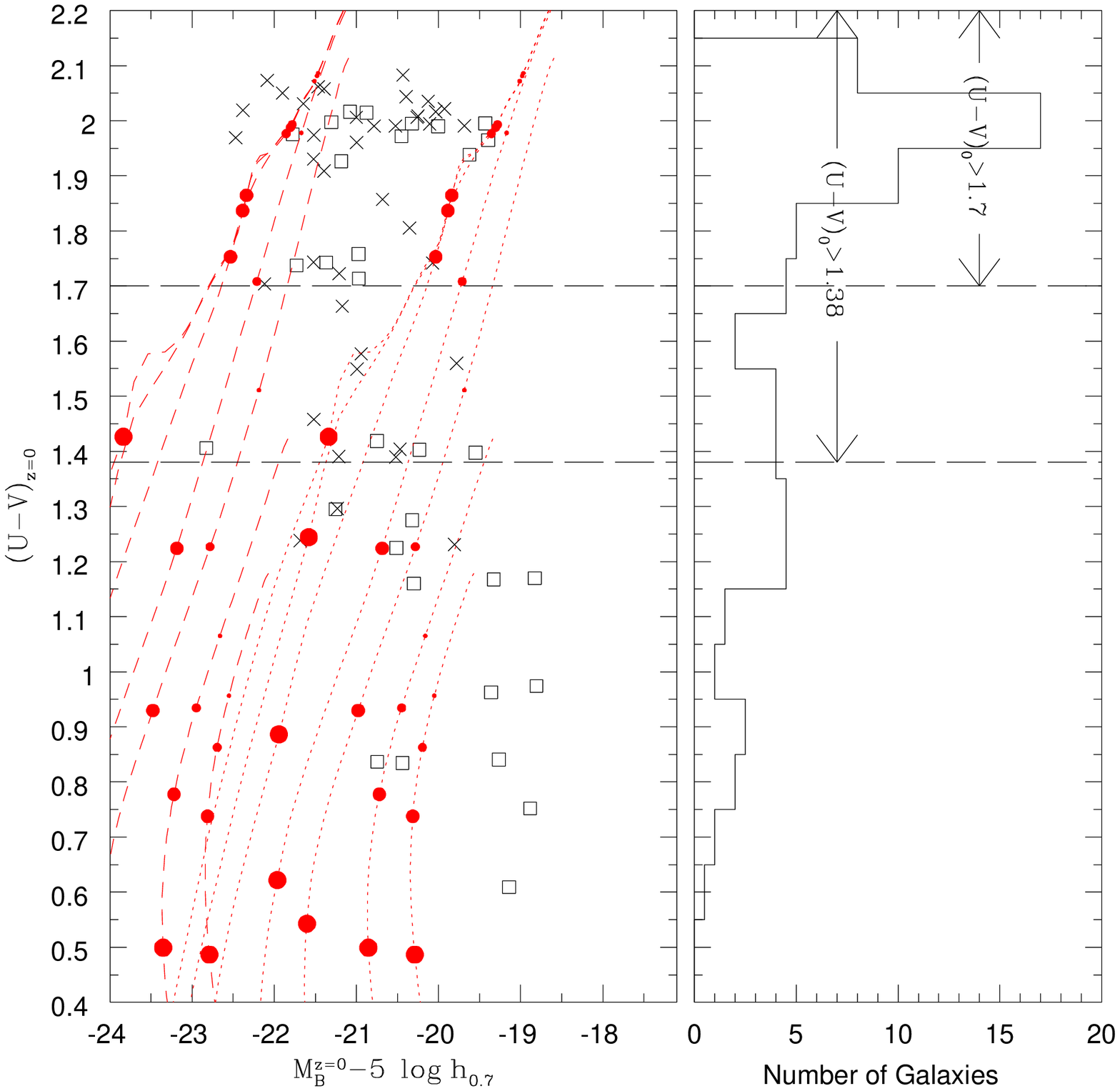}
\figcaption[restframe_color_test3.ps]{The left-hand panel shows the distribution of 
rest-frame $(U-V)_0$ color against $M_B^{z=0}$ for early-type galaxies with $0.5<z\le0.75$ 
(squares) and $0.75<z\le1.0$ (crosses). The dotted lines represent the Bruzual \&
Charlot (2003) evolutionary tracks for a $10^{11}{\cal M_{\odot}}$ galaxy with solar
metallicity and an exponentially decaying star-formation rate with decay timescales
$\tau=0.1, 0.2, 0.4, 1.0, 2.0, 5.0, 9.0$ Gyr, from left to right. The filled circles
represent the age of the galaxy in these models going from 1 Gyr (largest circle) 
to 7 Gyr (smallest circle) in steps of 2 Gyr. The short dashed lines represent the
same tracks for a $10^{12}{\cal M_{\odot}}$ galaxy. The tracks allow us to compare
the masses of blue galaxies and red galaxies. The long dashed lines represent the 
selection criteria used to mimic different color selections employed in the literature.  One subsample has 
$(U-V)_{z=0}>1.38$, to match the CFRS selection and the other has $(U-V)_{z=0}>1.7$ to mimic the selections by the COMBO-17 
and CADIS (Fried et al. 2001) surveys. The right-hand panel shows the histogram
in color for the combined sample. Note the significant number of E/S0s that are bluer than
any of the color selections shown.\label{fig:Mcol}}

\includegraphics[width=150mm,height=150mm]{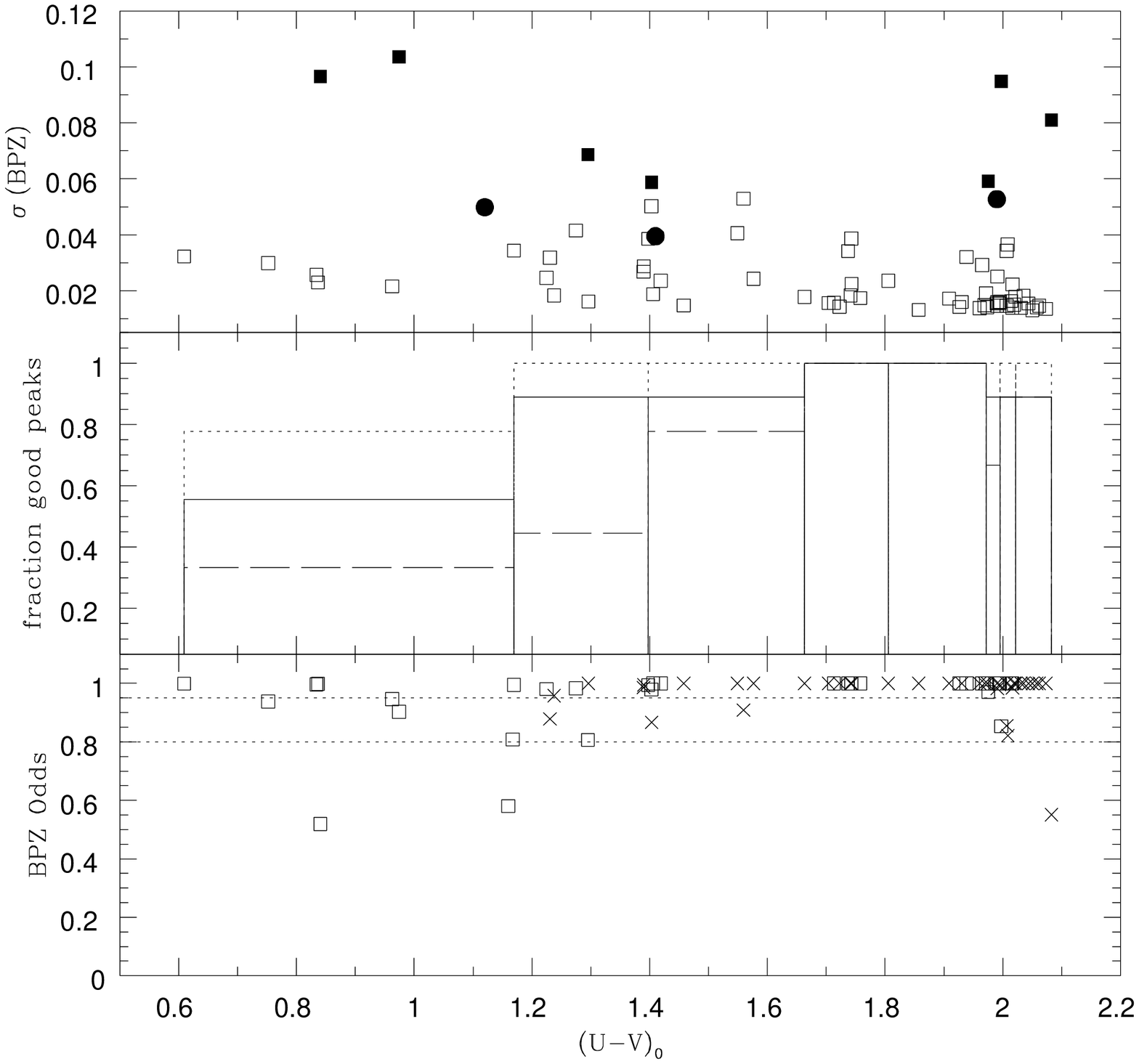}
\figcaption[UV_Odds.ps]{This plot shows how the redshift reliability depends 
on color. The lower panel shows the odds calculated from BPZ 
against the rest-frame $(U-V)_0$ color (see \S4, Eqn~\ref{eq:rfodds} for a 
definition of the odds). The squares represent $z<0.75$ 
galaxies and the crosses represent $z>0.75$ galaxies. This shows that 
$80\%$ of our objects have good odds, and that the reliability of the 
redshift does not vary significantly with color or redshift. The middle
panel splits the distribution into 8 bins of equal number and plots the
number of objects with a single peak in the probability density function
(dashed histogram); the solid histogram represents those with one narrow 
dominant peak (i.e one peak makes up $>90\%$ of the integrated probability)  
and the dotted histogram includes those with multiple overlapping peaks 
that are in effect a wider peak with $>90\%$ of the integrated probability.
While the bluer galaxies have more of the wider peaks objects with a single
dominant peak make up almost $90\%$ of objects for $(U-V)_0>1.2$. In the
top panel we show the widths of the peaks in the pdf. The open squares
show the single narrow peaks, the filled squares show the multiple dominant
peaks and the filled circles show the mean from the simulations for an
elliptical $(U-V)_0=1.99$, an Sbc $(U-V)_0=1.40$ and an Scd $(U-V)_0=1.12$.
\label{fig:UVodds}}

\includegraphics[width=150mm,height=150mm]{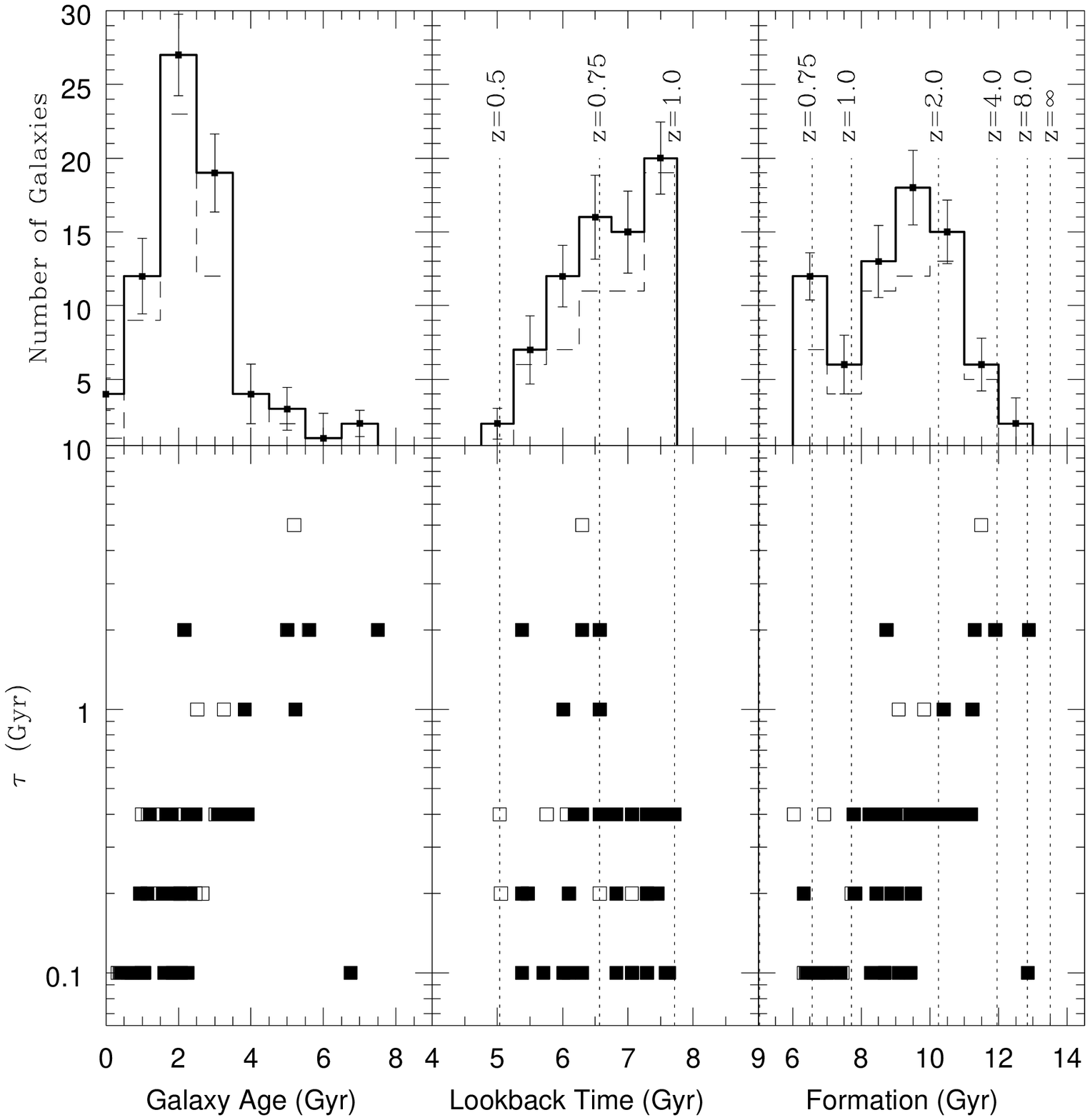}
\figcaption[age_hist.ps]{The age, lookback time and formation time of early-type galaxies 
assuming a star formation model with exponential decay 
timescales $\tau=0.1, 0.2, 0.4, 1.0, 2.0, 5.0, 9.0$ Gyr. The bottom left
hand panel shows the best fit $\tau$ versus the best fit age. The bottom 
middle panel shows the best fit $\tau$ versus the lookback time, and the
bottom right panel shows $\tau$ versus the formation time, which is the sum
of the lookback time and the age. The filled squares represent a volume 
limited sample. The top panels show the histograms of the
age (top left), the lookback time (top middle) and the formation time
(top right). The solid histograms show all the objects and the dashed 
histograms show a volume limited sample. The dotted lines in the lookback
time and formation time plots show the equivalent redshift. \label{fig:Ages}}

\includegraphics[width=150mm,height=150mm]{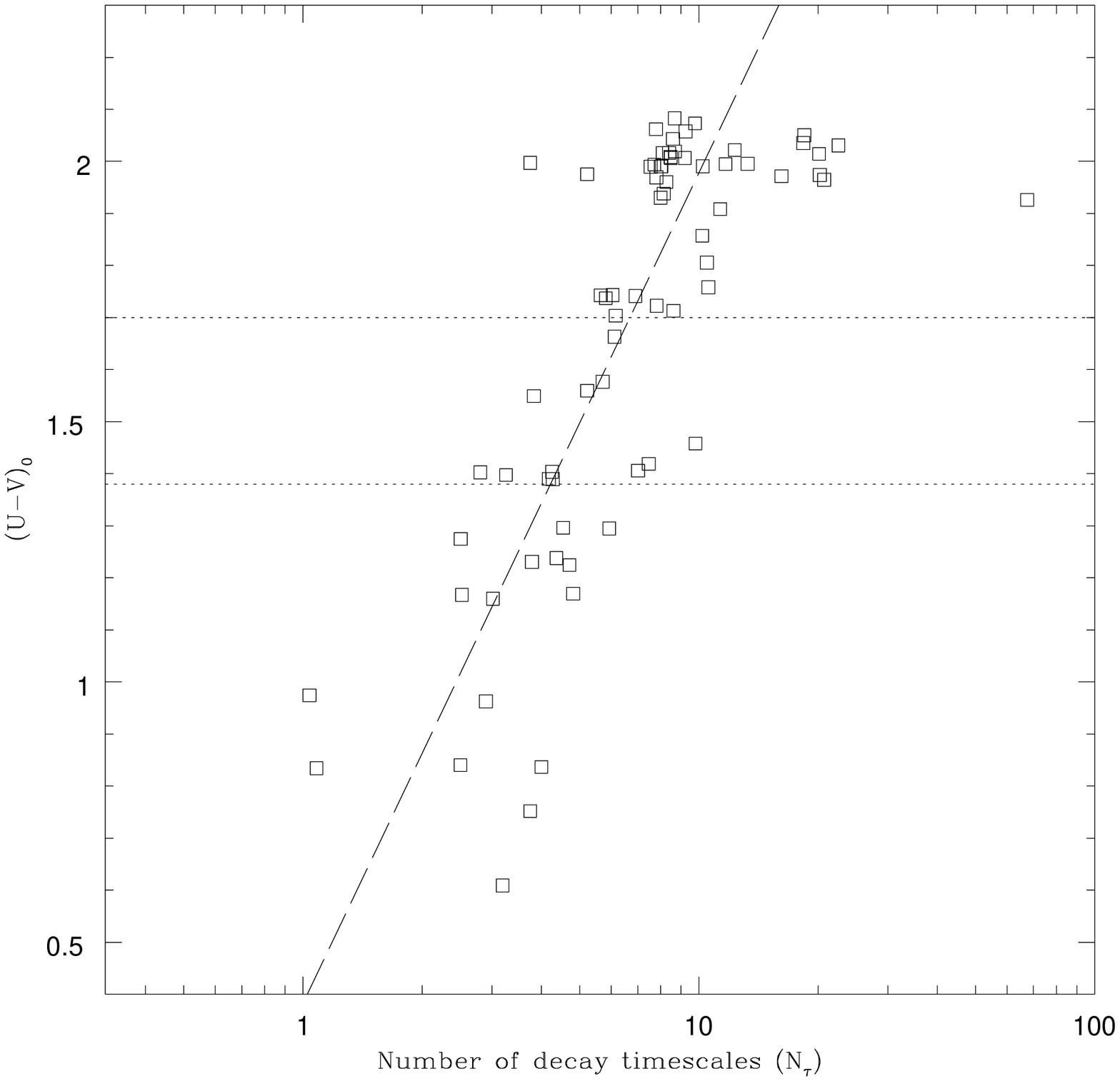}
\figcaption[notmsc_UV.ps]{This plot compares the rest-frame $(U-V)_0$ color to the 
number of decay timescales ($N_{\tau}=T/\tau$) derived from the Bruzual \& Charlot (2003) 
models. For $1<N_{\tau}<10$, there is a strong correlation with 
$(U-V)_{0}=1.59\log\,(N_{\tau})+0.38$. For $N_{\tau}>10$, 
$(U-V)_{0}\sim2.0$, the rest-frame color of the `El' template from 
Ben\'{\i}tez et al. 2004. The long-dashed line indicates the best fit linear 
correlation (from a simple least squares calculation). The dotted lines 
indicate the color cuts for CFRS ($(U-V)_{0}=1.38$) and COMBO-17 and 
CADIS ($(U-V)_{0}=1.7$).\label{fig:BCUV}}

\includegraphics[width=150mm,height=150mm]{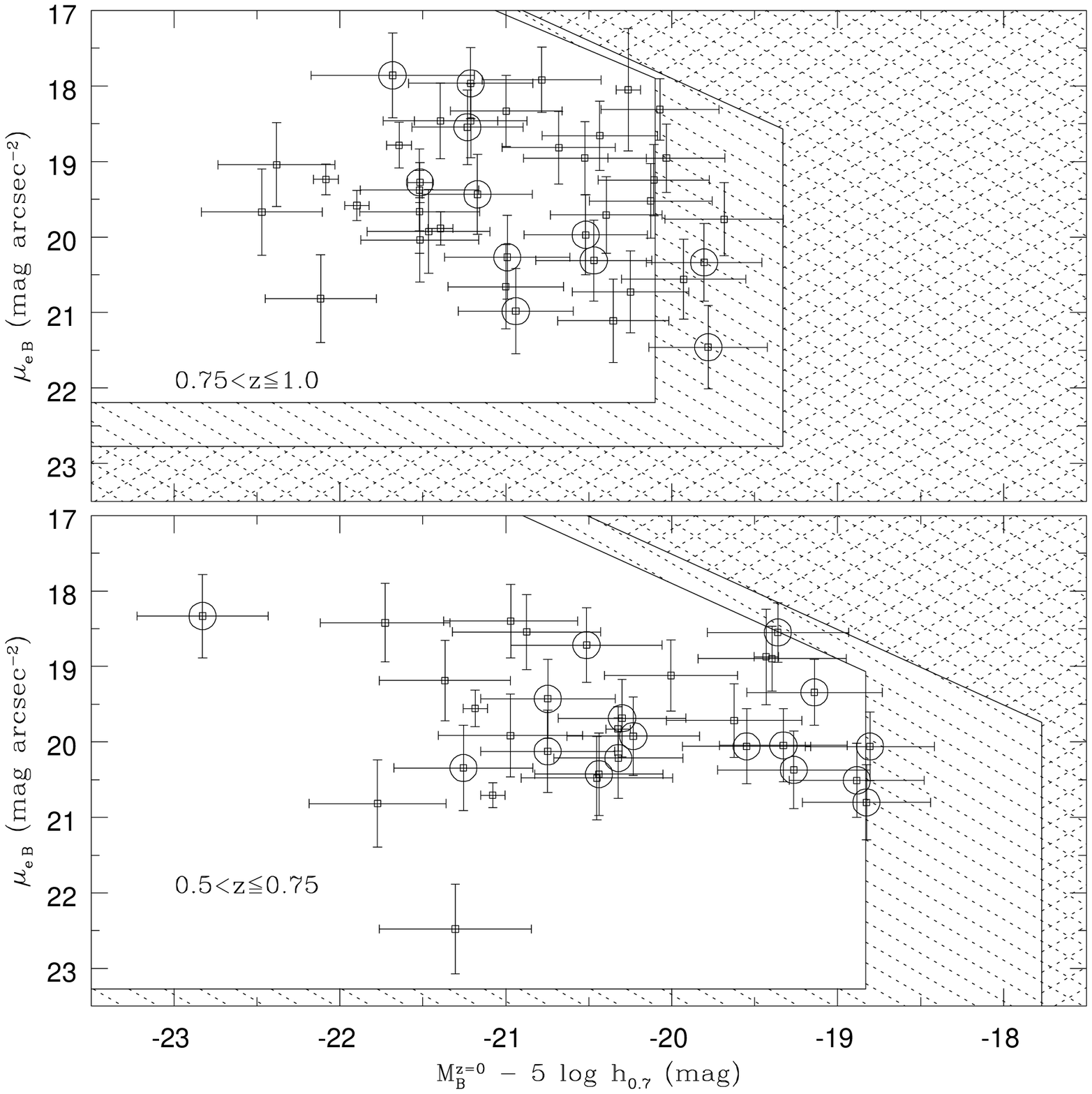}
\figcaption[BBD_both.ps]{The distribution of galaxies in the absolute 
magnitude and surface brightness plane. The size and magnitude limits are 
shown at
both the low and high redshift end of each sample. All objects in the
unshaded area are seen over the same volume. All objects in the 
cross-hatched area are outside the limits of the survey. The singly
shaded region 
denotes parameter space where galaxies cannot be seen to the maximum 
redshift. The lower plot shows the $0.5<z\le\,0.75$ sample 
and the upper plot shows the $0.75<z\le\,1.0$ sample. The blue E/S0s
are circled. They are not separated from the red E/S0s in $M_B$, $\mu_e$ 
space.\label{fig:BBD}}

\includegraphics[width=150mm,height=150mm]{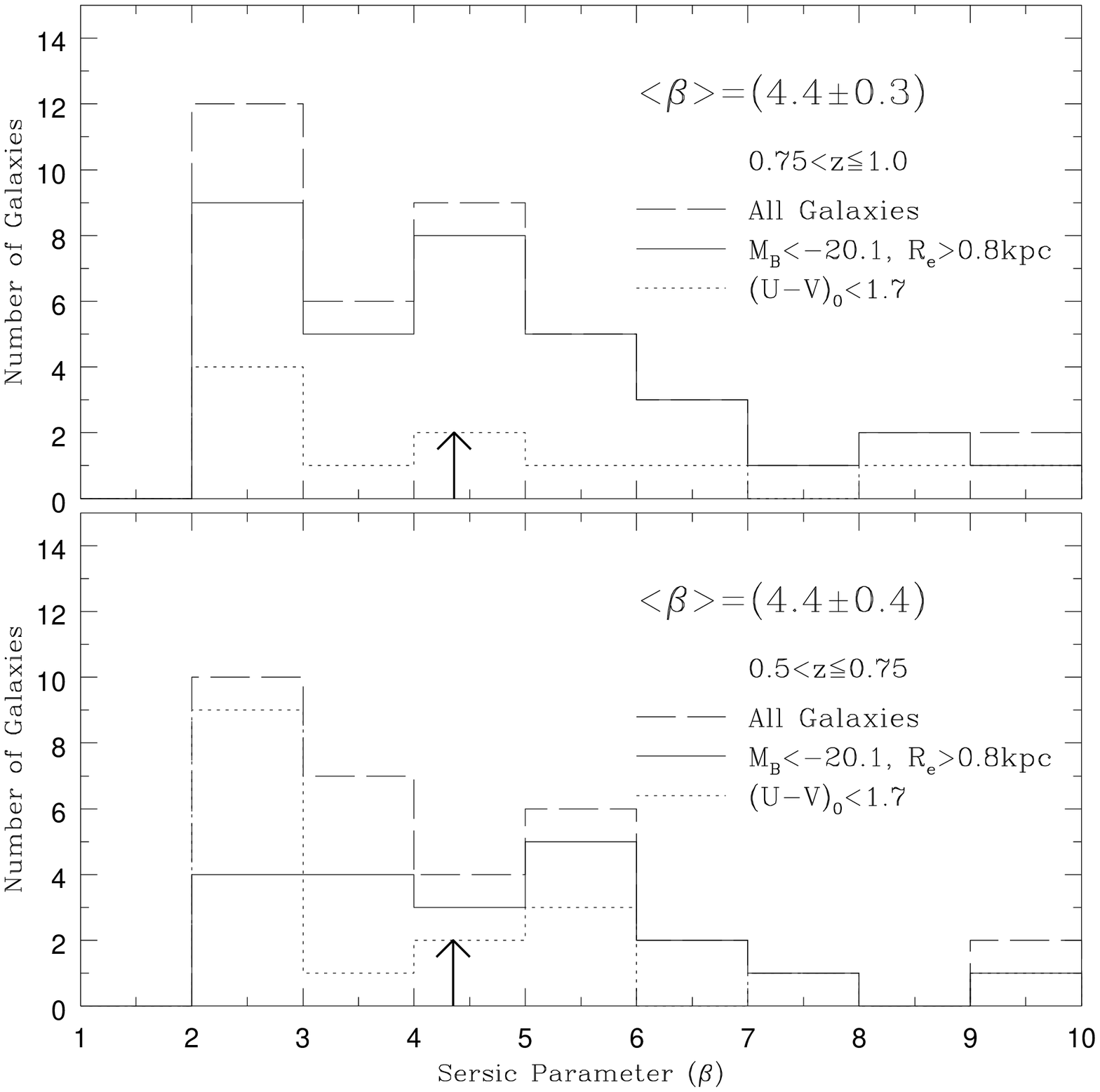}
\figcaption[hist_beta.ps]{We plot the distribution in the Sersic parameter $\beta$ 
for each sample (dashed histogram). The solid histogram in each panel is the
distribution for equivalent volume limited samples. The distribution of $\beta$ does 
not change significantly with redshift. The dotted histogram is the distribution 
of the `blue' E/S0s. The thick arrow in each panel marks the mean 
$\beta$.\label{fig:histbeta}}

\includegraphics[width=150mm,height=150mm]{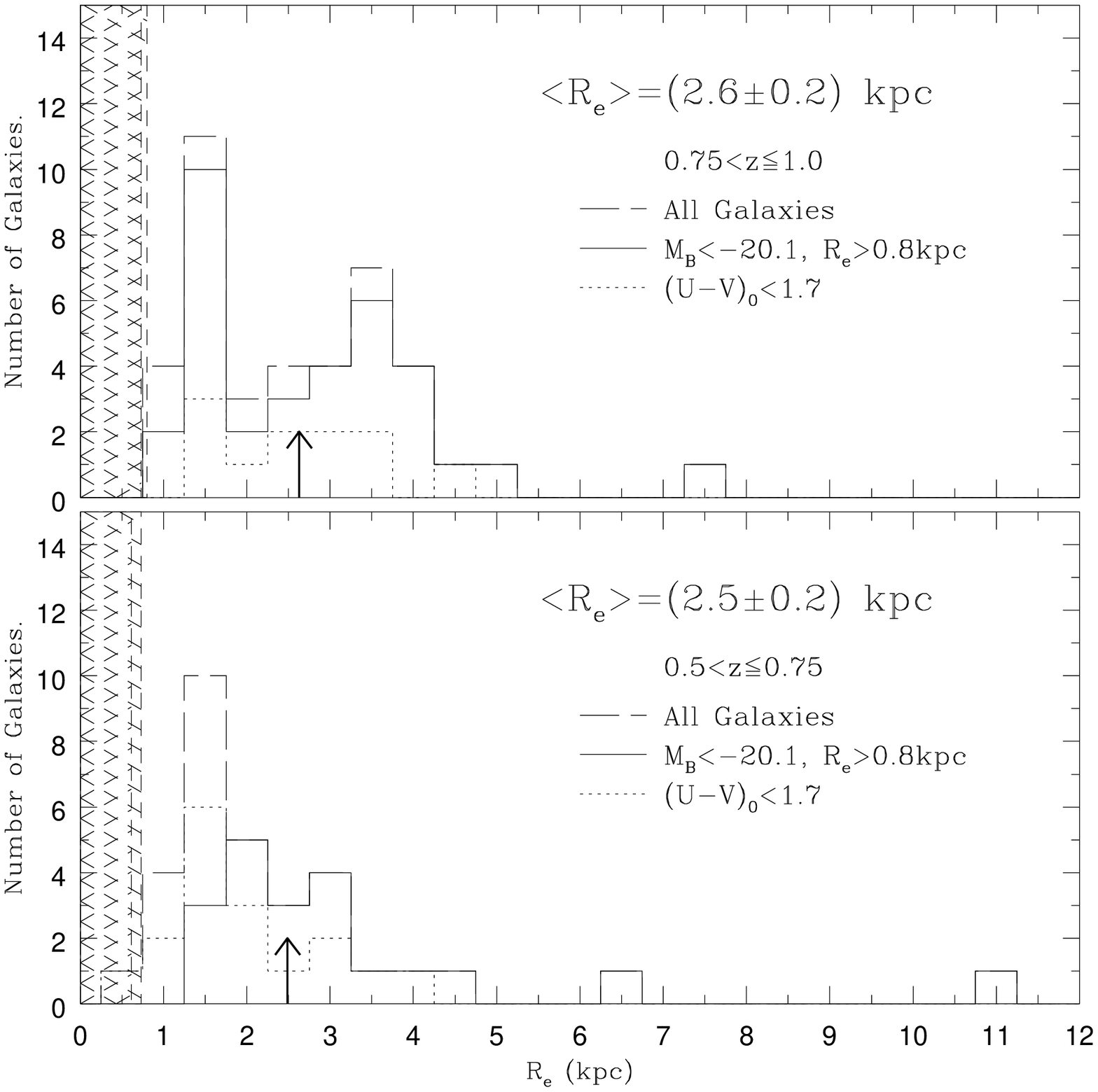}
\figcaption[hist_beta.ps]{The Sersic parameter $\beta$ distribution for each 
sample (dashed histogram). The solid histogram is the distribution of the 
volume limited samples and the dotted histogram is the distribution of
the `blue' E/S0s. The thick arrow shows the mean value of $R_e$.The shading
represents the selection limits in half light radius: the single shading begins
at the high-redshift limit of the sample and the cross-hatching begins at
the low-redshift limit.\label{fig:histRe}}

\includegraphics[width=150mm,height=150mm]{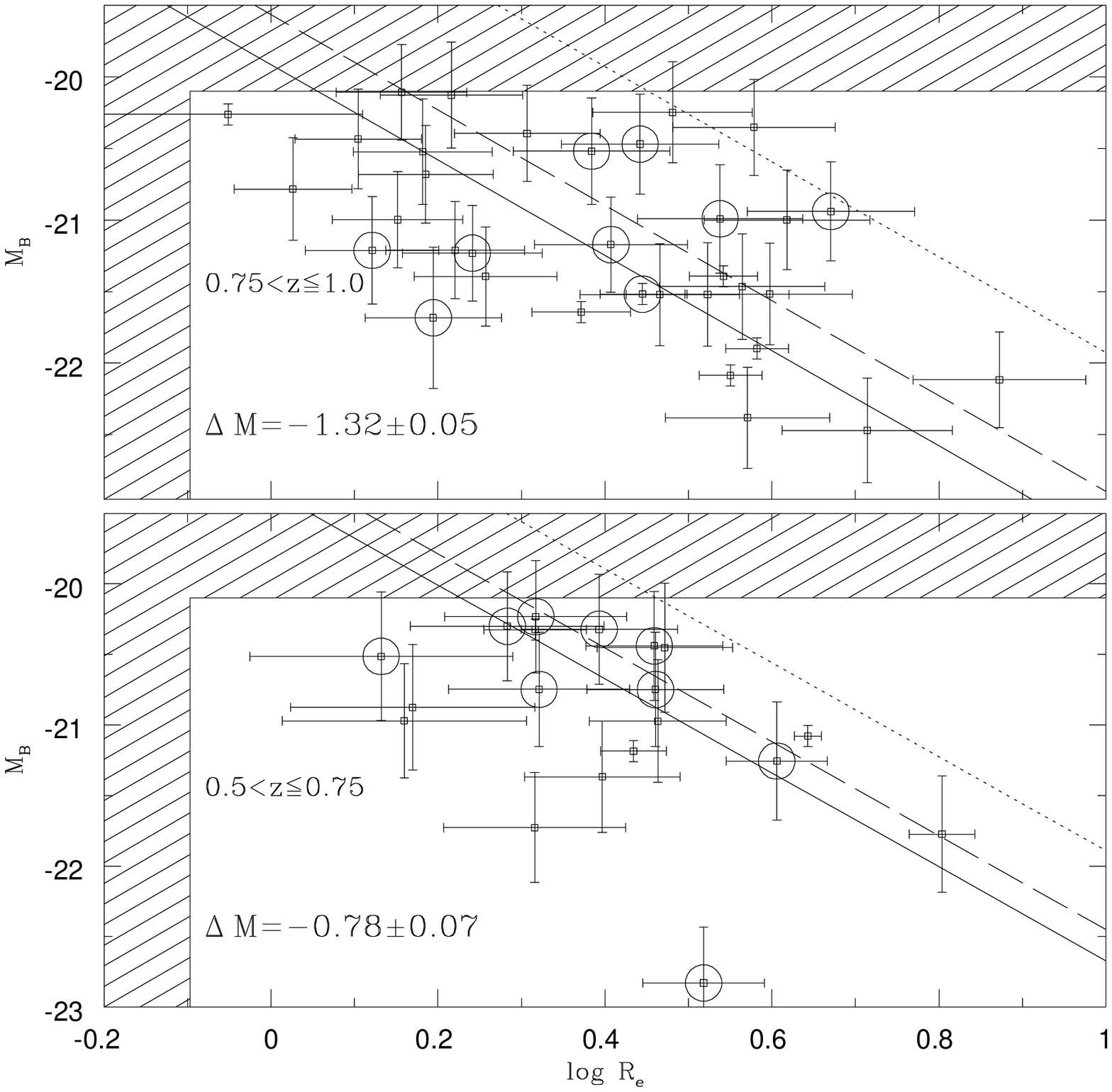}
\figcaption[Re_M.ps]{This plot shows the distribution of absolute magnitude 
($M_B$) versus the logarithm of half-light radius for both E/S0 subsamples 
considered in this study. The light shading represents the parts of 
the parameter space where galaxies cannot be seen out to the 
maximum redshift, and the heavy shading represents the parameter space where 
galaxies cannot be seen at all. The dotted line is the $z=0$ relationship 
from Schade et al. (1997) corrected to our cosmology. The dashed line is the 
expected fit from  Schade et al. (1999) for each sample and the solid line is our best 
fit in the volume limited region. The circles mark the positions of blue early-type 
galaxies. Neither type of galaxy shows strong evidence for a size-magnitude 
relationship. \label{fig:Re_M}}

\includegraphics[width=150mm,height=150mm]{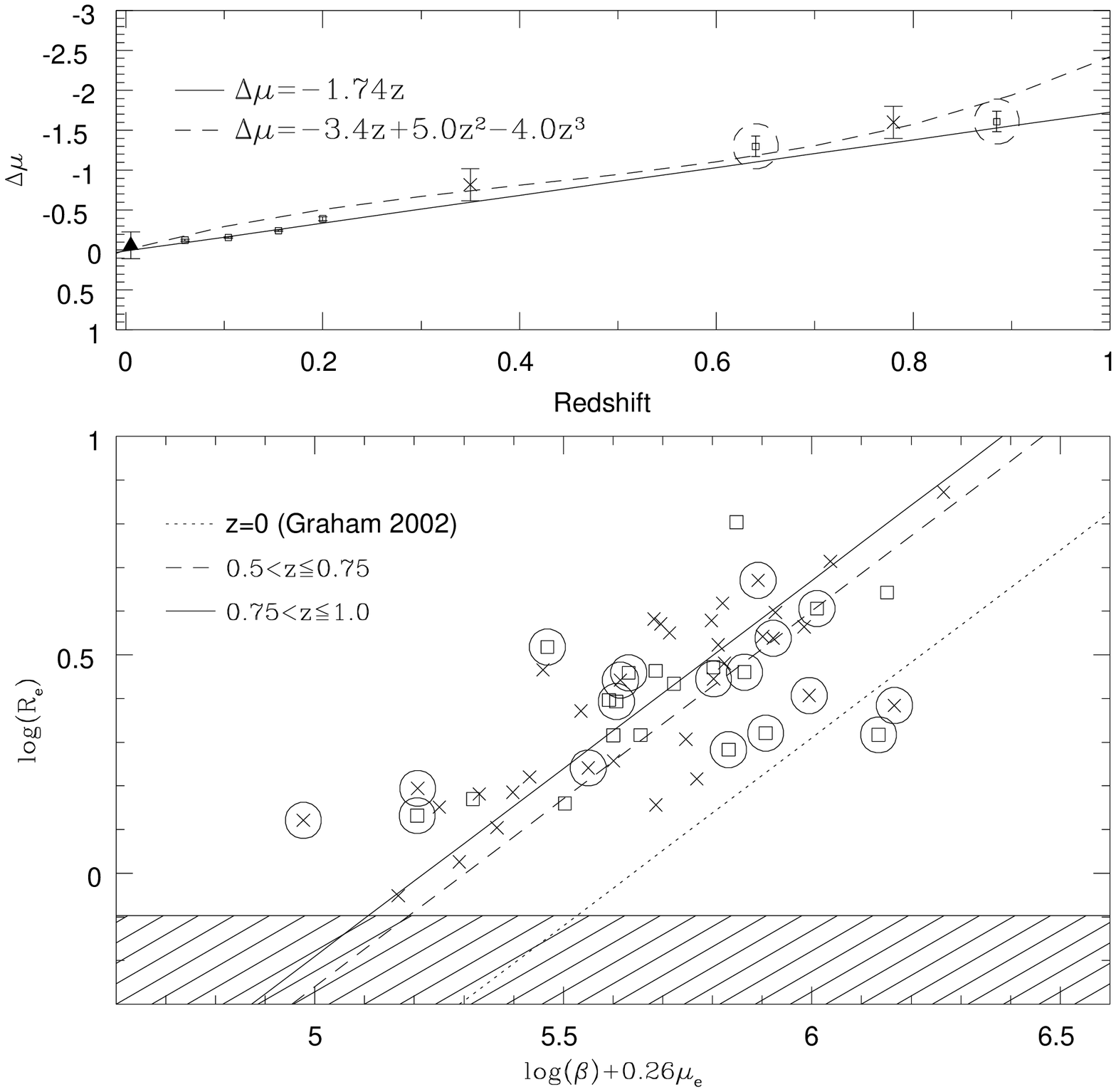}
\figcaption[nmu_re.ps]{The lower panel of this plot shows the photometric plane for 
elliptical galaxies in our fields. The squares denote the $0.5<z\le0.75$ 
sample and the crosses represent the $0.75<z\le1.0$ sample. Our best fit 
lines are
the solid one for $0.75<z\le1.0$ and the long-dashed for $0.5<z\le0.75$.
The $z=0$ fit from Graham (2002) is shown by the short dashed line. The
blue galaxies are marked by circles. We find good fits to the photometric plane 
although there are a few outliers amongst the blue E/S0 galaxies. 
In the top panel, we show the variation in surface-brightness with redshift,
calculated from this plot. Our points are marked by the squares ringed by 
large circles. The Schade et al. (1999) results are marked by crosses,
the Bernardi et al. (2003) results are marked by squares and the Graham 
(2002) result is marked with the triangle. The solid line shows our best fit 
to these results, and the dashed line is the Gebhardt et al. (2003) best 
fit.\label{fig:photplan}}

\includegraphics[width=150mm,height=150mm]{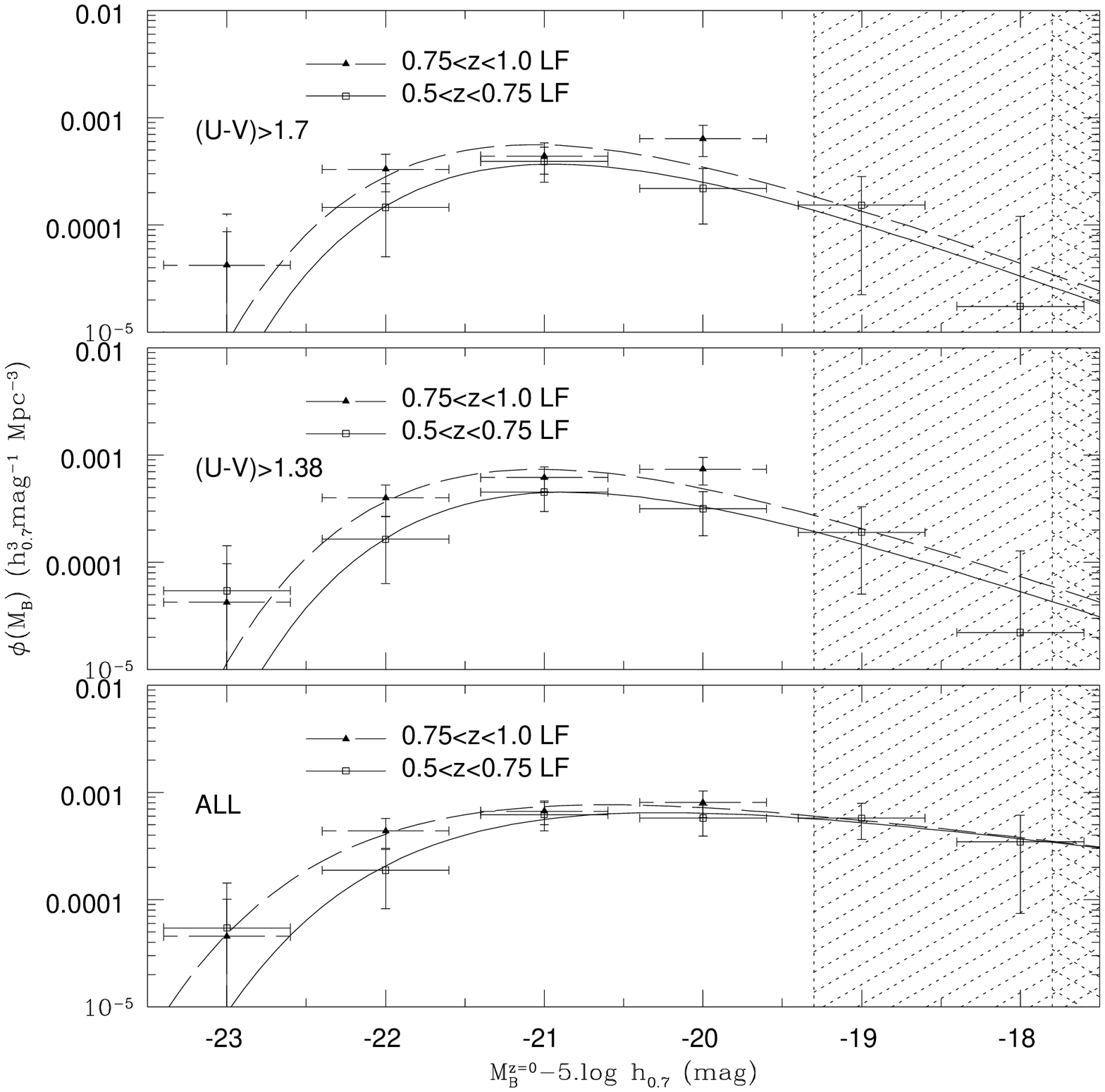}
\figcaption[lf_sim_ACS.ps]{The luminosity functions of the $0.5<z<0.75$ (squares with solid 
error bars) and the $0.75<z<1.0$ (triangles dashed error bars) samples. The bright ends are 
normalized to the volume limited samples and the lines show the Schechter function fits to each set 
of points. The single shaded hatching denotes the magnitude limit of the $0.75<z<1.0$ sample and the 
criss-cross hatching denotes the magnitude limit of the $0.5<z<0.75$ sample. The lower panel shows the 
morphologically selected samples, the middle panel shows the two $(U-V)_0>1.38$ color selected samples
and the upper panel shows the two $(U-V)_0>1.7$ color selected samples.\label{fig:LF}}

\includegraphics[width=150mm,height=150mm]{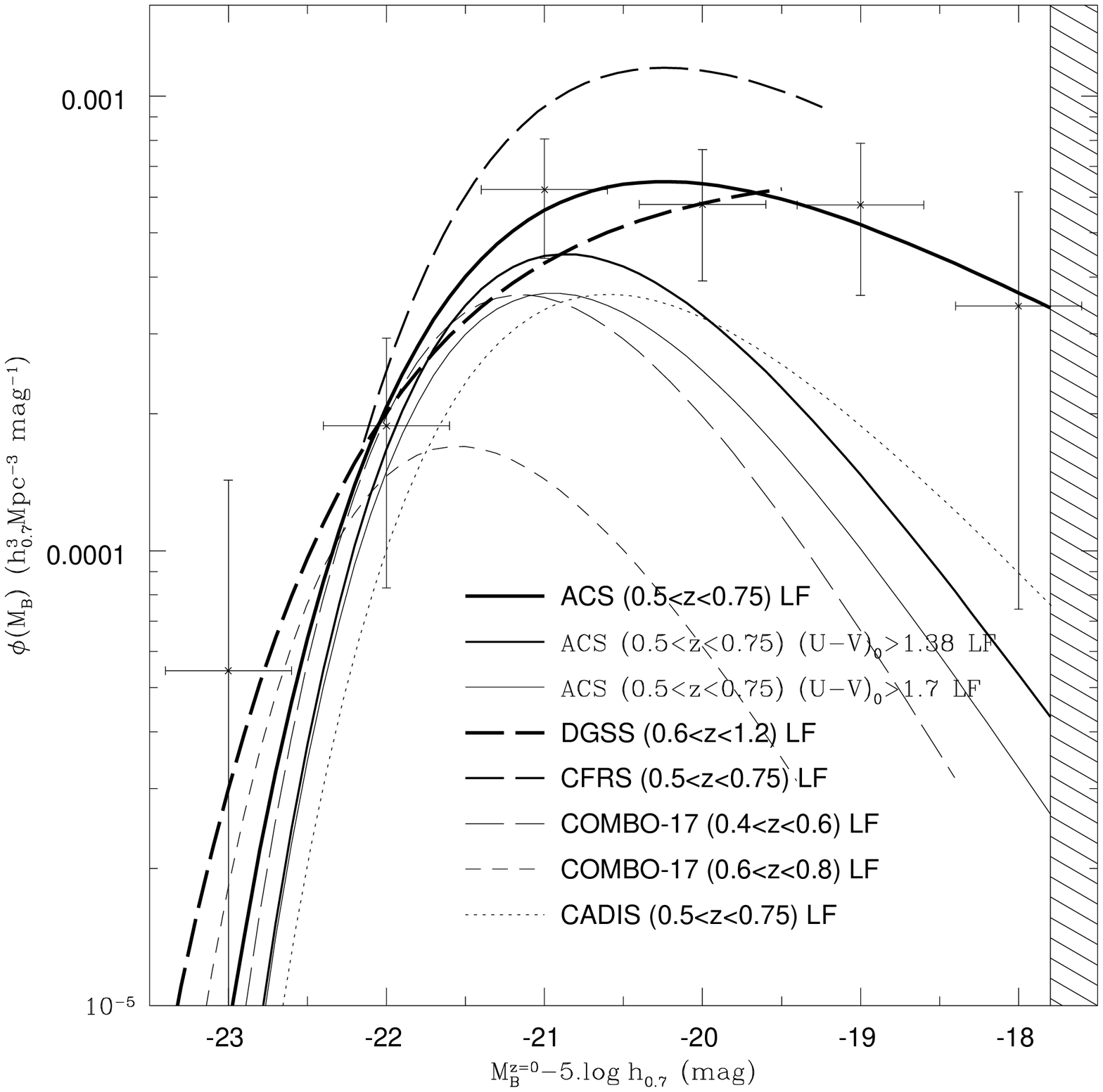}
\figcaption[lf_0.5z0.75.ps]{The luminosity functions of our $0.5<z<0.75$ 
early-types
compared to that from previous surveys. The ACS LFs are plotted with solid lines, with
the thickest showing the morphologically selected LF, the medium thick 
showing the $(U-V)_0>1.38$ LF and the thin line the $(U-V)_0>1.7$ LF. The 
points and errorbars are for the morphologically selected sample. The 
thick dashed line shows the morphologically selected DGSS LF, the medium 
thick dashed line shows the $(U-V)_0>1.38$ selected CFRS LF and the thin 
dotted or dashed lines show the SED selected COMBO-17 and CADIS LFs. 
All the luminosity functions have been converted to a $\Lambda$-CDM cosmology
with $H_0=70$ km s$^{-1}$ Mpc$^{-1}$.\label{fig:LF_0.5z0.75}}

\includegraphics[width=150mm,height=150mm]{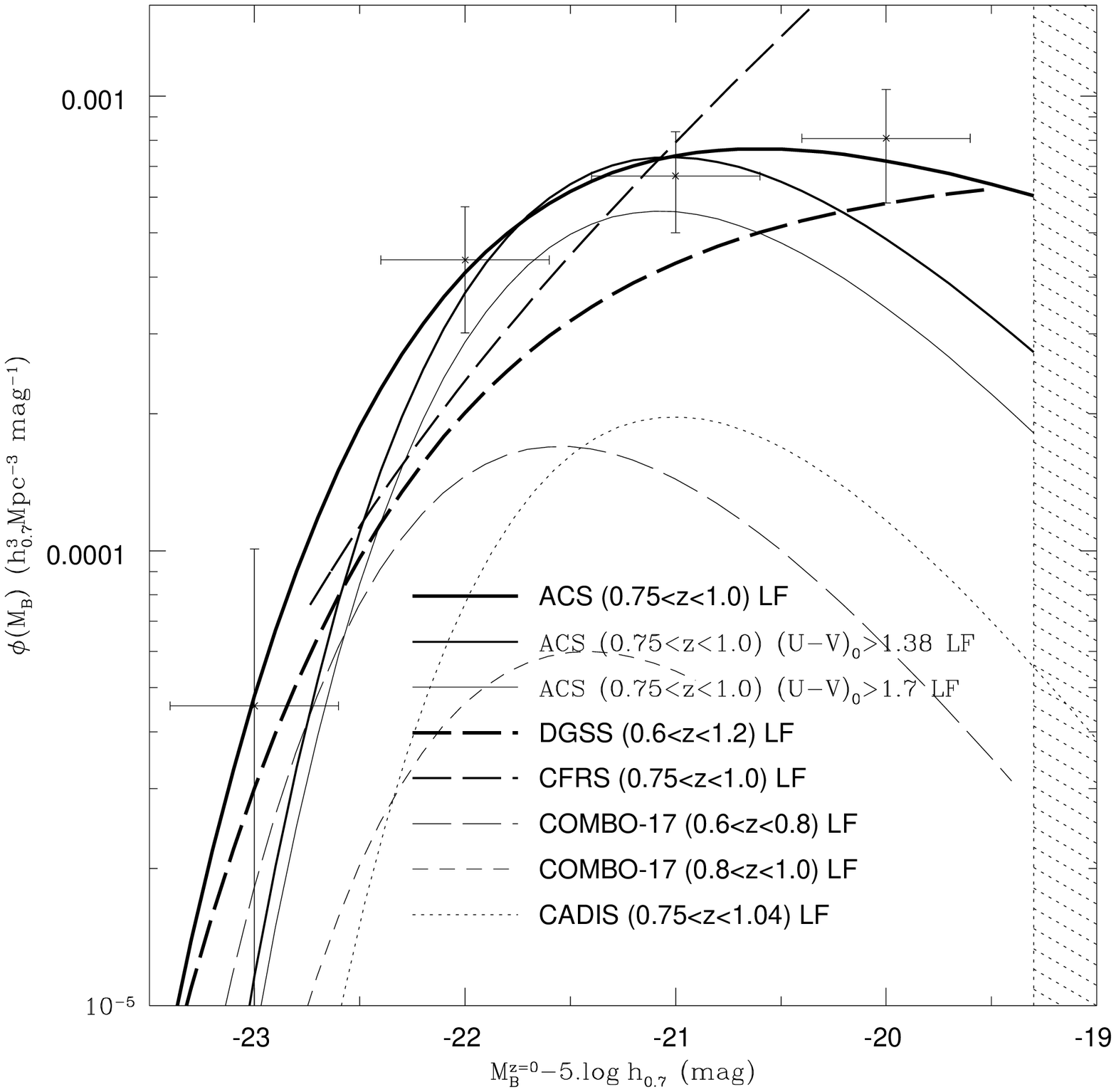}
\figcaption[lf_0.75z1.0.ps]{The luminosity functions of our $0.75<z<1.0$ early-types
compared to that from previous surveys. Otherwise, the same as Fig.~\ref{fig:LF_0.5z0.75}.\label{fig:LF_0.75z1.0}}

\includegraphics[width=150mm,height=150mm]{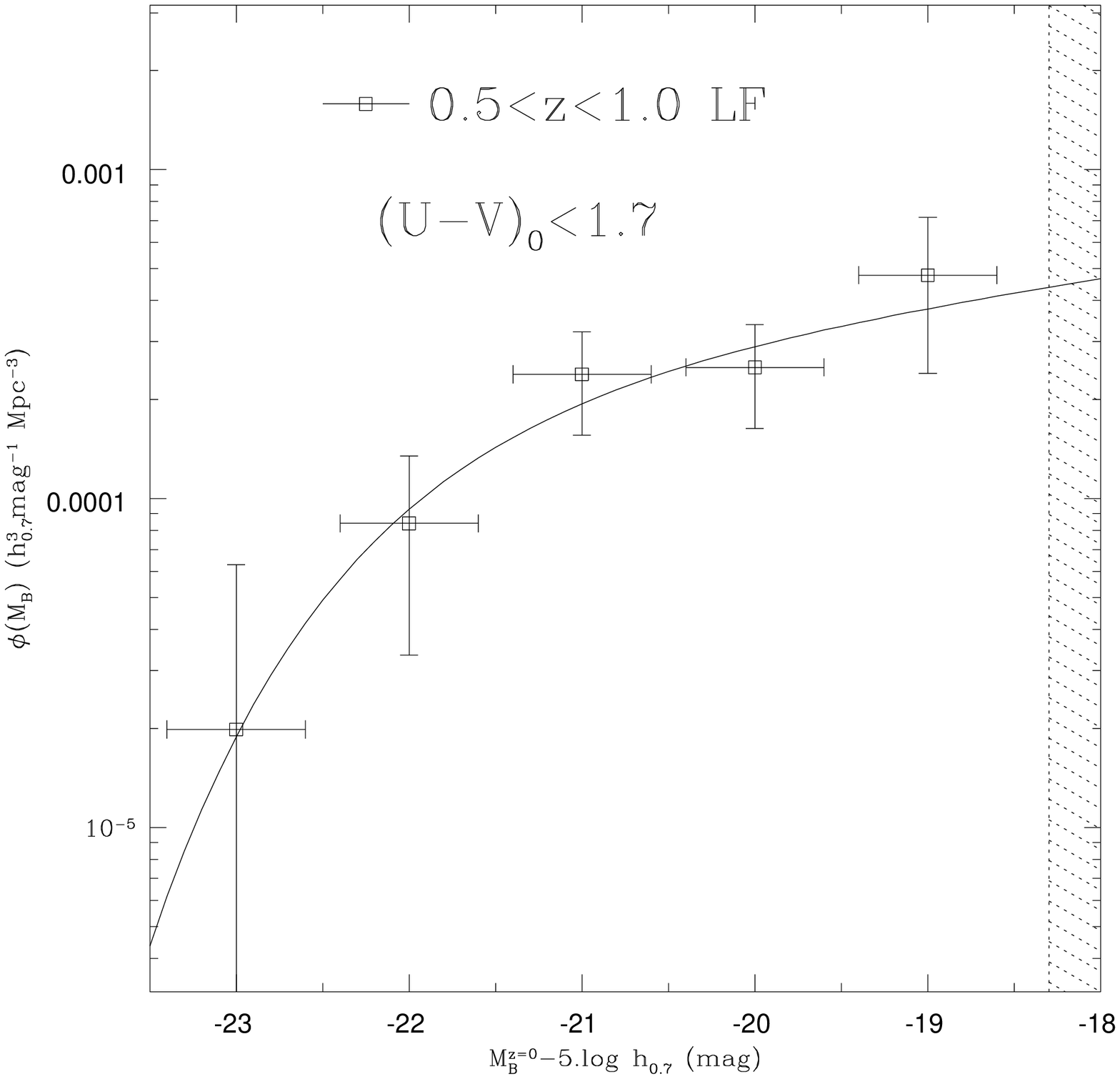}
\figcaption[lf_blue.ps]{The luminosity functions of our $(U-V)_0<1.7$ galaxies. The 
$0.5<z<1.0$ LF is shown by square points with the solid line representing
the best fit Schechter function, see Table~\ref{tab:lf}.\label{fig:LF_blue}}

\includegraphics[width=150mm,height=150mm]{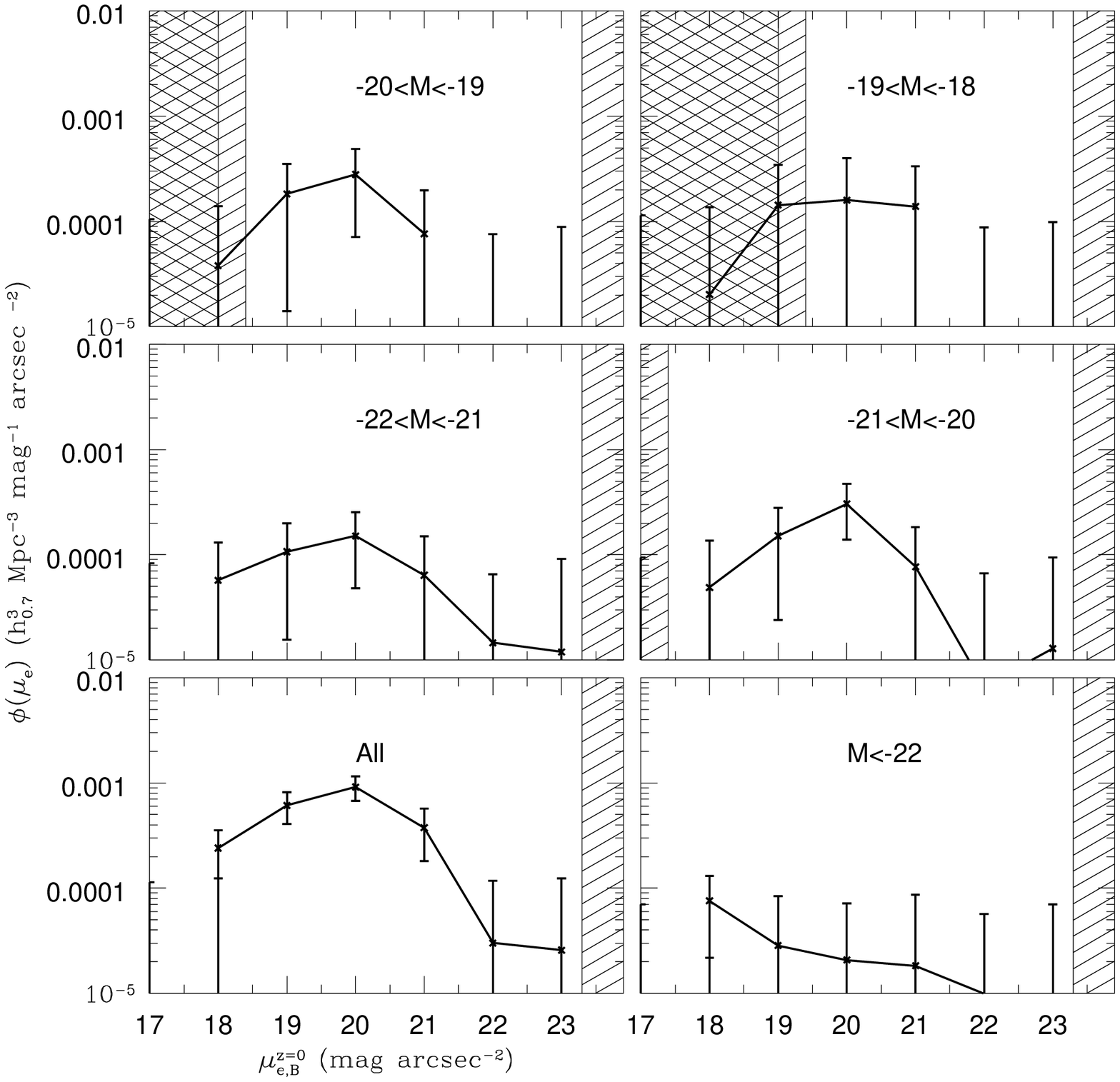}
\figcaption[sbd_sim_0.5z0.75.ps]{This plot shows the surface brightness distributions as a function
of absolute magnitude for the $0.5<z<0.75$ sample. The bottom left hand plot
shows the surface brightness distribution summed to $M_B=-18$, close to the
magnitude limit of the survey. The other 5 plots show the surface brightness
distribution in small ranges of absolute magnitude. The shading represents
the limits at the midpoint of all these magnitude ranges. The light shading 
is the region where the sample volume decreases from the maximum and the
dark shading shows where there sample volume is zero.\label{fig:Sbd1}}

\includegraphics[width=150mm,height=150mm]{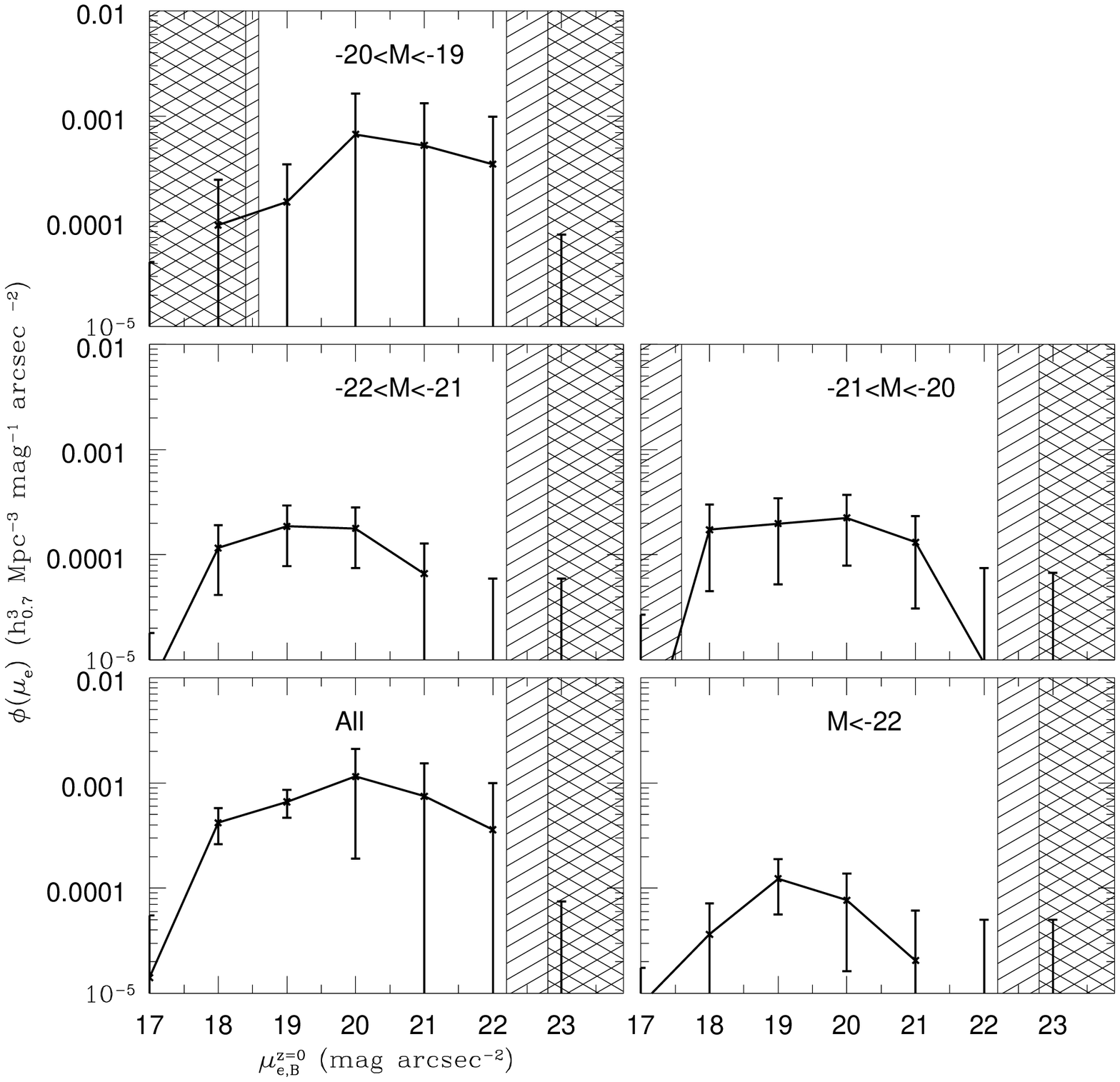}
\figcaption[sbd_sim_0.75z1.0.ps]{This plot shows the surface brightness 
distributions as a function
of absolute magnitude for the $0.75<z<1.0$ sample. The bottom left hand plot
shows the surface brightness distribution summed to $M_B=-19$. Otherwise, 
the same as Fig.~\ref{fig:Sbd1}.
\label{fig:Sbd2}}

\end{document}